\documentclass[prb,twocolumn,aps,floatfix,longbibliography,superscriptaddress]{revtex4-2}

\usepackage[utf8]{inputenc}
\usepackage{natbib}
\usepackage[normalem]{ulem}
\usepackage{graphicx}
\usepackage{xcolor}

\usepackage[tbtags]{amsmath}
\usepackage{physics}
\usepackage{hyperref}
\hypersetup{colorlinks,allcolors=black}
\usepackage{amssymb}
\usepackage{gensymb}
\usepackage{float}
\usepackage{amsmath}
\usepackage{tabularx,graphicx}
\usepackage{epstopdf}
\usepackage{latexsym}
\usepackage{color, colortbl}
\usepackage{psfrag}
\usepackage{bbm}
\usepackage{bm}
\usepackage{titlesec}
\usepackage{dsfont}
\usepackage{feynmp}
\usepackage{slashed}
\usepackage{multirow}
\usepackage{soul} 
\graphicspath{{image/}}
\usepackage{svg}
\usepackage[utf8]{inputenc}

\newcommand{\be}{\begin{equation}}
\newcommand{\ee}{\end{equation}}
\newcommand{\beq}{\begin{eqnarray}}
\newcommand{\eeq}{\end{eqnarray}}
\newcommand{\ba}{\[\begin{aligned}}
\newcommand{\ea}{\end{aligned}\]}

\newcommand{\p}{\partial}

\renewcommand{\hat}[1]{{\bf {\widehat #1}}}
\renewcommand{\phi}{\varphi}
\renewcommand{\epsilon}{\varepsilon}

\hypersetup{colorlinks=true,linkcolor=blue,citecolor=blue,urlcolor=blue}

\begin{document}

\title{Enhancing Shift Current via Virtual Multiband Transitions}
\author{Sihan Chen}
\email[]{sihanc@uchicago.edu}
\affiliation{Kadanoff Center for Theoretical Physics, University of Chicago, Chicago, Illinois 60637, USA}
\affiliation{Department of Physics, California Institute of Technology, Pasadena CA 91125, USA}

\author{Swati Chaudhary}
\affiliation{Department of Physics, The University of Texas at Austin, Austin, Texas 78712, USA}
\affiliation{Department of Physics, Northeastern University, Boston, Massachusetts 02115, USA}
\affiliation{Department of Physics, Massachusetts Institute of Technology, Cambridge, Massachusetts 02139, USA}
\author{Gil Refael}
\affiliation{Department of Physics, California Institute of Technology, Pasadena CA 91125, USA}
\affiliation{Institute for Quantum Information and Matter, California Institute of Technology, Pasadena CA 91125, USA}

\author{Cyprian Lewandowski}
\email[]{clewandowski@magnet.fsu.edu}
\affiliation{National High Magnetic Field Laboratory, Tallahassee, Florida, 32310, USA}
\affiliation{Department of Physics, Florida State University, Tallahassee, Florida 32306, USA}

\begin{abstract}
    Finding materials exhibiting substantial shift current holds the potential for designing shift current-based photovoltaics that outperform conventional solar cells. However, the myriad of factors governing shift current response poses significant challenges in designing devices that showcase large shift current.
    Here, we propose a general design principle that exploits inter-orbital mixing to excite virtual multiband transitions in materials with multiple flat bands to achieve enhanced shift current response. We further explicitly relate this design principle to maximizing Wannier function spread as expressed through the formalism of quantum geometry. We demonstrate the viability of our design using a 1D stacked Rice-Mele model. Then, we consider a concrete material realization - alternating angle twisted multilayer graphene (TMG) - a natural platform to experimentally realize such an effect. We identify a new set of twist angles at which the shift current response is maximized via virtual transitions for each multilayer graphene and highlight the importance of TMG as a promising material to achieve an enhanced shift current response at terahertz frequencies. Our proposed mechanism also applies to other 2D systems and can serve as a guiding principle for designing multiband systems that exhibit enhanced shift current response. 
\end{abstract}

\maketitle

\section{Introduction}
 
The bulk photovoltaic effect (BPVE) is a promising alternative source of photocurrent to conventional p-n junction based photovoltaics~\cite{Dai_Rappe_2023,sipe2000,Baltz1981,fridkin2001bulk,Nastos2006,pusch2023,Nagaosa2022}.   One of the microscopic mechanisms behind the BPVE is the shift current, which generates direct current upon electromagnetic radiation in noncentrosymmetric materials~\cite{sipe2000,Baltz1981}. Intuitively, the physical origins of shift current trace back to a real-space shift experienced by an electron wavepacket upon an interband excitation driven by linearly polarized light~\cite{Fregoso2017,wang2019}. Shift current is a second-order optical response whose contribution arises from direct optical transitions and transitions via a virtual state~\cite{sipe2000,cook2017design,Parker19,morimoto2016topological}. Multiple factors like interband velocity matrix elements, density of states, number of bands, band gap as well as the quantum geometry\cite{sipe2000,orenstein2021topology,morimoto2016topological,ma2021topology,ahn2022riemannian,morimoto2023geometric,wang2022generalized} determine the magnitude of the shift current response. The goal of establishing guiding principles to maximize the shift current response is an ongoing problem with direct technological ramifications~\cite{cook2017design,wang2016substantial,dong2023giant,grinberg2013perovskite,dong2023giant}.

\begin{figure}
    \centering
    \includegraphics[width=\linewidth]{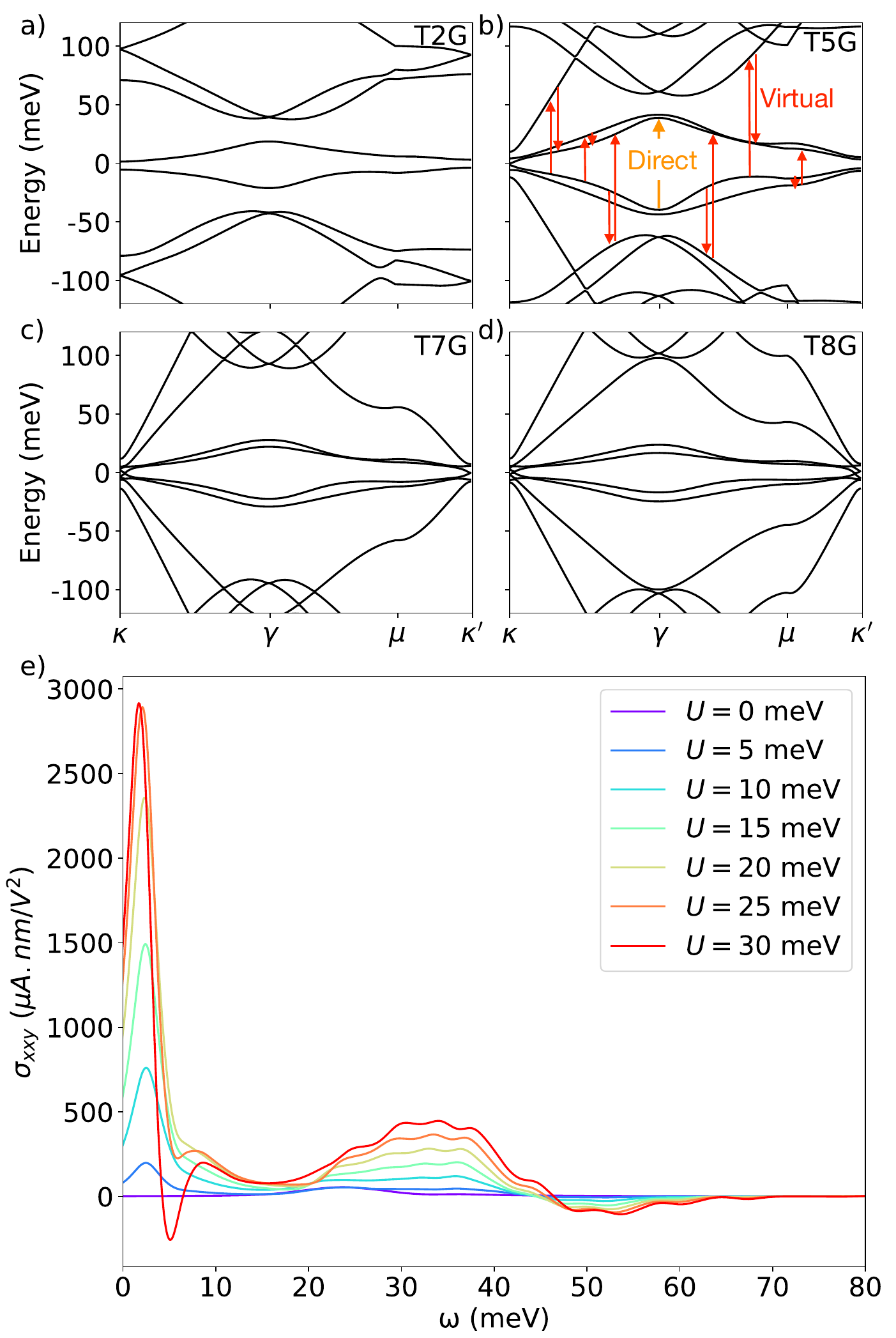}
    \caption{Bandstructure of T2G, T5G, T7G, T8G, and shift current conductivity of T5G with varying external displacement field strength. a-d) Band structure of TMG system with external displacement field strength $U=20$ meV at physical twist angles $\theta=0.8^\circ, 1.32^\circ, 1.63^\circ, 1.71^\circ$ for $N=2,5,7,8$, respectively. e) Shift current conductivity for T5G at $\theta=1.32^\circ$ with varying external displacement field strength. The result of increased external displacement field leads to band mixing, which leads to enhancement of shift current via virtual transition as well as a new peak at low frequency. Non-vanishing signal as $\omega\to0$ in this figure is a consequence of finite Lorentzian broadening (see Appendix \ref{lorentzian} for more details)}
    \label{fig1}
\end{figure}

In multi-band systems, virtual transitions through intermediary bands are an additional transition process that can contribute to the magnitude of nonlinear optical response. In addition to a direct transition from initial band to final band, virtual transition scatters through an intermediary band. This manifests in the structure of the shift current response, which can be decomposed into direct and virtual transition components~\cite{sipe2000,Parker19}. Previously, the seminal work~\cite{Fregoso2017} outlined the design principles to maximize the shift current from direct transitions. This work was inspired by ferroelectrics and orthorhombic monochalcogenides~\cite{singh2014computational,Gomes2015,young2012} (materials with large band edge responsivities which exhibit BPVE in the visible and far-infrared range). For the two-band model under consideration in the Ref. \citenum{Fregoso2017} work, the virtual transitions are not present, and thus the focus of Ref. \cite{Fregoso2017} lies primarily on maximizing singularities in the joint density of states (JDOS).  Conversely, in the case of materials with multiple bands, for example alternating angle moiré graphene\cite{khalaf2019magic,Zhang2022,Park2022} (see Fig.\ref{fig1}a-d for a typical bandstructure), it becomes imperative to explore an alternative approach to maximizing shift current response (e.g. Ref.~\cite{Dai_Rappe_2023,tan2019}), in particular based on the significance of contributions stemming from virtual transitions.

In this paper, we demonstrate a viable mechanism that leverages virtual transitions in multiband systems to significantly enhance shift current response, see Fig.\ref{fig1}e. The enhancement is achieved through designing materials with multiple bands close in energy, thereby increasing the number of possible virtual transitions through the increased number of intermediary bands. Increasing the number of closely spaced energy bands however can carry implications also for the quantum texture of the electron states as exemplified by the Berry connection sum rules~\cite{Xiao2010}. We demonstrate that shift current, whose magnitude is partly controlled by such quantum geometric matrix elements (e.g. Refs. \cite{shift_current,kaplan2021,ahn2022riemannian}), can take advantage of this mechanism. Specifically we find that the transition rate between the initial and final bands involved in the photo-absorption process is enhanced in the regime where wavefunction of the involved states (including the virtual state) become more delocalized. 

These two design principles are the key results of our paper, which underlying physical mechanisms we elucidate first using a 1D multi-chain Rice-Mele model and then we focus on the multilayer moiré graphene materals\cite{khalaf2019magic,Zhang2022,Park2022}. Moir\'e materials such as twisted multilayer graphene (TMG) have acquired much attention due to their multiple flat bands exhibiting nontrivial topology and exotic phenomena such as superconductivity and correlated-insulating behavior~\cite{Andrei2021,balents2020superconductivity,torma2022superconductivity}. Our focus on the moiré materials stems from the fact that they are expected to host a large shift current response\cite{shift_current,kaplan2021}, but we emphasize that the general principles we discuss apply to other materials well.

The paper is organized as follows. In Sec.~\ref{sec:theorySC}, we provide the theory for the shift current responses. Next, in Sec.~\ref{sec:RM}, we present a one-dimensional toy model to illustrate the  principles for enhancing the contribution of virtual transitions in the shift current response. We apply these ideas to TMG in Sec.~\ref{sec:TMG} and study the dependence of shift current on number of layers and the displacement field which controls the hybridization between different bands. In Sec.~\ref{sec:Discussion}, we discuss the implications of our results for TMG and other multilayer systems.

\section{Theory of Shift Current}
\label{sec:theorySC}
Shift current is a second-order DC response to linearly polarized AC fields~\cite{sipe2000,Parker19}. It is characterized by a third-rank conductivity tensor~\cite{sipe2000} given by 
\begin{align}
    \textbf{J}^\beta = 2\sigma_{\alpha\alpha}^\beta(0,-\omega, \omega)E^\alpha(\omega)E^\alpha(-\omega)
\end{align}
where $\textbf{J}^\beta$ is the $\beta$-th component of current density, $E^\alpha(\omega)$ is electric field with frequency $\omega$ in the $\alpha$ direction, with $\alpha,\beta$ denotes spatial indices $x,y$. The shift-current conductivity~\cite{Fregoso2017,sipe2000,Parker19} is given by the expression:
\begin{align}\label{sigma}
    \sigma_{\alpha\alpha}^\beta(0,-\omega, \omega) = \frac{\pi e^3}{\hbar^2}\sum_{m,n}\int d^2\textbf{k} f_{mn}  \textbf{S}_{mn}^{\beta\alpha} |\textbf{A}_{mn}^\alpha|^2 \delta(\omega-\epsilon_{mn}),
\end{align}
where it has the form of a shift vector $\mathbf{S}_{mn}^{\beta\alpha}= \mathbf{A}_{mm}^\beta - \textbf{A}_{nn}^\beta + \partial_{k_\beta}\text{Arg}(\textbf{A}_{mn}^\alpha) $ that characterizes the shift in the localization of Bloch wavefunction upon transition from state $\ket{u_n}$ to state $\ket{u_m}$, weighted by the transition amplitude given by the interband Berry connection $\textbf{A}_{mn}^\alpha = i \bra{u_m}\ket{\partial_{k_\alpha}u_n}$. The sum is over all energy bands, where $\epsilon_{mn}=\epsilon_m - \epsilon_n$ is the energy difference between two states, and $f_{mn}=f_m-f_n$ is the difference in occupancy of their energy levels.

To visualize the transition process and draw a distinction between direct and virtual transitions, we use generalized sum rules to replace wavefunction derivatives with sums over all states of matrix element derivatives ~\cite{sipe2000,cook2017design,Parker19}. The shift current integrand defined as $R_{mn}^{\alpha\alpha\beta}=|\textbf{A}_{mn}^\alpha|^2 \textbf{S}_{mn}^{\beta\alpha}$ is given by,
\begin{align}\label{integrand}
\begin{split}
    R_{mn}^{\alpha\alpha\beta} &= \frac{1}{\epsilon_{mn}^2}\text{Im} \left[\frac{h_{mn}^\alpha h_{nm}^\beta \Delta_{mn}^\alpha}{\epsilon_{mn}} + w_{mn}^{\alpha\beta} h_{mn}^\alpha \right]  \\
    &+ \frac{1}{\epsilon_{mn}^2}\text{Im}\left[\sum_{l\neq mn}\left(\frac{h_{nm}^\alpha h_{ml}^\beta h_{ln}^\alpha}{\epsilon_{ml}} - \frac{h_{nm}^\alpha h_{ml}^\alpha h_{ln}^\beta}{\epsilon_{ln}} \right)\right].
\end{split}
\end{align}
where $h_{mn}^\alpha = \mel{m}{\partial_\alpha H}{n}$, $\Delta_{mn}^\alpha = h_{mm}^\alpha - h_{nn}^\alpha$, and
$w_{mn}^{\alpha\beta} = \mel{m}{\partial_{k_\alpha}\partial_{k_\beta}H}{n}$ (See Appendix \ref{deriving_integrand} for derivation). The first term represents a direct transition from band $n$ to band $m$, and the second term represents virtual transitions through an intermediary band~\cite{Parker19}. In a two-band system in 1D, only $w_{mn}^{\alpha\beta}$ term contributes as the first term in the direct transition term becomes purely real and virtual transitions are not present in two-band model. In TMG system when we expand momentum to linear order near the $K/K'$ points, $w_{mn}^{\alpha\beta}$ vanishes.  Fig. \ref{fig1}b  shows a schematic depiction of direct and virtual transitions between different bands in TMG. The virtual transition sums over all intermediary bands, and we expect such a process to be the dominant contribution in systems containing multiple bands of similar energies. The coupling between these bands can be controlled by the external  displacement field which can be used to enhance the shift current conductivity as shown in Fig.~\ref{fig1}e. In addition to discerning the direct and virtual transition contributions, Eq. \ref{integrand} is numerically more amenable as it avoids dealing with the issue of gauge fixing when evaluating the derivative of the wavefunction directly~\cite{cook2017design,Parker19}.

\section{Multilayer Rice-Mele Model}
\label{sec:RM}
In order to study the role of virtual transitions in controlling the magnitude of the shift current response, we construct a toy model with multiple close spaced bands. Specifically, we consider a 1D stacked Rice-Mele (RM) model to demonstrate the interplay between virtual transitions and bandstructures in multilayer systems. The Rice-Mele model is a prototypical model for one-dimensional ferroelectrics~\cite{Fregoso2017} represented by the Hamiltonian
\begin{align}
    H = \sum_{i}\left[\left(\frac{t}{2}+(-1)^i \frac{\delta}{2}\right) (c_{i+1}^\dagger c_{i} + \text{h.c.}) + (-1)^i \Delta c_i^\dagger c_i\right],
\end{align}
where $t$ is the hopping potential, $\delta$ parameterizes the difference in hopping strength between the two neighboring sites with unit cell length $a$, and $\Delta$ is the staggered onsite potential. The canonical Bloch Hamiltonian in momentum space is given by~\cite{PhysRevLett.49.1455}
\begin{align}
    H(k) = \sigma_x t \cos{\frac{ka}{2}} - \sigma_y \delta \sin{\frac{ka}{2}} + \sigma_z \Delta.
\end{align}
To construct a multilayer Rice-Mele model, we denote the dependence of the momentum-space Bloch Hamiltonian on its parameters as $H_i = H_i(t_i, \delta_i, \Delta_i)$ where $i$ is the layer index, and with the full Hamiltonian for $N$ layers taking the following form:
\begin{align}\label{MRM}
    H_{\text{MRM},0}(k) =  H_1(k) \oplus H_2(k) \oplus \cdots \oplus H_N(k)
\end{align}

\begin{figure}[t]
    \centering
    \includegraphics[width=\linewidth]{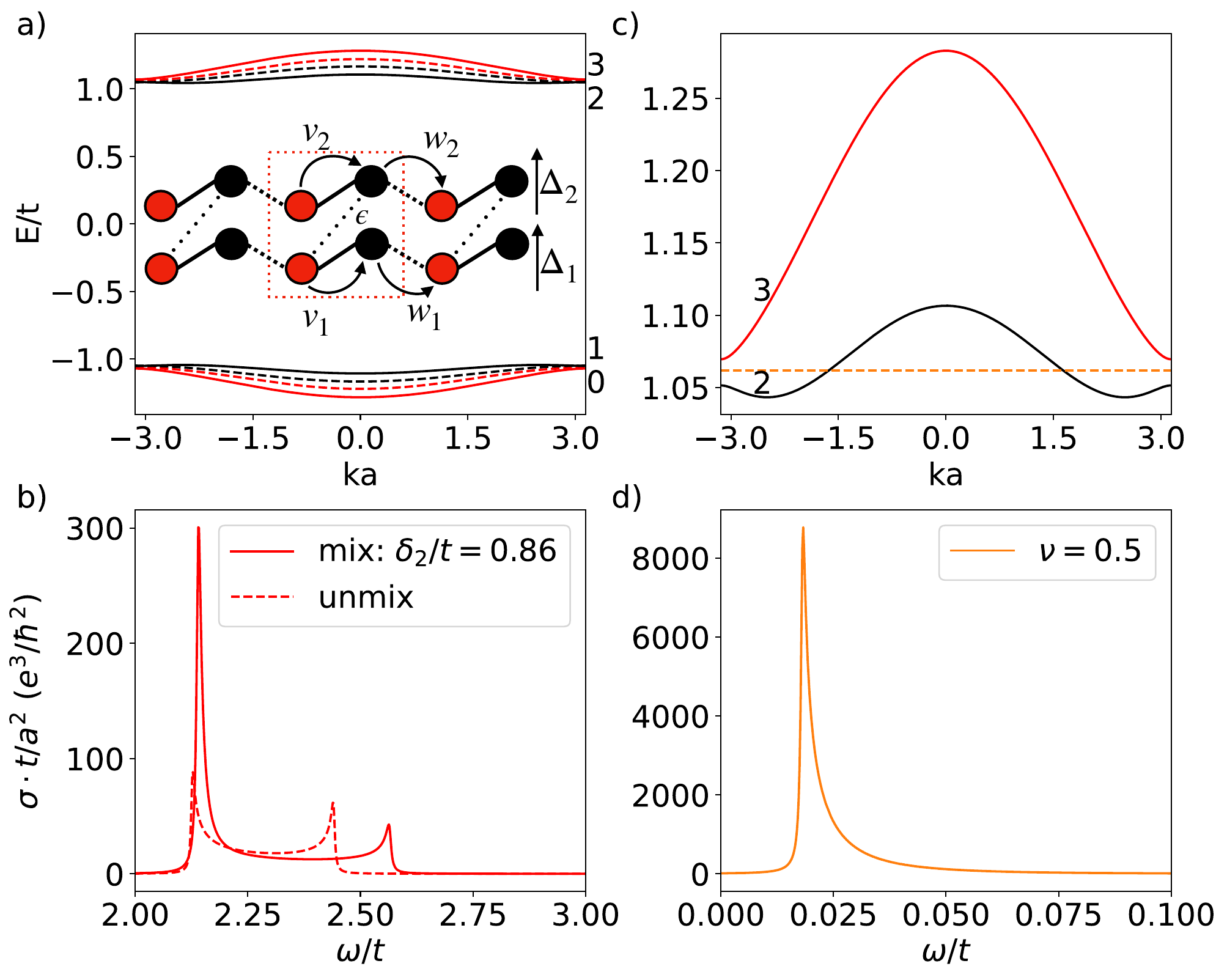}
    \caption{Bandstructure and shift current conductivity of 2RM model and its dependence on model parameters $\delta_2$. a) Band structure of 2RM model without band mixing (dashed line) and with band mixing of $\epsilon=0.1$ (solid line). The red/black band refers the first/second RM model, with intracell hopping strength $v_i = \frac{1}{2}(t_i+\delta_i)$, intercell hopping strength $w_i = \frac{1}{2}(t_i - \delta_i)$, and sublattice offset potential $\Delta_i$, where $i=1,2$ indexing the different RM layers.  A schematic description of the model is included, with $\epsilon$ coupling the different sublattice degree of freedom between the chain. The parameters of 2RM are $(t, \delta_1, \Delta_1, \delta_2, \Delta_2) = (1, 0.8, 0.7, 0.86, 0.6)$. b) Shift current conductivity due to transitions between the red band (first RM model) with and without band mixing. The presence of mixing enhances the peak due to transition at band edge via virtual transitions through the second RM bands. c) Bandstructure of the top two bands, with the orange dashed line denoting chemical potential at filling $\nu=0.5$. d) Shift current conductivity due to transitions from band $2$ to band $3$, showing a large response at low frequency determined by the energy gap of band $2$ and $3$ at $ka=\pm \pi$.  }
    \label{fig2}
\end{figure}

As written above in Eq. \ref{MRM}, the multilayer Rice-Mele Hamiltonian is block-diagonal thus transitions between the different blocks are impossible. To enable virtual transitions between the RM sectors belonging to different layers we introduce couplings between layers. Specifically, consider the case of $N=2$ with band mixing, the 2-layer Rice-Mele (2RM) model has Hamiltonian of the form 
\begin{align}
    H_{2\text{RM}}(k) = 
    \begin{pmatrix}
        H_1(k) & \epsilon V \\
        \epsilon V^{\dagger} & H_2(k)
    \end{pmatrix},
\end{align}
where $V$ is the operator that characterizes interactions and $\epsilon$ parameterize the mixing strength. Fig. \ref{fig2}a shows the bandstructure of 2RM model with $V = \sigma_x$, and $(t, \delta_1, \Delta_1, \delta_2, \Delta_2) = (1, 0.8, 0.7, 0.86, 0.6)$ with (without) mixing as depicted by solid (dashed) line. For this particular form of coupling, the two low (high) energy bands hybridize which widens the gap between black and red in low (high) energy sector. This effect manifest most strongly at $k a=\pm\pi$ where the bands flatten out significantly in comparison to the unmixed case as shown in Fig.~\ref{fig2}a. Here we fixed $t$ to be the same for both RM models and vary the hopping amplitude by tuning $\delta_i$.

The coupled bands after mixing enable transitions to occur among all four bands, allowing virtual transition to enhance the overall shift current response. As shown in Fig. \ref{fig2}b, the transition from valence to conduction band of the first RM model is enhanced after band mixing due to virtual transitions through a different RM sector. The peaks in the response are determined by the energy gaps at $k a=\pm \pi, 0$, where the joint density of state is the largest. Level repulsion as a result of band mixing widens the gap, consequently shifting the location of the peak to a slightly higher frequency. In addition to the selective enhancment at precise frequencies, the overall shift current conductivity $\sigma(\omega)$ over whole frequency range is increased. Specifically, using a ``figure of merit''
\begin{align} \label{M}
    M = \left|\int \sigma d\omega \right|
\end{align}
as a measure of the overall shift current response \cite{tan2019}, we find that $M_{\text{unmix}} \approx 9$ and $M_{\text{mix}} \approx 13.5$, demonstrating an enhanced response.

 The coupling-induced hybridization of bands not only flattens the dispersion around $ka = \pm\pi$ but also engenders additional transition pairs throughout the Brilliouin zone that further enhance the overall shift current response. The total shift current conductivity is now obtained by summing over all possible transition pairs, which include transitions from one RM sector to another. The sign of conductivity depends on the specific transition pairs. On the basis of quantity $M$, introduced in Eq. \ref{M} above, some transition pairs will contribute towards the cancellation of conductivity. Nevertheless, an overall enhancement is still observed compared to models without band mixing (See Appendix \ref{2RM_total_shift} for more details). Different transition pairs give rises to peaks in conductivity at different frequencies range. To avoid cancellation due to opposite signs, one could work at specific frequencies to excite the corresponding transition pairs.

Furthermore, by studying the shift current response of 2RM model at filling $\nu = 0.5$, where the chemical potential lies in between the band edge of the top two bands, see Fig.~\ref{fig2}c, we find an additional shift current response between the two bands 2 and 3 with $M_{32} \approx 42.4$ in the low frequency range as shown in Fig.~\ref{fig2}d. This response relies on band mixing and is much larger than the response at charge neutrality considered in the previous paragraph and in Fig. ~\ref{fig2}b. This response is dominated by direct transition as virtual transitions are suppressed by the large energy gap. Only one peak occurs at frequency determined by the energy gap difference at the band edge because the states around $k=0$ do not contribute to transitions. The reasons for such a large response are twofold: small energy gap and large joint density of state, which can be inferred from Eq. \ref{integrand} and Eq. \ref{sigma}, respectively. The exclusion of states around $k=0$ would not drastically reduce the shift current response, because the gap is much larger at $k=0$ in comparison to the gap at the band edge. 

We can understand the above results at $\nu = 0.5$ with help of the Ref.~\cite{tan2019}. As pointed out in Ref.~\cite{tan2019}, the energy gap is not the only criteria which dictates the shift current response but band dispersion also plays an important role. In a typical tight-binding two-band model, the structure of the valence band fully determines the structure of the conduction band. This unique relationship defines an upper limit on quantity $M$ in terms of energy gap, bandwidth, and the range of hopping as discussed in Ref.~\cite{tan2019}. These three factors also dictate the extent of delocalization of the wavefunction in a two-band model which was highlighted\cite{tan2019} to be the most important criteria for enhanced shift current response. 
On the other hand, for a system with energetically close multiple bands, there are additional physical principles  that control the limit of shift current response. In particular, many recent works have demonstrated how quantum geometry which manifests in multi-band systems can contribute to physical 
quantities, for example such as the superfluid stiffness, beyond their nominal limits inherent in single-band or two-band models~\cite{torma2022superconductivity,Torma2023,verma2021optical,Herzog2022,Xie2020}.

\begin{figure}
    \centering
    \includegraphics[width=\linewidth]{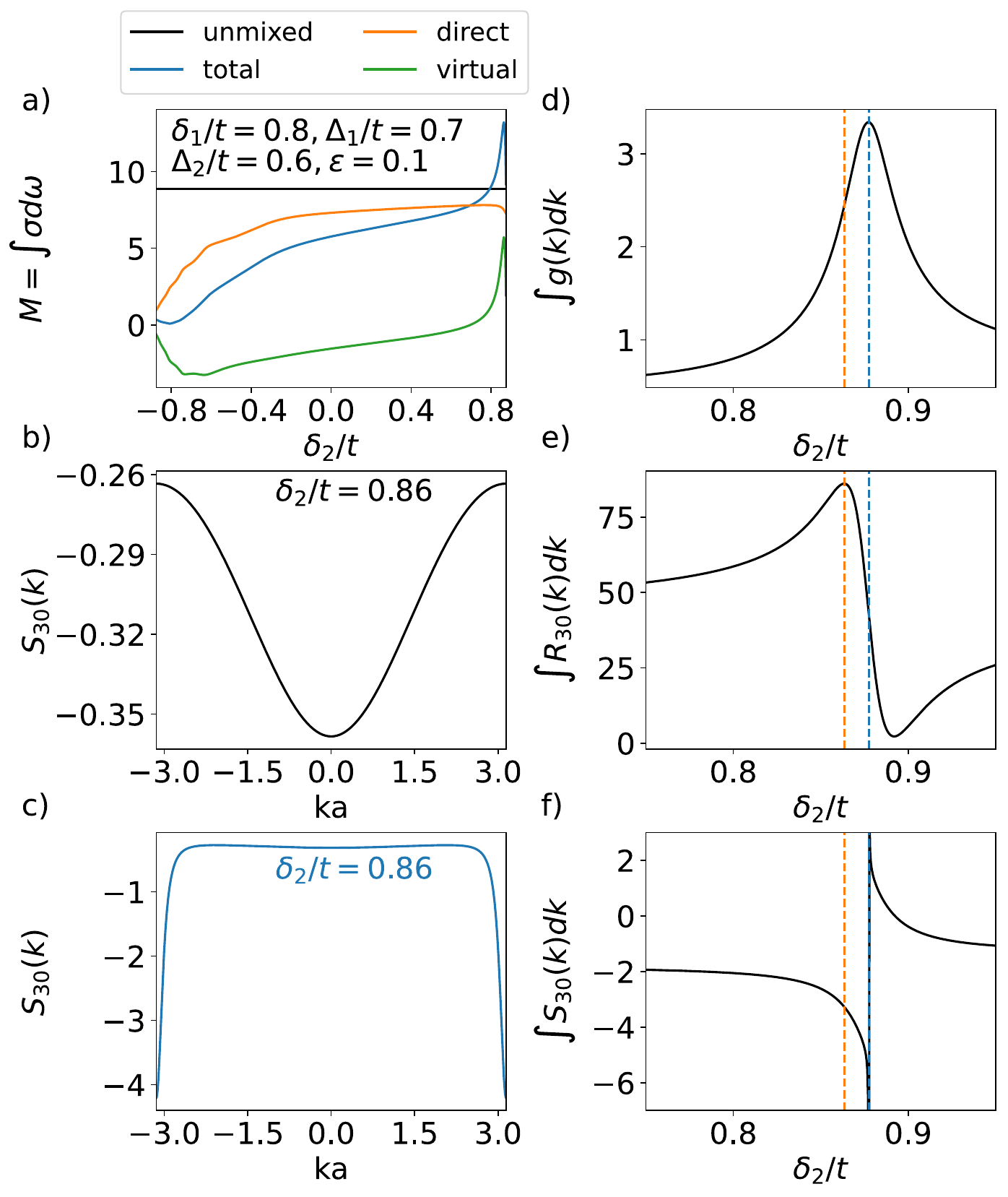}
    \caption{Parameter choice that leads to enhanced shift current response. a) Total shift current $M$ broken down into direct and virtual contributions. b) Shift vector before and c) after band mixing, demonstrating contribution mainly coming from band edge. d) FS metric, e) shift current integrand, and f) shift vector at $k a=-\pi$, showing that enhanced response corresponds to a larger spread in the Wannier wavefunction.  }
    \label{fig3}
\end{figure}

While the multi-band aspect of our system manifests explicitly through virtual transition contributions to shift-current response, the specific choice of model parameter is crucial in achieving enhanced response via virtual transitions. Fig. \ref{fig3}a compares the magnitude of total shift current response $M$ with and without band mixing of transition from band $0$ to band $3$ fixing $\delta_1, \Delta_1, \Delta_2, \epsilon$ and vary $\delta_2$, breaking down the contribution to total response into direct and virtual transition. We see that the effect of band mixing can lead to a reduction in direct transition contribution to shift current. Simultaneously however there can be a significant enhancement in the virtual transition contributions to the shift current. 

Specifically there exist parameter regime, where this increase in virtual transition component is enough to offset the reduction in direct transition. In this region, where we picked $\delta_2 \approx 0.86$, we observe an enhanced shift vector, $S_{30}(k)$, near the band edge, as shown in Fig. \ref{fig3}b,c. We attribute this enhancement effect to the sudden increase in the spatial delocalization of Wannier wavefunctions stemming from strong band hybridization at those band parameter values.

This Wannier function delocalization can be studied more systematically with help of the Fubini-Study (FS) metric~\cite{Provost1980, Torma2023}. (See Appendix \ref{FS} for more details) The FS metric defines a distance between momentum states, and its integral over the Brillouin zone serves as a lower bound for Wannier function localization \cite{Torma15, Xie2020,PhysRevB.101.060505}. Fig. \ref{fig3}d  shows a sweep across the region of $\delta_2$ that exhibits enhancement in total shift current via virtual transitions, and we find a corresponding increase in the FS metric integral $\int g(k) dk$, which implies a delocalization of Wannier function. This observation is in agreement with previous works that demonstrate a correspondence between larger delocalization and larger shift vector\cite{Dai_Rappe_2023}, and consequently large shift current response \cite{Tan_2019}. The blue dashed line denotes the maximum of the integrated FS metric and where $S_{30}$ from Fig. \ref{fig3}c diverges at the band edge. The blue dashed line is  where the band gap of the top/bottom is minimized, and it also corresponds to optical zero ($A_{mn}(k)$ = 0). Hence, we identify a competition between the increasing of $S_{mn}(k)$ and decreasing of $|A_{mn}(k)|$, leading to maximum of shift vector integrand $R_{mn}$ defined in Eq.~\ref{integrand} occurring at the parameter near optical zeros (marked by orange dashed line), where the shift vector is still large, see Fig. ~\ref{fig3}d-f.

We further note that the enhancement in FS metric $g(k)$ can be described by a Lorentzian function. (See Appendix \ref{resonance} for more details). This suggests that the sudden increase in FS metric, and consequently the spread of Wannier wavefunction, can be thought of as resonance in parameter space. It is this resonantly enhanced delocalization of Wannier wavefunction that leads to the enhanced shift current response. In momentum space, this enhancement is due to virtual transition via intermediary bands. Specifically, the enhancement comes from the small energy gap between the top two bands $E_{32}$ and the large matrix element $\text{Im}\{v_{20}\}$ with a pole occurring at $\delta_2 = \sqrt{\delta_1^2 + \Delta_1^2 - \Delta_2^2}$ at the band edge. (See Appendix \ref{2RM_ana} for more details).

The proposed mechanism above suggests to stack multiple 1D systems together and induce band mixing. By careful tuning the system's parameters, we find a region of parameters where the Wannier function becomes maximally delocalized. Although our analysis is constrained to the simple 1D model, we propose that the identified design principles can extend to more realistic models.Specifically, in the following section we demonstrate these principles with the help of a realistic material model.

\begin{figure*}
    \centering
    \includegraphics[width=\linewidth]{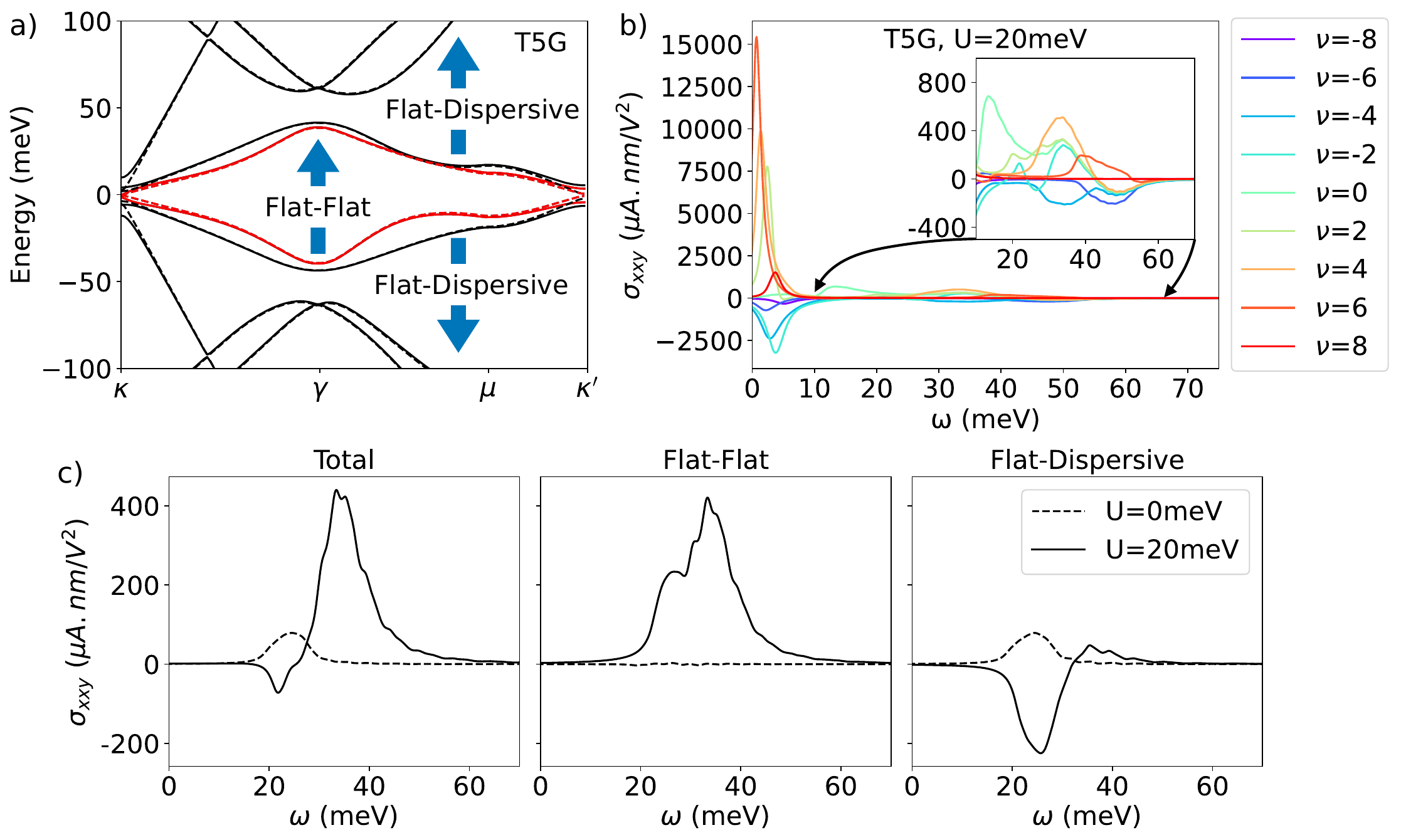}
    \caption{Bandstructure of T5G  and its total shift current conductivity plot at various filling, and breakdown of the enhancement effect via virtual transition of two flat bands at $\nu=2$. a) A schematic depiction of transition between flat to flat bands and from flat to dispersive bands, as well as bandstructure of T5G with $U=20$ meV  (solid line) and with $U=0meV$ (dashed line) at $\theta=1.32^\circ$. b) Shift current conductivity at various system fillings, demonstrating the ability to optimize conductivity at different frequency by tuning the system filling. c) Shift current conductivity due to transitions between bands colored in red in a) at $\nu=2$. Here we break down the total conductivity into virtual transition through flat bands and virtual transition through dispersive bands. Virtual transitions enhance both flat-flat and flat-dispersive contribution to shift current. }
    \label{fig4}
\end{figure*}

\section{Twisted Multilayer Graphene}
\label{sec:TMG}
Twisted bilayer graphene (TBG) has been proposed to exhibit large shift current response due to its nontrivial flatband topology \cite{shift_current,kaplan2021}. We thus propose to use TBG as a viable building block to construct the above proposed multilayer system. Specifically we will follow the stacking procedure introduced in Ref. \cite{khalaf2019magic}, that has been recently realized experimentally\cite{Zhang2022,Park2022}. The single-particle spectrum of TBG is described by an effective continuum model (e.g. Ref.~\cite{koshino2018maximally}). Consider Hamiltonian in the sublattice basis $(A_1, B_1, A_2, B_2)$, 
\begin{align}
    H_{\theta} = 
\begin{pmatrix}
H_1 & U^\dagger \\
U & H_2
\end{pmatrix}.
\end{align}
Let $l=1,2$ denote the bottom  and top layer, respectively. The intralayer Hamiltonian centered at Dirac cone $\mathbf{K}_\xi^{(l)}$ given by
\begin{align}
    H_{l}=-\hbar v [R(\pm \theta/2)(\mathbf{k} - \mathbf{K}_\xi^{(l)})] \cdot (\xi \sigma_x, \sigma_y) + \Delta_{l}\sigma_z
\end{align}
where $\theta$ is twist angle relative to the origin, $+/-$ corresponds to $l=1/2$, and $\xi = \pm 1$ labels the valley index of the original graphene Dirac cone. We include the sublattice offset term $\Delta_{l}\sigma_z$ to break inversion $C_{2z}$ symmetry, which is essential for non-zero shift current response\cite{Nagaosa2022}. Experimentally, this is achieved by coupling TBG to top/bottom layer substrate. The effective interlayer coupling $U$ has the form
\begin{align}
    U &= 
\begin{pmatrix}
U_{A_2,A_1} & U_{A_2, B_1} \\
U_{B_2,A_1} & U_{B_2,B_1}
\end{pmatrix}
\\&=
\begin{pmatrix}
u & u' \\
u' & u
\end{pmatrix}+\begin{pmatrix}
u & u{'} w^{-\xi}\\
u{'} w^{\xi} & u
\end{pmatrix} e^{i\xi \mathbf{G}_1^M \cdot \mathbf{r}}\\
&+ 
\begin{pmatrix}
u & u{'} w^\xi \\
u{'} w^{-\xi} & u
\end{pmatrix} 
e^{i\xi (\mathbf{G}_1^M + \mathbf{G}_2^M) \cdot \mathbf{r}},
\end{align}
where $w=e^{i 2\pi/3}$, and $\mathbf{G} = n_1 \mathbf{G}_1^M + n_2 \mathbf{G}_2^M$ is the moir\'e reciprocal lattice vector, for $n_1, n_2 \in \mathbb{Z}$. We choose $\mathbf{G}_i^M = R(-\theta/2)\mathbf{G}_i - R(\theta/2)\mathbf{G}_i$, with the $\mathbf{G}_1 = (2\pi/a)(1, -1/\sqrt{3})$ and $\mathbf{G}_2 = (2\pi/a)(0, 2/\sqrt{3})$. Following the work done in \cite{shift_current}, we choose $u{'}=90meV$ and $u=0.4u'$, sublattice offset $\Delta = 5 meV$, and set $\hbar v/a=2.1354eV$ and $a=0.246 nm$ however the qualitative behaviour of the shift current that will be discussed below does not depend on these specific parameters of the model.

We alternatively twist each layer of graphene by an angle $\theta/2$ with respect to each adjacent layer to extend the bilayer to a multilayer system \cite{khalaf2019magic,Zhang2022,Park2022}. Coupling is only considered between adjacent layers. The Hamiltonian in the layer basis is given by
\begin{align}
H_{n} = 
\begin{pmatrix}
H_1 & U^\dagger & 0 & 0&\cdots\\
U & H_2 & U & 0 &\cdots\\
0 & U^\dagger & H_1 & U^\dagger &\cdots\\
\vdots & \vdots & \vdots & \vdots & \ddots
\end{pmatrix}
\end{align}
where we treated tunneling terms between all layers to be the same, and the angle of $H_n$ is determined by $\theta_n = 2 \cos(\frac{\pi}{n+1})\theta_{\text{TBG}}$. In the multilayer model, sublattice offset term is added at the top and bottom layers to incorporate the effect of coupling to hBN substrate. This multilayer Hamiltonian can be block-diagonalized into a direct sum of effective TBG pairs under a unitary transformation \cite{khalaf2019magic}. To induce virtual transition among different Hamiltonian layers, we consider applying an external displacement field on the top and bottom gates. (See Appendix \ref{band_mixing} for details) The bandstructure of $N=2, 5, 7, 8$ is shown in \ref{fig1}a-d, where we see $\lfloor n/2 \rfloor$ number of TBG-like bands with an additional Dirac cone if $n$ is odd. Bandstructure of T5G is reproduced in Fig. \ref{fig4}a with an external displacement field, including a schematic description of flat-to-flat band transition and flat-to-dispersive band transition. We study the transition between the lowest flat bands (colored in red) and observe an enhancement in shift current via virtual transitions to nearby flat bands once the external displacement field is included.  To note, our TMG model has $C_{3z}$ symmetry, and by group theoretical method we can show that only two components of the conductivity tensor are independent (See Appendix \ref{symmetry} for details).

In our analysis, we neglect the role of electron-electron interactions, as our goal is to demonstrate just the principle of the physical mechanism that causes the shift current enhancement. The role of interactions in TBG and moiré materials is extensive as studied both theoretically (e.g. Refs.~\cite{cea2021coulomb,guineaElectrostaticEffectsBand2018,pantale2022,goodwinHartreeTheoryCalculations2020,PhysRevB.100.205114,PhysRevB.103.195127,PhysRevB.98.235158,PhysRevB.102.155149, PhysRevLett.124.097601,PhysRevB.103.205415,PhysRevLett.122.246401,PhysRevX.10.031034,PhysRevB.103.205414,PhysRevX.11.041063,kolar2023}) and experimentally (e.g. Refs.~\cite{cao1,cao2,yankowitz2019tuning,efetov2019,Oh2021,Sharpe605,Serlin2020,Nuckolls2020,10.1038/s41586-020-2473-8,SNP19,Choi2021,Choi2021,Kim2023,Nuckolls2023}). However, the exchange-driven physics, particularly at high temperatures ($T>10$ K), should not affect the qualitative ideas discussed in this manuscript. Moreover, as argued in Ref.~\cite{kolar2023}, interactions can, in fact, give rise to a self-generated ``displacement'' field, which will induce hybridization between different energy sectors even in the absence of an external field. For simplicity of the discussion in what follows, we focus on the non-interacting model, leaving a detailed study of interactions on the shift-current photoresponse of a multilayer moiré system to future works.

The physical twist angle in TMG determines the effective twist angles of TBG-like bands, which in turn controls the energy difference between two flat band pairs in each energy sector of the model. The magnitude and direction of virtual transition are determined by the velocity matrix element and the energy gap of bands involved in the optical transition process as implied by Eq. \ref{integrand}. To receive maximum enhancement from virtual transitions, we choose twist angles such that the energy gap between two flat bands is as small as possible while the velocity matrix element is nonzero. The interplay between velocity matrix element and energy gap determines the optimal twist angle for each multilayer system. Crucially here, since our goal is not to achieve a flat band, but rather to obtain closely spaced energy bands, the relevant twist angles differ from the magic angles of these multilayer systems. For example, from our numerics, we found $\theta_5 = 1.32^\circ$ for T5G helps optimize the shift current reponse, while the magic angle for T5G occurs at $\theta_{5,m} = 1.73^\circ$, where perfect flat band emerges. (See Appendix \ref{twist_angle} for more details) Correspondingly, there is an increase in shift current response due to reduced band gap and large joint density of states, an enhancement in direct transition. At the magic angle, the strongly interacting nature of the system requires us to incorporate Coulomb interaction, which enhances the resonance at non-interacting characteristic frequency due to filling-dependent band renormalization \cite{shift_current}. In TBG away from the magic angle, the shift current response is qualitatively the same, scales according to band gap and moir\'e length. In T5G, however, an enhancement via virtual transition to nearby flat bands occurs at angles where the two flat bands become close in energy, in addition to a new resonance peak at low frequency.

 In contrast to the simple 4-band model (the 2RM model) described in last section, the TMG system has a contribution to shift current conductivity from both direct transition from flat to flat bands and virtual transitions from flat to dispersive bands coming from both the same and different TBG-like sectors. In the absence of external displacement field, it is the virtual transition to dispersive bands within the same TBG sector that dominates the shift current response. With external displacement field, band mixing enables transitions to access different TBG-like sectors in the form of virtual transitions to either flat or dispersive bands of the other TBG-like sectors. Shift current conductivity $\sigma_{xxy}$ at $\nu=2$ is shown in Fig. \ref{fig4}c. After band mixing, virtual transition through intermediary bands significantly enhances the response, in both flat-to-flat and flat-to-dispersive bands. We notice a shift in the peak of shift current response to higher $\omega$ with band mixing. Without mixing, the dominant contribution to shift current comes from flat-to-dispersive band transition within the same TBG sector, with peak determined by the energy difference between the flat and dispersive band at the $\gamma$ point. After band mixing, virtual transition via nearby flat bands becomes the dominant contribution, which enhances the response occurring that peak near $\mu$, where van Hove singularity occurs. Our choice of T5G angles gives rise to a larger energy gap at $\mu$ point between the two flat bands than the energy gap between flat and dispersive band at $\gamma$ point, resulting in a shift to higher $\omega$ in shift current peak.

Summing up all transition pairs reveals a giant peak at low frequency in shift current conductivity due to transitions to nearby flat bands and Dirac cone around the $\kappa$ points. As shown in Fig. \ref{fig4}b) for various fillings at $U=20meV$, the peak of the response occurs at a frequency determined by the energy gap of those bands at the $\kappa$ points, roughly determined by the sublattice offset term $\Delta$. Previous work \cite{low-freq_divergence} also pointed out the effect of low-frequency divergence in topological semi-metals, where they attributed the divergence to singularity in quantum geometry. At a given filling, in addition to the giant peak at low frequency, there is a secondary peak occurring at frequency proportional to the energy gap of the two flat bands at $\mu$, which corresponds to transitions among the flat bands pairs. The appearance of two distinct peaks is a result of summing over different transition pairs. More explicitly, transitions between top/bottom pairs of flat band give rise to low-frequency divergence, whereas transition from bottom to top flat bands give rise to the secondary peaks. This is another unique feature to TMG system compared to TBG system studied in previous works \cite{shift_current,kaplan2021}.

\section{Discussion}
\label{sec:Discussion}
In this work, we discussed new design principles for optimizing shift current response in multi-band systems. In particular, we demonstrated a mechanism based on enhancement of virtual transition contributions in systems with multiple layers. The enhancement arises mainly from the mixing of bands belonging to different layers. The mechanism is explained using a 1D toy model where the stacking order provides an additional ``knob'' to tune the shift-current response.

We have shown that in TMG systems, shift current response is enhanced by both virtual transitions among nearby bands and the additional transition pairs between additional bands. Virtual transitions enhance resonance at the original peaks whereas transitioning among nearby bands around $\kappa$ point produces another peak in conductivity at low frequency. Both phenomena of enhancement occur over a wide range of fillings, with different fillings corresponding to a slightly different enhanced peak in conductivity. The effect of virtual transitions found in this paper relies on the existence of multiple flat bands and applies when the number of layers $N \geq 5$. Compared to TBG, T3G only has an additional Dirac cone at $\kappa$ point, and although T4G has two flat band pairs, the second pair is too dispersive for the enhancement effect to occur. This makes shift current response in TMG for $N\geq 5$ qualitatively different from that of $N < 5$. As we increase $N$, the existence of additional flat band pairs can lead to further enhancement via virtual transitions. The physical angle at which we would observe enhanced shift current response is also expected to increase, making it potentially easier to realize experimentally.

Much of the experimental attention in the studies of TMG has been focused on the system's behavior near magic angles (e.g. Refs.~\cite{cao1,cao2,yankowitz2019tuning,efetov2019,Oh2021,Sharpe605,Serlin2020,Nuckolls2020,10.1038/s41586-020-2473-8,SNP19,Choi2021,Choi2021,Kim2023,Nuckolls2023}), where superconductivity and correlated insulating behavior are revealed. In TMG systems, in addition to the set of magic angles that give rise to superconductivity and correlated insulating behavior, we find a different set of angles that will lead to a giant shift current response. (See Appendix \ref{twist_angle} for details) At these angles, we find enhanced virtual transitions via nearby bands, as well as more delocalized wavefunctions. These angles are relevant in order to achieve large nonlinear optical response. These findings make TMG a promising platform for designing devices that produce giant photocurrents at terahertz frequency.

 Finally, we believe that this procedure of constructing multilayer systems can be used as a general guide to enhance shift current response in other 2D systems. Given a generic 1D or 2D system that already exhibits a large shift current response, we can further enhance its response through a stacking procedure that induces virtual transition between neighbouring energy bands. It is particularly appealing to twisted materials as this procedure of twist and stack is the commonly used technique of constructing this class of materials. By varying the twist angle, one can find a setup where shift current response is enhanced.

 Motivated by Eq. \ref{integrand}, we see that to achieve a large shift current, we want to minimize the band gap $\epsilon_{mn}$. However, simply engineering bands with a small band gap will also shift the frequency peak to small $\omega$, making the current susceptible to thermal fluctuations at finite temperature \cite{low-freq_divergence}. In addition, the derivation of shift current expression via perturbation theory assumes small $|E/\omega|$, leading to worries concerning the breakdown of perturbation theory in the small $\omega$ regime. By keeping the large band gap between band $m$ and $n$ and engineering nearby bands $l$ close in energy such that $\epsilon_{ml}$ is small, we leverage virtual transitions to enhance shift current response while maintaining transition at relatively large $\omega$. Furthermore, the FS metric is also inversely proportional to the energy separation of two bands. When a pair of bands with non-zero interband Berry connection have a small energy gap, it leads to a large FS metric, which also corresponds to a large localization length of Wannier wavefunction and a large shift vector. This confirms the intuition that a large shift current arises when there is a large shift in the center of charge of the wavefunction~\cite{Fregoso2017,wang2019} in real space upon excitation to the conduction band. 
 Recently, nonlinear optical response has been formulated in the language of quantum geometry \cite{Ahn2021}, and it has been shown that shift current is related to the quantum geometric connection. From this perspective, the meticulous adjustment of model parameters positions the material in close proximity to the geometric singularity in the band structure, leading to divergence of the geometric quantities that govern the shift current response.

 The intricate interplay between shift current and quantum geometry has enabled its use as a probe of quantum geometry and interactions in topological materials \cite{shift_current}. Here we show two examples of enhanced shift current response via virtual transition near the band closing point. In topological materials, gap closing is associated with topological phase transition. It would be interesting to study shift current response in noncentrosymmetric multiband topological materials near critical point to see if we observe a corresponding enhanced response. This could potentially allow us to use shift current as a probe of topological phase transitions.

 One could also imagine our design principle to be utilized to enhance other optical responses that demonstrate a geometric origin. Another common second-order bulk photovoltaic response is the injection current, which is shown to be related to quantum geometric tensor \cite{Ahn2021, low-freq_divergence}. Engineering bands near geometrical singularity might also lead to large injection current response. 
 
 Furthermore, quantum geometric effects are not limited to optical responses but also lead to many other exotic phenomena like flat-band superconductivity~\cite{verma2021optical,Herzog2022,Xie2020,Huhtinen2022}, undamped plasmons~\cite{lewandowski2019intrinsically}, non-reciprocity in plasmonics~\cite{arora2022quantum,Dutta2023} and in Landau-Zener tunneling~\cite{Kitamura2020}. The mechanism illustrated here for enhancement of shift-current response  using multi-band systems would potentially be applicable in a wide variety of quantum geometric effects.

\section*{Acknowledgement}
We thank Roshan Krishna Kumar for useful discussions and collaboration on a related project. S.C. acknowledges support from the Summer Undergraduate Research Fellowship at Caltech. Swati Chaudhary acknowledges support from National Science Foundation through the Center for Dynamics and Control of Materials: an NSF MRSEC under Cooperative Agreement No.\ DMR-1720595. G.R. expresses gratitude for the support by the Simons Foundation, and the ARO MURI Grant No. W911NF-16-1-
0361 and the Institute of Quantum Information and
Matter.  C.L. was supported by start-up funds from Florida State University and the National High Magnetic Field Laboratory. The National High Magnetic Field Laboratory is supported by the National Science Foundation through NSF/DMR-2128556 and the State of Florida.

\bibliography{bibliography.bib}

\begin{thebibliography}{81}%
\makeatletter
\providecommand \@ifxundefined [1]{%
 \@ifx{#1\undefined}
}%
\providecommand \@ifnum [1]{%
 \ifnum #1\expandafter \@firstoftwo
 \else \expandafter \@secondoftwo
 \fi
}%
\providecommand \@ifx [1]{%
 \ifx #1\expandafter \@firstoftwo
 \else \expandafter \@secondoftwo
 \fi
}%
\providecommand \natexlab [1]{#1}%
\providecommand \enquote  [1]{``#1''}%
\providecommand \bibnamefont  [1]{#1}%
\providecommand \bibfnamefont [1]{#1}%
\providecommand \citenamefont [1]{#1}%
\providecommand \href@noop [0]{\@secondoftwo}%
\providecommand \href [0]{\begingroup \@sanitize@url \@href}%
\providecommand \@href[1]{\@@startlink{#1}\@@href}%
\providecommand \@@href[1]{\endgroup#1\@@endlink}%
\providecommand \@sanitize@url [0]{\catcode `\\12\catcode `\$12\catcode
  `\&12\catcode `\#12\catcode `\^12\catcode `\_12\catcode `\%12\relax}%
\providecommand \@@startlink[1]{}%
\providecommand \@@endlink[0]{}%
\providecommand \url  [0]{\begingroup\@sanitize@url \@url }%
\providecommand \@url [1]{\endgroup\@href {#1}{\urlprefix }}%
\providecommand \urlprefix  [0]{URL }%
\providecommand \Eprint [0]{\href }%
\providecommand \doibase [0]{https://doi.org/}%
\providecommand \selectlanguage [0]{\@gobble}%
\providecommand \bibinfo  [0]{\@secondoftwo}%
\providecommand \bibfield  [0]{\@secondoftwo}%
\providecommand \translation [1]{[#1]}%
\providecommand \BibitemOpen [0]{}%
\providecommand \bibitemStop [0]{}%
\providecommand \bibitemNoStop [0]{.\EOS\space}%
\providecommand \EOS [0]{\spacefactor3000\relax}%
\providecommand \BibitemShut  [1]{\csname bibitem#1\endcsname}%
\let\auto@bib@innerbib\@empty
\bibitem [{\citenamefont {Dai}\ and\ \citenamefont
  {Rappe}(2023)}]{Dai_Rappe_2023}%
  \BibitemOpen
  \bibfield  {author} {\bibinfo {author} {\bibfnamefont {Z.}~\bibnamefont
  {Dai}}\ and\ \bibinfo {author} {\bibfnamefont {A.~M.}\ \bibnamefont
  {Rappe}},\ }\bibfield  {title} {\bibinfo {title} {Recent progress in the
  theory of bulk photovoltaic effect},\ }\bibfield  {journal} {\bibinfo
  {journal} {Chemical Physics Reviews}\ }\textbf {\bibinfo {volume} {4}},\
  \href {https://doi.org/10.1063/5.0101513} {10.1063/5.0101513} (\bibinfo
  {year} {2023})\BibitemShut {NoStop}%
\bibitem [{\citenamefont {Sipe}\ and\ \citenamefont
  {Shkrebtii}(2000)}]{sipe2000}%
  \BibitemOpen
  \bibfield  {author} {\bibinfo {author} {\bibfnamefont {J.~E.}\ \bibnamefont
  {Sipe}}\ and\ \bibinfo {author} {\bibfnamefont {A.~I.}\ \bibnamefont
  {Shkrebtii}},\ }\bibfield  {title} {\bibinfo {title} {Second-order optical
  response in semiconductors},\ }\href
  {https://doi.org/10.1103/PhysRevB.61.5337} {\bibfield  {journal} {\bibinfo
  {journal} {Phys. Rev. B}\ }\textbf {\bibinfo {volume} {61}},\ \bibinfo
  {pages} {5337} (\bibinfo {year} {2000})}\BibitemShut {NoStop}%
\bibitem [{\citenamefont {von Baltz}\ and\ \citenamefont
  {Kraut}(1981)}]{Baltz1981}%
  \BibitemOpen
  \bibfield  {author} {\bibinfo {author} {\bibfnamefont {R.}~\bibnamefont {von
  Baltz}}\ and\ \bibinfo {author} {\bibfnamefont {W.}~\bibnamefont {Kraut}},\
  }\bibfield  {title} {\bibinfo {title} {Theory of the bulk photovoltaic effect
  in pure crystals},\ }\href {https://doi.org/10.1103/PhysRevB.23.5590}
  {\bibfield  {journal} {\bibinfo  {journal} {Phys. Rev. B}\ }\textbf {\bibinfo
  {volume} {23}},\ \bibinfo {pages} {5590} (\bibinfo {year}
  {1981})}\BibitemShut {NoStop}%
\bibitem [{\citenamefont {Fridkin}(2001)}]{fridkin2001bulk}%
  \BibitemOpen
  \bibfield  {author} {\bibinfo {author} {\bibfnamefont {V.}~\bibnamefont
  {Fridkin}},\ }\bibfield  {title} {\bibinfo {title} {Bulk photovoltaic effect
  in noncentrosymmetric crystals},\ }\href
  {https://link.springer.com/article/10.1134/1.1387133} {\bibfield  {journal}
  {\bibinfo  {journal} {Crystallography Reports}\ }\textbf {\bibinfo {volume}
  {46}},\ \bibinfo {pages} {654} (\bibinfo {year} {2001})}\BibitemShut
  {NoStop}%
\bibitem [{\citenamefont {Nastos}\ and\ \citenamefont
  {Sipe}(2006)}]{Nastos2006}%
  \BibitemOpen
  \bibfield  {author} {\bibinfo {author} {\bibfnamefont {F.}~\bibnamefont
  {Nastos}}\ and\ \bibinfo {author} {\bibfnamefont {J.~E.}\ \bibnamefont
  {Sipe}},\ }\bibfield  {title} {\bibinfo {title} {Optical rectification and
  shift currents in gaas and gap response: Below and above the band gap},\
  }\href {https://doi.org/10.1103/PhysRevB.74.035201} {\bibfield  {journal}
  {\bibinfo  {journal} {Phys. Rev. B}\ }\textbf {\bibinfo {volume} {74}},\
  \bibinfo {pages} {035201} (\bibinfo {year} {2006})}\BibitemShut {NoStop}%
\bibitem [{\citenamefont {Pusch}\ \emph {et~al.}(2023)\citenamefont {Pusch},
  \citenamefont {R\"omer}, \citenamefont {Culcer},\ and\ \citenamefont
  {Ekins-Daukes}}]{pusch2023}%
  \BibitemOpen
  \bibfield  {author} {\bibinfo {author} {\bibfnamefont {A.}~\bibnamefont
  {Pusch}}, \bibinfo {author} {\bibfnamefont {U.}~\bibnamefont {R\"omer}},
  \bibinfo {author} {\bibfnamefont {D.}~\bibnamefont {Culcer}},\ and\ \bibinfo
  {author} {\bibfnamefont {N.~J.}\ \bibnamefont {Ekins-Daukes}},\ }\bibfield
  {title} {\bibinfo {title} {Energy conversion efficiency of the bulk
  photovoltaic effect},\ }\href {https://doi.org/10.1103/PRXEnergy.2.013006}
  {\bibfield  {journal} {\bibinfo  {journal} {PRX Energy}\ }\textbf {\bibinfo
  {volume} {2}},\ \bibinfo {pages} {013006} (\bibinfo {year}
  {2023})}\BibitemShut {NoStop}%
\bibitem [{\citenamefont {Nagaosa}(2022)}]{Nagaosa2022}%
  \BibitemOpen
  \bibfield  {author} {\bibinfo {author} {\bibfnamefont {N.}~\bibnamefont
  {Nagaosa}},\ }\bibfield  {title} {\bibinfo {title} {Nonlinear optical
  responses in noncentrosymmetric quantum materials},\ }\href
  {https://doi.org/10.1016/j.aop.2022.169146} {\bibfield  {journal} {\bibinfo
  {journal} {Annals of Physics}\ }\textbf {\bibinfo {volume} {447}},\ \bibinfo
  {pages} {169146} (\bibinfo {year} {2022})}\BibitemShut {NoStop}%
\bibitem [{\citenamefont {Fregoso}\ \emph {et~al.}(2017)\citenamefont
  {Fregoso}, \citenamefont {Morimoto},\ and\ \citenamefont
  {Moore}}]{Fregoso2017}%
  \BibitemOpen
  \bibfield  {author} {\bibinfo {author} {\bibfnamefont {B.~M.}\ \bibnamefont
  {Fregoso}}, \bibinfo {author} {\bibfnamefont {T.}~\bibnamefont {Morimoto}},\
  and\ \bibinfo {author} {\bibfnamefont {J.~E.}\ \bibnamefont {Moore}},\
  }\bibfield  {title} {\bibinfo {title} {Quantitative relationship between
  polarization differences and the zone-averaged shift photocurrent},\ }\href
  {https://doi.org/10.1103/PhysRevB.96.075421} {\bibfield  {journal} {\bibinfo
  {journal} {Phys. Rev. B}\ }\textbf {\bibinfo {volume} {96}},\ \bibinfo
  {pages} {075421} (\bibinfo {year} {2017})}\BibitemShut {NoStop}%
\bibitem [{\citenamefont {Wang}\ and\ \citenamefont {Qian}(2019)}]{wang2019}%
  \BibitemOpen
  \bibfield  {author} {\bibinfo {author} {\bibfnamefont {H.}~\bibnamefont
  {Wang}}\ and\ \bibinfo {author} {\bibfnamefont {X.}~\bibnamefont {Qian}},\
  }\bibfield  {title} {\bibinfo {title} {Ferroicity-driven nonlinear
  photocurrent switching in time-reversal invariant ferroic materials},\ }\href
  {https://www.science.org/doi/10.1126/sciadv.aav9743} {\bibfield  {journal}
  {\bibinfo  {journal} {Science advances}\ }\textbf {\bibinfo {volume} {5}},\
  \bibinfo {pages} {eaav9743} (\bibinfo {year} {2019})}\BibitemShut {NoStop}%
\bibitem [{\citenamefont {Cook}\ \emph {et~al.}(2017)\citenamefont {Cook},
  \citenamefont {Fregoso}, \citenamefont {De~Juan}, \citenamefont {Coh},\ and\
  \citenamefont {Moore}}]{cook2017design}%
  \BibitemOpen
  \bibfield  {author} {\bibinfo {author} {\bibfnamefont {A.~M.}\ \bibnamefont
  {Cook}}, \bibinfo {author} {\bibfnamefont {B.~M.}\ \bibnamefont {Fregoso}},
  \bibinfo {author} {\bibfnamefont {F.}~\bibnamefont {De~Juan}}, \bibinfo
  {author} {\bibfnamefont {S.}~\bibnamefont {Coh}},\ and\ \bibinfo {author}
  {\bibfnamefont {J.~E.}\ \bibnamefont {Moore}},\ }\bibfield  {title} {\bibinfo
  {title} {Design principles for shift current photovoltaics},\ }\href
  {https://www.nature.com/articles/ncomms14176} {\bibfield  {journal} {\bibinfo
   {journal} {Nature communications}\ }\textbf {\bibinfo {volume} {8}},\
  \bibinfo {pages} {1} (\bibinfo {year} {2017})}\BibitemShut {NoStop}%
\bibitem [{\citenamefont {Parker}\ \emph {et~al.}(2019)\citenamefont {Parker},
  \citenamefont {Morimoto}, \citenamefont {Orenstein},\ and\ \citenamefont
  {Moore}}]{Parker19}%
  \BibitemOpen
  \bibfield  {author} {\bibinfo {author} {\bibfnamefont {D.~E.}\ \bibnamefont
  {Parker}}, \bibinfo {author} {\bibfnamefont {T.}~\bibnamefont {Morimoto}},
  \bibinfo {author} {\bibfnamefont {J.}~\bibnamefont {Orenstein}},\ and\
  \bibinfo {author} {\bibfnamefont {J.~E.}\ \bibnamefont {Moore}},\ }\bibfield
  {title} {\bibinfo {title} {Diagrammatic approach to nonlinear optical
  response with application to weyl semimetals},\ }\href
  {https://doi.org/10.1103/PhysRevB.99.045121} {\bibfield  {journal} {\bibinfo
  {journal} {Phys. Rev. B}\ }\textbf {\bibinfo {volume} {99}},\ \bibinfo
  {pages} {045121} (\bibinfo {year} {2019})}\BibitemShut {NoStop}%
\bibitem [{\citenamefont {Morimoto}\ and\ \citenamefont
  {Nagaosa}(2016)}]{morimoto2016topological}%
  \BibitemOpen
  \bibfield  {author} {\bibinfo {author} {\bibfnamefont {T.}~\bibnamefont
  {Morimoto}}\ and\ \bibinfo {author} {\bibfnamefont {N.}~\bibnamefont
  {Nagaosa}},\ }\bibfield  {title} {\bibinfo {title} {Topological nature of
  nonlinear optical effects in solids},\ }\href
  {https://advances.sciencemag.org/content/2/5/e1501524?intcmp=trendmd-adv}
  {\bibfield  {journal} {\bibinfo  {journal} {Science advances}\ }\textbf
  {\bibinfo {volume} {2}},\ \bibinfo {pages} {e1501524} (\bibinfo {year}
  {2016})}\BibitemShut {NoStop}%
\bibitem [{\citenamefont {Orenstein}\ \emph {et~al.}(2021)\citenamefont
  {Orenstein}, \citenamefont {Moore}, \citenamefont {Morimoto}, \citenamefont
  {Torchinsky}, \citenamefont {Harter},\ and\ \citenamefont
  {Hsieh}}]{orenstein2021topology}%
  \BibitemOpen
  \bibfield  {author} {\bibinfo {author} {\bibfnamefont {J.}~\bibnamefont
  {Orenstein}}, \bibinfo {author} {\bibfnamefont {J.}~\bibnamefont {Moore}},
  \bibinfo {author} {\bibfnamefont {T.}~\bibnamefont {Morimoto}}, \bibinfo
  {author} {\bibfnamefont {D.}~\bibnamefont {Torchinsky}}, \bibinfo {author}
  {\bibfnamefont {J.}~\bibnamefont {Harter}},\ and\ \bibinfo {author}
  {\bibfnamefont {D.}~\bibnamefont {Hsieh}},\ }\bibfield  {title} {\bibinfo
  {title} {Topology and symmetry of quantum materials via nonlinear optical
  responses},\ }\href
  {https://www.annualreviews.org/doi/full/10.1146/annurev-conmatphys-031218-013712}
  {\bibfield  {journal} {\bibinfo  {journal} {Annual Review of Condensed Matter
  Physics}\ }\textbf {\bibinfo {volume} {12}},\ \bibinfo {pages} {247}
  (\bibinfo {year} {2021})}\BibitemShut {NoStop}%
\bibitem [{\citenamefont {Ma}\ \emph {et~al.}(2021)\citenamefont {Ma},
  \citenamefont {Grushin},\ and\ \citenamefont {Burch}}]{ma2021topology}%
  \BibitemOpen
  \bibfield  {author} {\bibinfo {author} {\bibfnamefont {Q.}~\bibnamefont
  {Ma}}, \bibinfo {author} {\bibfnamefont {A.~G.}\ \bibnamefont {Grushin}},\
  and\ \bibinfo {author} {\bibfnamefont {K.~S.}\ \bibnamefont {Burch}},\
  }\bibfield  {title} {\bibinfo {title} {Topology and geometry under the
  nonlinear electromagnetic spotlight},\ }\href
  {https://www.nature.com/articles/s41563-021-00992-7} {\bibfield  {journal}
  {\bibinfo  {journal} {Nature materials}\ }\textbf {\bibinfo {volume} {20}},\
  \bibinfo {pages} {1601} (\bibinfo {year} {2021})}\BibitemShut {NoStop}%
\bibitem [{\citenamefont {Ahn}\ \emph {et~al.}(2022)\citenamefont {Ahn},
  \citenamefont {Guo}, \citenamefont {Nagaosa},\ and\ \citenamefont
  {Vishwanath}}]{ahn2022riemannian}%
  \BibitemOpen
  \bibfield  {author} {\bibinfo {author} {\bibfnamefont {J.}~\bibnamefont
  {Ahn}}, \bibinfo {author} {\bibfnamefont {G.-Y.}\ \bibnamefont {Guo}},
  \bibinfo {author} {\bibfnamefont {N.}~\bibnamefont {Nagaosa}},\ and\ \bibinfo
  {author} {\bibfnamefont {A.}~\bibnamefont {Vishwanath}},\ }\bibfield  {title}
  {\bibinfo {title} {Riemannian geometry of resonant optical responses},\
  }\href {https://www.nature.com/articles/s41567-021-01465-z} {\bibfield
  {journal} {\bibinfo  {journal} {Nature Physics}\ }\textbf {\bibinfo {volume}
  {18}},\ \bibinfo {pages} {290} (\bibinfo {year} {2022})}\BibitemShut
  {NoStop}%
\bibitem [{\citenamefont {Morimoto}\ \emph {et~al.}(2023)\citenamefont
  {Morimoto}, \citenamefont {Kitamura},\ and\ \citenamefont
  {Nagaosa}}]{morimoto2023geometric}%
  \BibitemOpen
  \bibfield  {author} {\bibinfo {author} {\bibfnamefont {T.}~\bibnamefont
  {Morimoto}}, \bibinfo {author} {\bibfnamefont {S.}~\bibnamefont {Kitamura}},\
  and\ \bibinfo {author} {\bibfnamefont {N.}~\bibnamefont {Nagaosa}},\
  }\bibfield  {title} {\bibinfo {title} {Geometric aspects of nonlinear and
  nonequilibrium phenomena},\ }\href
  {https://journals.jps.jp/doi/full/10.7566/JPSJ.92.072001} {\bibfield
  {journal} {\bibinfo  {journal} {Journal of the Physical Society of Japan}\
  }\textbf {\bibinfo {volume} {92}},\ \bibinfo {pages} {072001} (\bibinfo
  {year} {2023})}\BibitemShut {NoStop}%
\bibitem [{\citenamefont {Wang}\ \emph {et~al.}(2022)\citenamefont {Wang},
  \citenamefont {Tang}, \citenamefont {Xu}, \citenamefont {Li},\ and\
  \citenamefont {Qian}}]{wang2022generalized}%
  \BibitemOpen
  \bibfield  {author} {\bibinfo {author} {\bibfnamefont {H.}~\bibnamefont
  {Wang}}, \bibinfo {author} {\bibfnamefont {X.}~\bibnamefont {Tang}}, \bibinfo
  {author} {\bibfnamefont {H.}~\bibnamefont {Xu}}, \bibinfo {author}
  {\bibfnamefont {J.}~\bibnamefont {Li}},\ and\ \bibinfo {author}
  {\bibfnamefont {X.}~\bibnamefont {Qian}},\ }\bibfield  {title} {\bibinfo
  {title} {Generalized wilson loop method for nonlinear light-matter
  interaction},\ }\href {https://www.nature.com/articles/s41535-022-00472-4}
  {\bibfield  {journal} {\bibinfo  {journal} {npj Quantum Materials}\ }\textbf
  {\bibinfo {volume} {7}},\ \bibinfo {pages} {61} (\bibinfo {year}
  {2022})}\BibitemShut {NoStop}%
\bibitem [{\citenamefont {Wang}\ \emph {et~al.}(2016)\citenamefont {Wang},
  \citenamefont {Young}, \citenamefont {Zheng}, \citenamefont {Grinberg},\ and\
  \citenamefont {Rappe}}]{wang2016substantial}%
  \BibitemOpen
  \bibfield  {author} {\bibinfo {author} {\bibfnamefont {F.}~\bibnamefont
  {Wang}}, \bibinfo {author} {\bibfnamefont {S.~M.}\ \bibnamefont {Young}},
  \bibinfo {author} {\bibfnamefont {F.}~\bibnamefont {Zheng}}, \bibinfo
  {author} {\bibfnamefont {I.}~\bibnamefont {Grinberg}},\ and\ \bibinfo
  {author} {\bibfnamefont {A.~M.}\ \bibnamefont {Rappe}},\ }\bibfield  {title}
  {\bibinfo {title} {Substantial bulk photovoltaic effect enhancement via
  nanolayering},\ }\href {https://www.nature.com/articles/ncomms10419}
  {\bibfield  {journal} {\bibinfo  {journal} {Nature communications}\ }\textbf
  {\bibinfo {volume} {7}},\ \bibinfo {pages} {10419} (\bibinfo {year}
  {2016})}\BibitemShut {NoStop}%
\bibitem [{\citenamefont {Dong}\ \emph {et~al.}(2023)\citenamefont {Dong},
  \citenamefont {Yang}, \citenamefont {Yoshii}, \citenamefont {Matsuoka},
  \citenamefont {Kitamura}, \citenamefont {Hasegawa}, \citenamefont {Ogawa},
  \citenamefont {Morimoto}, \citenamefont {Ideue},\ and\ \citenamefont
  {Iwasa}}]{dong2023giant}%
  \BibitemOpen
  \bibfield  {author} {\bibinfo {author} {\bibfnamefont {Y.}~\bibnamefont
  {Dong}}, \bibinfo {author} {\bibfnamefont {M.-M.}\ \bibnamefont {Yang}},
  \bibinfo {author} {\bibfnamefont {M.}~\bibnamefont {Yoshii}}, \bibinfo
  {author} {\bibfnamefont {S.}~\bibnamefont {Matsuoka}}, \bibinfo {author}
  {\bibfnamefont {S.}~\bibnamefont {Kitamura}}, \bibinfo {author}
  {\bibfnamefont {T.}~\bibnamefont {Hasegawa}}, \bibinfo {author}
  {\bibfnamefont {N.}~\bibnamefont {Ogawa}}, \bibinfo {author} {\bibfnamefont
  {T.}~\bibnamefont {Morimoto}}, \bibinfo {author} {\bibfnamefont
  {T.}~\bibnamefont {Ideue}},\ and\ \bibinfo {author} {\bibfnamefont
  {Y.}~\bibnamefont {Iwasa}},\ }\bibfield  {title} {\bibinfo {title} {Giant
  bulk piezophotovoltaic effect in 3r-mos2},\ }\href
  {https://www.nature.com/articles/s41565-022-01252-8} {\bibfield  {journal}
  {\bibinfo  {journal} {Nature nanotechnology}\ }\textbf {\bibinfo {volume}
  {18}},\ \bibinfo {pages} {36} (\bibinfo {year} {2023})}\BibitemShut {NoStop}%
\bibitem [{\citenamefont {Grinberg}\ \emph {et~al.}(2013)\citenamefont
  {Grinberg}, \citenamefont {West}, \citenamefont {Torres}, \citenamefont
  {Gou}, \citenamefont {Stein}, \citenamefont {Wu}, \citenamefont {Chen},
  \citenamefont {Gallo}, \citenamefont {Akbashev}, \citenamefont {Davies} \emph
  {et~al.}}]{grinberg2013perovskite}%
  \BibitemOpen
  \bibfield  {author} {\bibinfo {author} {\bibfnamefont {I.}~\bibnamefont
  {Grinberg}}, \bibinfo {author} {\bibfnamefont {D.~V.}\ \bibnamefont {West}},
  \bibinfo {author} {\bibfnamefont {M.}~\bibnamefont {Torres}}, \bibinfo
  {author} {\bibfnamefont {G.}~\bibnamefont {Gou}}, \bibinfo {author}
  {\bibfnamefont {D.~M.}\ \bibnamefont {Stein}}, \bibinfo {author}
  {\bibfnamefont {L.}~\bibnamefont {Wu}}, \bibinfo {author} {\bibfnamefont
  {G.}~\bibnamefont {Chen}}, \bibinfo {author} {\bibfnamefont {E.~M.}\
  \bibnamefont {Gallo}}, \bibinfo {author} {\bibfnamefont {A.~R.}\ \bibnamefont
  {Akbashev}}, \bibinfo {author} {\bibfnamefont {P.~K.}\ \bibnamefont
  {Davies}}, \emph {et~al.},\ }\bibfield  {title} {\bibinfo {title} {Perovskite
  oxides for visible-light-absorbing ferroelectric and photovoltaic
  materials},\ }\href {https://www.nature.com/articles/nature12622} {\bibfield
  {journal} {\bibinfo  {journal} {Nature}\ }\textbf {\bibinfo {volume} {503}},\
  \bibinfo {pages} {509} (\bibinfo {year} {2013})}\BibitemShut {NoStop}%
\bibitem [{\citenamefont {Singh}\ and\ \citenamefont
  {Hennig}(2014)}]{singh2014computational}%
  \BibitemOpen
  \bibfield  {author} {\bibinfo {author} {\bibfnamefont {A.~K.}\ \bibnamefont
  {Singh}}\ and\ \bibinfo {author} {\bibfnamefont {R.~G.}\ \bibnamefont
  {Hennig}},\ }\bibfield  {title} {\bibinfo {title} {Computational prediction
  of two-dimensional group-iv mono-chalcogenides},\ }\href
  {https://pubs.aip.org/aip/apl/article/105/4/042103/376621/Computational-prediction-of-two-dimensional-group}
  {\bibfield  {journal} {\bibinfo  {journal} {Applied Physics Letters}\
  }\textbf {\bibinfo {volume} {105}} (\bibinfo {year} {2014})}\BibitemShut
  {NoStop}%
\bibitem [{\citenamefont {Gomes}\ and\ \citenamefont
  {Carvalho}(2015)}]{Gomes2015}%
  \BibitemOpen
  \bibfield  {author} {\bibinfo {author} {\bibfnamefont {L.~C.}\ \bibnamefont
  {Gomes}}\ and\ \bibinfo {author} {\bibfnamefont {A.}~\bibnamefont
  {Carvalho}},\ }\bibfield  {title} {\bibinfo {title} {Phosphorene analogues:
  Isoelectronic two-dimensional group-iv monochalcogenides with orthorhombic
  structure},\ }\href {https://doi.org/10.1103/PhysRevB.92.085406} {\bibfield
  {journal} {\bibinfo  {journal} {Phys. Rev. B}\ }\textbf {\bibinfo {volume}
  {92}},\ \bibinfo {pages} {085406} (\bibinfo {year} {2015})}\BibitemShut
  {NoStop}%
\bibitem [{\citenamefont {Young}\ and\ \citenamefont
  {Rappe}(2012)}]{young2012}%
  \BibitemOpen
  \bibfield  {author} {\bibinfo {author} {\bibfnamefont {S.~M.}\ \bibnamefont
  {Young}}\ and\ \bibinfo {author} {\bibfnamefont {A.~M.}\ \bibnamefont
  {Rappe}},\ }\bibfield  {title} {\bibinfo {title} {First principles
  calculation of the shift current photovoltaic effect in ferroelectrics},\
  }\href {https://doi.org/10.1103/PhysRevLett.109.116601} {\bibfield  {journal}
  {\bibinfo  {journal} {Phys. Rev. Lett.}\ }\textbf {\bibinfo {volume} {109}},\
  \bibinfo {pages} {116601} (\bibinfo {year} {2012})}\BibitemShut {NoStop}%
\bibitem [{\citenamefont {Khalaf}\ \emph {et~al.}(2019)\citenamefont {Khalaf},
  \citenamefont {Kruchkov}, \citenamefont {Tarnopolsky},\ and\ \citenamefont
  {Vishwanath}}]{khalaf2019magic}%
  \BibitemOpen
  \bibfield  {author} {\bibinfo {author} {\bibfnamefont {E.}~\bibnamefont
  {Khalaf}}, \bibinfo {author} {\bibfnamefont {A.~J.}\ \bibnamefont
  {Kruchkov}}, \bibinfo {author} {\bibfnamefont {G.}~\bibnamefont
  {Tarnopolsky}},\ and\ \bibinfo {author} {\bibfnamefont {A.}~\bibnamefont
  {Vishwanath}},\ }\bibfield  {title} {\bibinfo {title} {Magic angle hierarchy
  in twisted graphene multilayers},\ }\href
  {https://journals.aps.org/prb/abstract/10.1103/PhysRevB.100.085109}
  {\bibfield  {journal} {\bibinfo  {journal} {Physical Review B}\ }\textbf
  {\bibinfo {volume} {100}},\ \bibinfo {pages} {085109} (\bibinfo {year}
  {2019})}\BibitemShut {NoStop}%
\bibitem [{\citenamefont {Zhang}\ \emph {et~al.}(2022)\citenamefont {Zhang},
  \citenamefont {Polski}, \citenamefont {Lewandowski}, \citenamefont {Thomson},
  \citenamefont {Peng}, \citenamefont {Choi}, \citenamefont {Kim},
  \citenamefont {Watanabe}, \citenamefont {Taniguchi}, \citenamefont {Alicea},
  \citenamefont {von Oppen}, \citenamefont {Refael},\ and\ \citenamefont
  {Nadj-Perge}}]{Zhang2022}%
  \BibitemOpen
  \bibfield  {author} {\bibinfo {author} {\bibfnamefont {Y.}~\bibnamefont
  {Zhang}}, \bibinfo {author} {\bibfnamefont {R.}~\bibnamefont {Polski}},
  \bibinfo {author} {\bibfnamefont {C.}~\bibnamefont {Lewandowski}}, \bibinfo
  {author} {\bibfnamefont {A.}~\bibnamefont {Thomson}}, \bibinfo {author}
  {\bibfnamefont {Y.}~\bibnamefont {Peng}}, \bibinfo {author} {\bibfnamefont
  {Y.}~\bibnamefont {Choi}}, \bibinfo {author} {\bibfnamefont {H.}~\bibnamefont
  {Kim}}, \bibinfo {author} {\bibfnamefont {K.}~\bibnamefont {Watanabe}},
  \bibinfo {author} {\bibfnamefont {T.}~\bibnamefont {Taniguchi}}, \bibinfo
  {author} {\bibfnamefont {J.}~\bibnamefont {Alicea}}, \bibinfo {author}
  {\bibfnamefont {F.}~\bibnamefont {von Oppen}}, \bibinfo {author}
  {\bibfnamefont {G.}~\bibnamefont {Refael}},\ and\ \bibinfo {author}
  {\bibfnamefont {S.}~\bibnamefont {Nadj-Perge}},\ }\bibfield  {title}
  {\bibinfo {title} {Promotion of superconductivity in magic-angle graphene
  multilayers},\ }\href {https://doi.org/10.1126/science.abn8585} {\bibfield
  {journal} {\bibinfo  {journal} {Science}\ }\textbf {\bibinfo {volume}
  {377}},\ \bibinfo {pages} {1538–1543} (\bibinfo {year} {2022})}\BibitemShut
  {NoStop}%
\bibitem [{\citenamefont {Park}\ \emph {et~al.}(2022)\citenamefont {Park},
  \citenamefont {Cao}, \citenamefont {Xia}, \citenamefont {Sun}, \citenamefont
  {Watanabe}, \citenamefont {Taniguchi},\ and\ \citenamefont
  {Jarillo-Herrero}}]{Park2022}%
  \BibitemOpen
  \bibfield  {author} {\bibinfo {author} {\bibfnamefont {J.~M.}\ \bibnamefont
  {Park}}, \bibinfo {author} {\bibfnamefont {Y.}~\bibnamefont {Cao}}, \bibinfo
  {author} {\bibfnamefont {L.-Q.}\ \bibnamefont {Xia}}, \bibinfo {author}
  {\bibfnamefont {S.}~\bibnamefont {Sun}}, \bibinfo {author} {\bibfnamefont
  {K.}~\bibnamefont {Watanabe}}, \bibinfo {author} {\bibfnamefont
  {T.}~\bibnamefont {Taniguchi}},\ and\ \bibinfo {author} {\bibfnamefont
  {P.}~\bibnamefont {Jarillo-Herrero}},\ }\bibfield  {title} {\bibinfo {title}
  {Robust superconductivity in magic-angle multilayer graphene family},\ }\href
  {https://doi.org/10.1038/s41563-022-01287-1} {\bibfield  {journal} {\bibinfo
  {journal} {Nature Materials}\ }\textbf {\bibinfo {volume} {21}},\ \bibinfo
  {pages} {877–883} (\bibinfo {year} {2022})}\BibitemShut {NoStop}%
\bibitem [{\citenamefont {Tan}\ and\ \citenamefont
  {Rappe}(2019{\natexlab{a}})}]{tan2019}%
  \BibitemOpen
  \bibfield  {author} {\bibinfo {author} {\bibfnamefont {L.~Z.}\ \bibnamefont
  {Tan}}\ and\ \bibinfo {author} {\bibfnamefont {A.~M.}\ \bibnamefont
  {Rappe}},\ }\bibfield  {title} {\bibinfo {title} {Upper limit on shift
  current generation in extended systems},\ }\href
  {https://doi.org/10.1103/PhysRevB.100.085102} {\bibfield  {journal} {\bibinfo
   {journal} {Phys. Rev. B}\ }\textbf {\bibinfo {volume} {100}},\ \bibinfo
  {pages} {085102} (\bibinfo {year} {2019}{\natexlab{a}})}\BibitemShut
  {NoStop}%
\bibitem [{\citenamefont {Xiao}\ \emph {et~al.}(2010)\citenamefont {Xiao},
  \citenamefont {Chang},\ and\ \citenamefont {Niu}}]{Xiao2010}%
  \BibitemOpen
  \bibfield  {author} {\bibinfo {author} {\bibfnamefont {D.}~\bibnamefont
  {Xiao}}, \bibinfo {author} {\bibfnamefont {M.-C.}\ \bibnamefont {Chang}},\
  and\ \bibinfo {author} {\bibfnamefont {Q.}~\bibnamefont {Niu}},\ }\bibfield
  {title} {\bibinfo {title} {Berry phase effects on electronic properties},\
  }\href {https://doi.org/10.1103/RevModPhys.82.1959} {\bibfield  {journal}
  {\bibinfo  {journal} {Rev. Mod. Phys.}\ }\textbf {\bibinfo {volume} {82}},\
  \bibinfo {pages} {1959} (\bibinfo {year} {2010})}\BibitemShut {NoStop}%
\bibitem [{\citenamefont {Chaudhary}\ \emph {et~al.}(2022)\citenamefont
  {Chaudhary}, \citenamefont {Lewandowski},\ and\ \citenamefont
  {Refael}}]{shift_current}%
  \BibitemOpen
  \bibfield  {author} {\bibinfo {author} {\bibfnamefont {S.}~\bibnamefont
  {Chaudhary}}, \bibinfo {author} {\bibfnamefont {C.}~\bibnamefont
  {Lewandowski}},\ and\ \bibinfo {author} {\bibfnamefont {G.}~\bibnamefont
  {Refael}},\ }\bibfield  {title} {\bibinfo {title} {Shift-current response as
  a probe of quantum geometry and electron-electron interactions in twisted
  bilayer graphene},\ }\href {https://doi.org/10.1103/PhysRevResearch.4.013164}
  {\bibfield  {journal} {\bibinfo  {journal} {Phys. Rev. Res.}\ }\textbf
  {\bibinfo {volume} {4}},\ \bibinfo {pages} {013164} (\bibinfo {year}
  {2022})}\BibitemShut {NoStop}%
\bibitem [{\citenamefont {Kaplan}\ \emph {et~al.}(2022)\citenamefont {Kaplan},
  \citenamefont {Holder},\ and\ \citenamefont {Yan}}]{kaplan2021}%
  \BibitemOpen
  \bibfield  {author} {\bibinfo {author} {\bibfnamefont {D.}~\bibnamefont
  {Kaplan}}, \bibinfo {author} {\bibfnamefont {T.}~\bibnamefont {Holder}},\
  and\ \bibinfo {author} {\bibfnamefont {B.}~\bibnamefont {Yan}},\ }\bibfield
  {title} {\bibinfo {title} {Twisted photovoltaics at terahertz frequencies
  from momentum shift current},\ }\href
  {https://doi.org/10.1103/PhysRevResearch.4.013209} {\bibfield  {journal}
  {\bibinfo  {journal} {Phys. Rev. Research}\ }\textbf {\bibinfo {volume}
  {4}},\ \bibinfo {pages} {013209} (\bibinfo {year} {2022})}\BibitemShut
  {NoStop}%
\bibitem [{\citenamefont {Andrei}\ \emph {et~al.}(2021)\citenamefont {Andrei},
  \citenamefont {Efetov}, \citenamefont {Jarillo-Herrero}, \citenamefont
  {MacDonald}, \citenamefont {Mak}, \citenamefont {Senthil}, \citenamefont
  {Tutuc}, \citenamefont {Yazdani},\ and\ \citenamefont {Young}}]{Andrei2021}%
  \BibitemOpen
  \bibfield  {author} {\bibinfo {author} {\bibfnamefont {E.~Y.}\ \bibnamefont
  {Andrei}}, \bibinfo {author} {\bibfnamefont {D.~K.}\ \bibnamefont {Efetov}},
  \bibinfo {author} {\bibfnamefont {P.}~\bibnamefont {Jarillo-Herrero}},
  \bibinfo {author} {\bibfnamefont {A.~H.}\ \bibnamefont {MacDonald}}, \bibinfo
  {author} {\bibfnamefont {K.~F.}\ \bibnamefont {Mak}}, \bibinfo {author}
  {\bibfnamefont {T.}~\bibnamefont {Senthil}}, \bibinfo {author} {\bibfnamefont
  {E.}~\bibnamefont {Tutuc}}, \bibinfo {author} {\bibfnamefont
  {A.}~\bibnamefont {Yazdani}},\ and\ \bibinfo {author} {\bibfnamefont {A.~F.}\
  \bibnamefont {Young}},\ }\bibfield  {title} {\bibinfo {title} {The marvels of
  moir{\'e}materials},\ }\href {https://doi.org/10.1038/s41578-021-00284-1}
  {\bibfield  {journal} {\bibinfo  {journal} {Nature Reviews Materials}\
  }\textbf {\bibinfo {volume} {6}},\ \bibinfo {pages} {201} (\bibinfo {year}
  {2021})}\BibitemShut {NoStop}%
\bibitem [{\citenamefont {Balents}\ \emph {et~al.}(2020)\citenamefont
  {Balents}, \citenamefont {Dean}, \citenamefont {Efetov},\ and\ \citenamefont
  {Young}}]{balents2020superconductivity}%
  \BibitemOpen
  \bibfield  {author} {\bibinfo {author} {\bibfnamefont {L.}~\bibnamefont
  {Balents}}, \bibinfo {author} {\bibfnamefont {C.~R.}\ \bibnamefont {Dean}},
  \bibinfo {author} {\bibfnamefont {D.~K.}\ \bibnamefont {Efetov}},\ and\
  \bibinfo {author} {\bibfnamefont {A.~F.}\ \bibnamefont {Young}},\ }\bibfield
  {title} {\bibinfo {title} {Superconductivity and strong correlations in
  moir{\'e} flat bands},\ }\href {superconductivity and strong correlations in
  moire materials} {\bibfield  {journal} {\bibinfo  {journal} {Nature Physics}\
  }\textbf {\bibinfo {volume} {16}},\ \bibinfo {pages} {725} (\bibinfo {year}
  {2020})}\BibitemShut {NoStop}%
\bibitem [{\citenamefont {T{\"o}rm{\"a}}\ \emph {et~al.}(2022)\citenamefont
  {T{\"o}rm{\"a}}, \citenamefont {Peotta},\ and\ \citenamefont
  {Bernevig}}]{torma2022superconductivity}%
  \BibitemOpen
  \bibfield  {author} {\bibinfo {author} {\bibfnamefont {P.}~\bibnamefont
  {T{\"o}rm{\"a}}}, \bibinfo {author} {\bibfnamefont {S.}~\bibnamefont
  {Peotta}},\ and\ \bibinfo {author} {\bibfnamefont {B.~A.}\ \bibnamefont
  {Bernevig}},\ }\bibfield  {title} {\bibinfo {title} {Superconductivity,
  superfluidity and quantum geometry in twisted multilayer systems},\ }\href
  {https://www.nature.com/articles/s42254-022-00466-y} {\bibfield  {journal}
  {\bibinfo  {journal} {Nature Reviews Physics}\ }\textbf {\bibinfo {volume}
  {4}},\ \bibinfo {pages} {528} (\bibinfo {year} {2022})}\BibitemShut {NoStop}%
\bibitem [{\citenamefont {Rice}\ and\ \citenamefont
  {Mele}(1982)}]{PhysRevLett.49.1455}%
  \BibitemOpen
  \bibfield  {author} {\bibinfo {author} {\bibfnamefont {M.~J.}\ \bibnamefont
  {Rice}}\ and\ \bibinfo {author} {\bibfnamefont {E.~J.}\ \bibnamefont
  {Mele}},\ }\bibfield  {title} {\bibinfo {title} {Elementary excitations of a
  linearly conjugated diatomic polymer},\ }\href
  {https://doi.org/10.1103/PhysRevLett.49.1455} {\bibfield  {journal} {\bibinfo
   {journal} {Phys. Rev. Lett.}\ }\textbf {\bibinfo {volume} {49}},\ \bibinfo
  {pages} {1455} (\bibinfo {year} {1982})}\BibitemShut {NoStop}%
\bibitem [{\citenamefont {T\"orm\"a}(2023)}]{Torma2023}%
  \BibitemOpen
  \bibfield  {author} {\bibinfo {author} {\bibfnamefont {P.}~\bibnamefont
  {T\"orm\"a}},\ }\bibfield  {title} {\bibinfo {title} {Essay: Where can
  quantum geometry lead us?},\ }\href
  {https://doi.org/10.1103/PhysRevLett.131.240001} {\bibfield  {journal}
  {\bibinfo  {journal} {Phys. Rev. Lett.}\ }\textbf {\bibinfo {volume} {131}},\
  \bibinfo {pages} {240001} (\bibinfo {year} {2023})}\BibitemShut {NoStop}%
\bibitem [{\citenamefont {Verma}\ \emph {et~al.}(2021)\citenamefont {Verma},
  \citenamefont {Hazra},\ and\ \citenamefont {Randeria}}]{verma2021optical}%
  \BibitemOpen
  \bibfield  {author} {\bibinfo {author} {\bibfnamefont {N.}~\bibnamefont
  {Verma}}, \bibinfo {author} {\bibfnamefont {T.}~\bibnamefont {Hazra}},\ and\
  \bibinfo {author} {\bibfnamefont {M.}~\bibnamefont {Randeria}},\ }\bibfield
  {title} {\bibinfo {title} {Optical spectral weight, phase stiffness, and t c
  bounds for trivial and topological flat band superconductors},\ }\href
  {https://www.pnas.org/doi/abs/10.1073/pnas.2106744118} {\bibfield  {journal}
  {\bibinfo  {journal} {Proceedings of the National Academy of Sciences}\
  }\textbf {\bibinfo {volume} {118}},\ \bibinfo {pages} {e2106744118} (\bibinfo
  {year} {2021})}\BibitemShut {NoStop}%
\bibitem [{\citenamefont {Herzog-Arbeitman}\ \emph {et~al.}(2022)\citenamefont
  {Herzog-Arbeitman}, \citenamefont {Peri}, \citenamefont {Schindler},
  \citenamefont {Huber},\ and\ \citenamefont {Bernevig}}]{Herzog2022}%
  \BibitemOpen
  \bibfield  {author} {\bibinfo {author} {\bibfnamefont {J.}~\bibnamefont
  {Herzog-Arbeitman}}, \bibinfo {author} {\bibfnamefont {V.}~\bibnamefont
  {Peri}}, \bibinfo {author} {\bibfnamefont {F.}~\bibnamefont {Schindler}},
  \bibinfo {author} {\bibfnamefont {S.~D.}\ \bibnamefont {Huber}},\ and\
  \bibinfo {author} {\bibfnamefont {B.~A.}\ \bibnamefont {Bernevig}},\
  }\bibfield  {title} {\bibinfo {title} {Superfluid weight bounds from symmetry
  and quantum geometry in flat bands},\ }\href
  {https://doi.org/10.1103/PhysRevLett.128.087002} {\bibfield  {journal}
  {\bibinfo  {journal} {Phys. Rev. Lett.}\ }\textbf {\bibinfo {volume} {128}},\
  \bibinfo {pages} {087002} (\bibinfo {year} {2022})}\BibitemShut {NoStop}%
\bibitem [{\citenamefont {Xie}\ \emph {et~al.}(2020{\natexlab{a}})\citenamefont
  {Xie}, \citenamefont {Song}, \citenamefont {Lian},\ and\ \citenamefont
  {Bernevig}}]{Xie2020}%
  \BibitemOpen
  \bibfield  {author} {\bibinfo {author} {\bibfnamefont {F.}~\bibnamefont
  {Xie}}, \bibinfo {author} {\bibfnamefont {Z.}~\bibnamefont {Song}}, \bibinfo
  {author} {\bibfnamefont {B.}~\bibnamefont {Lian}},\ and\ \bibinfo {author}
  {\bibfnamefont {B.~A.}\ \bibnamefont {Bernevig}},\ }\bibfield  {title}
  {\bibinfo {title} {Topology-bounded superfluid weight in twisted bilayer
  graphene},\ }\href {https://doi.org/10.1103/PhysRevLett.124.167002}
  {\bibfield  {journal} {\bibinfo  {journal} {Phys. Rev. Lett.}\ }\textbf
  {\bibinfo {volume} {124}},\ \bibinfo {pages} {167002} (\bibinfo {year}
  {2020}{\natexlab{a}})}\BibitemShut {NoStop}%
\bibitem [{\citenamefont {Provost}\ and\ \citenamefont
  {Vallee}(1980)}]{Provost1980}%
  \BibitemOpen
  \bibfield  {author} {\bibinfo {author} {\bibfnamefont {J.~P.}\ \bibnamefont
  {Provost}}\ and\ \bibinfo {author} {\bibfnamefont {G.}~\bibnamefont
  {Vallee}},\ }\bibfield  {title} {\bibinfo {title} {Riemannian structure on
  manifolds of quantum states},\ }\href {https://doi.org/10.1007/bf02193559}
  {\bibfield  {journal} {\bibinfo  {journal} {Communications in Mathematical
  Physics}\ }\textbf {\bibinfo {volume} {76}},\ \bibinfo {pages} {289–301}
  (\bibinfo {year} {1980})}\BibitemShut {NoStop}%
\bibitem [{\citenamefont {Peotta}\ and\ \citenamefont
  {T{\"o}rm{\"a}}(2015)}]{Torma15}%
  \BibitemOpen
  \bibfield  {author} {\bibinfo {author} {\bibfnamefont {S.}~\bibnamefont
  {Peotta}}\ and\ \bibinfo {author} {\bibfnamefont {P.}~\bibnamefont
  {T{\"o}rm{\"a}}},\ }\bibfield  {title} {\bibinfo {title} {Superfluidity in
  topologically nontrivial flat bands},\ }\href
  {https://doi.org/10.1038/ncomms9944} {\bibfield  {journal} {\bibinfo
  {journal} {Nature Communications}\ }\textbf {\bibinfo {volume} {6}},\
  \bibinfo {pages} {8944} (\bibinfo {year} {2015})}\BibitemShut {NoStop}%
\bibitem [{\citenamefont {Julku}\ \emph {et~al.}(2020)\citenamefont {Julku},
  \citenamefont {Peltonen}, \citenamefont {Liang}, \citenamefont {Heikkil\"a},\
  and\ \citenamefont {T\"orm\"a}}]{PhysRevB.101.060505}%
  \BibitemOpen
  \bibfield  {author} {\bibinfo {author} {\bibfnamefont {A.}~\bibnamefont
  {Julku}}, \bibinfo {author} {\bibfnamefont {T.~J.}\ \bibnamefont {Peltonen}},
  \bibinfo {author} {\bibfnamefont {L.}~\bibnamefont {Liang}}, \bibinfo
  {author} {\bibfnamefont {T.~T.}\ \bibnamefont {Heikkil\"a}},\ and\ \bibinfo
  {author} {\bibfnamefont {P.}~\bibnamefont {T\"orm\"a}},\ }\bibfield  {title}
  {\bibinfo {title} {Superfluid weight and berezinskii-kosterlitz-thouless
  transition temperature of twisted bilayer graphene},\ }\href
  {https://doi.org/10.1103/PhysRevB.101.060505} {\bibfield  {journal} {\bibinfo
   {journal} {Phys. Rev. B}\ }\textbf {\bibinfo {volume} {101}},\ \bibinfo
  {pages} {060505} (\bibinfo {year} {2020})}\BibitemShut {NoStop}%
\bibitem [{\citenamefont {Tan}\ and\ \citenamefont
  {Rappe}(2019{\natexlab{b}})}]{Tan_2019}%
  \BibitemOpen
  \bibfield  {author} {\bibinfo {author} {\bibfnamefont {L.~Z.}\ \bibnamefont
  {Tan}}\ and\ \bibinfo {author} {\bibfnamefont {A.~M.}\ \bibnamefont
  {Rappe}},\ }\bibfield  {title} {\bibinfo {title} {Effect of wavefunction
  delocalization on shift current generation},\ }\href
  {https://doi.org/10.1088/1361-648X/aaf74b} {\bibfield  {journal} {\bibinfo
  {journal} {Journal of Physics: Condensed Matter}\ }\textbf {\bibinfo {volume}
  {31}},\ \bibinfo {pages} {084002} (\bibinfo {year}
  {2019}{\natexlab{b}})}\BibitemShut {NoStop}%
\bibitem [{\citenamefont {Koshino}\ \emph {et~al.}(2018)\citenamefont
  {Koshino}, \citenamefont {Yuan}, \citenamefont {Koretsune}, \citenamefont
  {Ochi}, \citenamefont {Kuroki},\ and\ \citenamefont
  {Fu}}]{koshino2018maximally}%
  \BibitemOpen
  \bibfield  {author} {\bibinfo {author} {\bibfnamefont {M.}~\bibnamefont
  {Koshino}}, \bibinfo {author} {\bibfnamefont {N.~F.}\ \bibnamefont {Yuan}},
  \bibinfo {author} {\bibfnamefont {T.}~\bibnamefont {Koretsune}}, \bibinfo
  {author} {\bibfnamefont {M.}~\bibnamefont {Ochi}}, \bibinfo {author}
  {\bibfnamefont {K.}~\bibnamefont {Kuroki}},\ and\ \bibinfo {author}
  {\bibfnamefont {L.}~\bibnamefont {Fu}},\ }\bibfield  {title} {\bibinfo
  {title} {Maximally localized wannier orbitals and the extended hubbard model
  for twisted bilayer graphene},\ }\href
  {https://journals.aps.org/prx/abstract/10.1103/PhysRevX.8.031087} {\bibfield
  {journal} {\bibinfo  {journal} {Physical Review X}\ }\textbf {\bibinfo
  {volume} {8}},\ \bibinfo {pages} {031087} (\bibinfo {year}
  {2018})}\BibitemShut {NoStop}%
\bibitem [{\citenamefont {Cea}\ and\ \citenamefont
  {Guinea}(2021)}]{cea2021coulomb}%
  \BibitemOpen
  \bibfield  {author} {\bibinfo {author} {\bibfnamefont {T.}~\bibnamefont
  {Cea}}\ and\ \bibinfo {author} {\bibfnamefont {F.}~\bibnamefont {Guinea}},\
  }\bibfield  {title} {\bibinfo {title} {Coulomb interaction, phonons, and
  superconductivity in twisted bilayer graphene},\ }\href
  {https://arxiv.org/abs/2103.16132} {\bibfield  {journal} {\bibinfo  {journal}
  {arXiv:2103.01815}\ } (\bibinfo {year} {2021})}\BibitemShut {NoStop}%
\bibitem [{\citenamefont {Guinea}\ and\ \citenamefont
  {Walet}(2018)}]{guineaElectrostaticEffectsBand2018}%
  \BibitemOpen
  \bibfield  {author} {\bibinfo {author} {\bibfnamefont {F.}~\bibnamefont
  {Guinea}}\ and\ \bibinfo {author} {\bibfnamefont {N.~R.}\ \bibnamefont
  {Walet}},\ }\bibfield  {title} {\bibinfo {title} {Electrostatic effects, band
  distortions, and superconductivity in twisted graphene bilayers},\ }\href
  {https://doi.org/10.1073/pnas.1810947115} {\bibfield  {journal} {\bibinfo
  {journal} {Proceedings of the National Academy of Sciences}\ }\textbf
  {\bibinfo {volume} {115}},\ \bibinfo {pages} {13174} (\bibinfo {year}
  {2018})}\BibitemShut {NoStop}%
\bibitem [{\citenamefont {Pantale\'on}\ \emph {et~al.}(2022)\citenamefont
  {Pantale\'on}, \citenamefont {Phong}, \citenamefont {Naumis},\ and\
  \citenamefont {Guinea}}]{pantale2022}%
  \BibitemOpen
  \bibfield  {author} {\bibinfo {author} {\bibfnamefont {P.~A.}\ \bibnamefont
  {Pantale\'on}}, \bibinfo {author} {\bibfnamefont {V.~o.~T.}\ \bibnamefont
  {Phong}}, \bibinfo {author} {\bibfnamefont {G.~G.}\ \bibnamefont {Naumis}},\
  and\ \bibinfo {author} {\bibfnamefont {F.}~\bibnamefont {Guinea}},\
  }\bibfield  {title} {\bibinfo {title} {Interaction-enhanced topological hall
  effects in strained twisted bilayer graphene},\ }\href
  {https://doi.org/10.1103/PhysRevB.106.L161101} {\bibfield  {journal}
  {\bibinfo  {journal} {Phys. Rev. B}\ }\textbf {\bibinfo {volume} {106}},\
  \bibinfo {pages} {L161101} (\bibinfo {year} {2022})}\BibitemShut {NoStop}%
\bibitem [{\citenamefont {Goodwin}\ \emph {et~al.}(2020)\citenamefont
  {Goodwin}, \citenamefont {Vitale}, \citenamefont {Liang}, \citenamefont
  {Mostofi},\ and\ \citenamefont
  {Lischner}}]{goodwinHartreeTheoryCalculations2020}%
  \BibitemOpen
  \bibfield  {author} {\bibinfo {author} {\bibfnamefont {Z.~A.}\ \bibnamefont
  {Goodwin}}, \bibinfo {author} {\bibfnamefont {V.}~\bibnamefont {Vitale}},
  \bibinfo {author} {\bibfnamefont {X.}~\bibnamefont {Liang}}, \bibinfo
  {author} {\bibfnamefont {A.~A.}\ \bibnamefont {Mostofi}},\ and\ \bibinfo
  {author} {\bibfnamefont {J.}~\bibnamefont {Lischner}},\ }\bibfield  {title}
  {\bibinfo {title} {Hartree theory calculations of quasiparticle properties in
  twisted bilayer graphene},\ }\href
  {https://iopscience.iop.org/article/10.1088/2516-1075/ab9f94} {\bibfield
  {journal} {\bibinfo  {journal} {Electronic Structure}\ }\textbf {\bibinfo
  {volume} {2}},\ \bibinfo {pages} {034001} (\bibinfo {year}
  {2020})}\BibitemShut {NoStop}%
\bibitem [{\citenamefont {Rademaker}\ \emph {et~al.}(2019)\citenamefont
  {Rademaker}, \citenamefont {Abanin},\ and\ \citenamefont
  {Mellado}}]{PhysRevB.100.205114}%
  \BibitemOpen
  \bibfield  {author} {\bibinfo {author} {\bibfnamefont {L.}~\bibnamefont
  {Rademaker}}, \bibinfo {author} {\bibfnamefont {D.~A.}\ \bibnamefont
  {Abanin}},\ and\ \bibinfo {author} {\bibfnamefont {P.}~\bibnamefont
  {Mellado}},\ }\bibfield  {title} {\bibinfo {title} {Charge smoothening and
  band flattening due to hartree corrections in twisted bilayer graphene},\
  }\href {https://doi.org/10.1103/PhysRevB.100.205114} {\bibfield  {journal}
  {\bibinfo  {journal} {Phys. Rev. B}\ }\textbf {\bibinfo {volume} {100}},\
  \bibinfo {pages} {205114} (\bibinfo {year} {2019})}\BibitemShut {NoStop}%
\bibitem [{\citenamefont {Klebl}\ \emph {et~al.}(2021)\citenamefont {Klebl},
  \citenamefont {Goodwin}, \citenamefont {Mostofi}, \citenamefont {Kennes},\
  and\ \citenamefont {Lischner}}]{PhysRevB.103.195127}%
  \BibitemOpen
  \bibfield  {author} {\bibinfo {author} {\bibfnamefont {L.}~\bibnamefont
  {Klebl}}, \bibinfo {author} {\bibfnamefont {Z.~A.~H.}\ \bibnamefont
  {Goodwin}}, \bibinfo {author} {\bibfnamefont {A.~A.}\ \bibnamefont
  {Mostofi}}, \bibinfo {author} {\bibfnamefont {D.~M.}\ \bibnamefont
  {Kennes}},\ and\ \bibinfo {author} {\bibfnamefont {J.}~\bibnamefont
  {Lischner}},\ }\bibfield  {title} {\bibinfo {title} {Importance of
  long-ranged electron-electron interactions for the magnetic phase diagram of
  twisted bilayer graphene},\ }\href
  {https://doi.org/10.1103/PhysRevB.103.195127} {\bibfield  {journal} {\bibinfo
   {journal} {Phys. Rev. B}\ }\textbf {\bibinfo {volume} {103}},\ \bibinfo
  {pages} {195127} (\bibinfo {year} {2021})}\BibitemShut {NoStop}%
\bibitem [{\citenamefont {Rademaker}\ and\ \citenamefont
  {Mellado}(2018)}]{PhysRevB.98.235158}%
  \BibitemOpen
  \bibfield  {author} {\bibinfo {author} {\bibfnamefont {L.}~\bibnamefont
  {Rademaker}}\ and\ \bibinfo {author} {\bibfnamefont {P.}~\bibnamefont
  {Mellado}},\ }\bibfield  {title} {\bibinfo {title} {Charge-transfer
  insulation in twisted bilayer graphene},\ }\href
  {https://doi.org/10.1103/PhysRevB.98.235158} {\bibfield  {journal} {\bibinfo
  {journal} {Phys. Rev. B}\ }\textbf {\bibinfo {volume} {98}},\ \bibinfo
  {pages} {235158} (\bibinfo {year} {2018})}\BibitemShut {NoStop}%
\bibitem [{\citenamefont {Calder\'on}\ and\ \citenamefont
  {Bascones}(2020)}]{PhysRevB.102.155149}%
  \BibitemOpen
  \bibfield  {author} {\bibinfo {author} {\bibfnamefont {M.~J.}\ \bibnamefont
  {Calder\'on}}\ and\ \bibinfo {author} {\bibfnamefont {E.}~\bibnamefont
  {Bascones}},\ }\bibfield  {title} {\bibinfo {title} {Interactions in the
  8-orbital model for twisted bilayer graphene},\ }\href
  {https://doi.org/10.1103/PhysRevB.102.155149} {\bibfield  {journal} {\bibinfo
   {journal} {Phys. Rev. B}\ }\textbf {\bibinfo {volume} {102}},\ \bibinfo
  {pages} {155149} (\bibinfo {year} {2020})}\BibitemShut {NoStop}%
\bibitem [{\citenamefont {Xie}\ and\ \citenamefont
  {MacDonald}(2020)}]{PhysRevLett.124.097601}%
  \BibitemOpen
  \bibfield  {author} {\bibinfo {author} {\bibfnamefont {M.}~\bibnamefont
  {Xie}}\ and\ \bibinfo {author} {\bibfnamefont {A.~H.}\ \bibnamefont
  {MacDonald}},\ }\bibfield  {title} {\bibinfo {title} {Nature of the
  correlated insulator states in twisted bilayer graphene},\ }\href
  {https://doi.org/10.1103/PhysRevLett.124.097601} {\bibfield  {journal}
  {\bibinfo  {journal} {Phys. Rev. Lett.}\ }\textbf {\bibinfo {volume} {124}},\
  \bibinfo {pages} {097601} (\bibinfo {year} {2020})}\BibitemShut {NoStop}%
\bibitem [{\citenamefont {Bernevig}\ \emph {et~al.}(2021)\citenamefont
  {Bernevig}, \citenamefont {Lian}, \citenamefont {Cowsik}, \citenamefont
  {Xie}, \citenamefont {Regnault},\ and\ \citenamefont
  {Song}}]{PhysRevB.103.205415}%
  \BibitemOpen
  \bibfield  {author} {\bibinfo {author} {\bibfnamefont {B.~A.}\ \bibnamefont
  {Bernevig}}, \bibinfo {author} {\bibfnamefont {B.}~\bibnamefont {Lian}},
  \bibinfo {author} {\bibfnamefont {A.}~\bibnamefont {Cowsik}}, \bibinfo
  {author} {\bibfnamefont {F.}~\bibnamefont {Xie}}, \bibinfo {author}
  {\bibfnamefont {N.}~\bibnamefont {Regnault}},\ and\ \bibinfo {author}
  {\bibfnamefont {Z.-D.}\ \bibnamefont {Song}},\ }\bibfield  {title} {\bibinfo
  {title} {Twisted bilayer graphene. v. exact analytic many-body excitations in
  coulomb hamiltonians: Charge gap, goldstone modes, and absence of cooper
  pairing},\ }\href {https://doi.org/10.1103/PhysRevB.103.205415} {\bibfield
  {journal} {\bibinfo  {journal} {Phys. Rev. B}\ }\textbf {\bibinfo {volume}
  {103}},\ \bibinfo {pages} {205415} (\bibinfo {year} {2021})}\BibitemShut
  {NoStop}%
\bibitem [{\citenamefont {Kang}\ and\ \citenamefont
  {Vafek}(2019)}]{PhysRevLett.122.246401}%
  \BibitemOpen
  \bibfield  {author} {\bibinfo {author} {\bibfnamefont {J.}~\bibnamefont
  {Kang}}\ and\ \bibinfo {author} {\bibfnamefont {O.}~\bibnamefont {Vafek}},\
  }\bibfield  {title} {\bibinfo {title} {Strong coupling phases of partially
  filled twisted bilayer graphene narrow bands},\ }\href
  {https://doi.org/10.1103/PhysRevLett.122.246401} {\bibfield  {journal}
  {\bibinfo  {journal} {Phys. Rev. Lett.}\ }\textbf {\bibinfo {volume} {122}},\
  \bibinfo {pages} {246401} (\bibinfo {year} {2019})}\BibitemShut {NoStop}%
\bibitem [{\citenamefont {Bultinck}\ \emph {et~al.}(2020)\citenamefont
  {Bultinck}, \citenamefont {Khalaf}, \citenamefont {Liu}, \citenamefont
  {Chatterjee}, \citenamefont {Vishwanath},\ and\ \citenamefont
  {Zaletel}}]{PhysRevX.10.031034}%
  \BibitemOpen
  \bibfield  {author} {\bibinfo {author} {\bibfnamefont {N.}~\bibnamefont
  {Bultinck}}, \bibinfo {author} {\bibfnamefont {E.}~\bibnamefont {Khalaf}},
  \bibinfo {author} {\bibfnamefont {S.}~\bibnamefont {Liu}}, \bibinfo {author}
  {\bibfnamefont {S.}~\bibnamefont {Chatterjee}}, \bibinfo {author}
  {\bibfnamefont {A.}~\bibnamefont {Vishwanath}},\ and\ \bibinfo {author}
  {\bibfnamefont {M.~P.}\ \bibnamefont {Zaletel}},\ }\bibfield  {title}
  {\bibinfo {title} {Ground state and hidden symmetry of magic-angle graphene
  at even integer filling},\ }\href
  {https://doi.org/10.1103/PhysRevX.10.031034} {\bibfield  {journal} {\bibinfo
  {journal} {Phys. Rev. X}\ }\textbf {\bibinfo {volume} {10}},\ \bibinfo
  {pages} {031034} (\bibinfo {year} {2020})}\BibitemShut {NoStop}%
\bibitem [{\citenamefont {Lian}\ \emph {et~al.}(2021)\citenamefont {Lian},
  \citenamefont {Song}, \citenamefont {Regnault}, \citenamefont {Efetov},
  \citenamefont {Yazdani},\ and\ \citenamefont
  {Bernevig}}]{PhysRevB.103.205414}%
  \BibitemOpen
  \bibfield  {author} {\bibinfo {author} {\bibfnamefont {B.}~\bibnamefont
  {Lian}}, \bibinfo {author} {\bibfnamefont {Z.-D.}\ \bibnamefont {Song}},
  \bibinfo {author} {\bibfnamefont {N.}~\bibnamefont {Regnault}}, \bibinfo
  {author} {\bibfnamefont {D.~K.}\ \bibnamefont {Efetov}}, \bibinfo {author}
  {\bibfnamefont {A.}~\bibnamefont {Yazdani}},\ and\ \bibinfo {author}
  {\bibfnamefont {B.~A.}\ \bibnamefont {Bernevig}},\ }\bibfield  {title}
  {\bibinfo {title} {Twisted bilayer graphene. iv. exact insulator ground
  states and phase diagram},\ }\href
  {https://doi.org/10.1103/PhysRevB.103.205414} {\bibfield  {journal} {\bibinfo
   {journal} {Phys. Rev. B}\ }\textbf {\bibinfo {volume} {103}},\ \bibinfo
  {pages} {205414} (\bibinfo {year} {2021})}\BibitemShut {NoStop}%
\bibitem [{\citenamefont {Kwan}\ \emph {et~al.}(2021)\citenamefont {Kwan},
  \citenamefont {Wagner}, \citenamefont {Soejima}, \citenamefont {Zaletel},
  \citenamefont {Simon}, \citenamefont {Parameswaran},\ and\ \citenamefont
  {Bultinck}}]{PhysRevX.11.041063}%
  \BibitemOpen
  \bibfield  {author} {\bibinfo {author} {\bibfnamefont {Y.~H.}\ \bibnamefont
  {Kwan}}, \bibinfo {author} {\bibfnamefont {G.}~\bibnamefont {Wagner}},
  \bibinfo {author} {\bibfnamefont {T.}~\bibnamefont {Soejima}}, \bibinfo
  {author} {\bibfnamefont {M.~P.}\ \bibnamefont {Zaletel}}, \bibinfo {author}
  {\bibfnamefont {S.~H.}\ \bibnamefont {Simon}}, \bibinfo {author}
  {\bibfnamefont {S.~A.}\ \bibnamefont {Parameswaran}},\ and\ \bibinfo {author}
  {\bibfnamefont {N.}~\bibnamefont {Bultinck}},\ }\bibfield  {title} {\bibinfo
  {title} {Kekul\'e spiral order at all nonzero integer fillings in twisted
  bilayer graphene},\ }\href {https://doi.org/10.1103/PhysRevX.11.041063}
  {\bibfield  {journal} {\bibinfo  {journal} {Phys. Rev. X}\ }\textbf {\bibinfo
  {volume} {11}},\ \bibinfo {pages} {041063} (\bibinfo {year}
  {2021})}\BibitemShut {NoStop}%
\bibitem [{\citenamefont {Kol\'a\ifmmode~\check{r}\else \v{r}\fi{}}\ \emph
  {et~al.}(2023)\citenamefont {Kol\'a\ifmmode~\check{r}\else \v{r}\fi{}},
  \citenamefont {Zhang}, \citenamefont {Nadj-Perge}, \citenamefont {von
  Oppen},\ and\ \citenamefont {Lewandowski}}]{kolar2023}%
  \BibitemOpen
  \bibfield  {author} {\bibinfo {author} {\bibfnamefont {K.~c.~v.}\
  \bibnamefont {Kol\'a\ifmmode~\check{r}\else \v{r}\fi{}}}, \bibinfo {author}
  {\bibfnamefont {Y.}~\bibnamefont {Zhang}}, \bibinfo {author} {\bibfnamefont
  {S.}~\bibnamefont {Nadj-Perge}}, \bibinfo {author} {\bibfnamefont
  {F.}~\bibnamefont {von Oppen}},\ and\ \bibinfo {author} {\bibfnamefont
  {C.}~\bibnamefont {Lewandowski}},\ }\bibfield  {title} {\bibinfo {title}
  {Electrostatic fate of $n$-layer moir\'e graphene},\ }\href
  {https://doi.org/10.1103/PhysRevB.108.195148} {\bibfield  {journal} {\bibinfo
   {journal} {Phys. Rev. B}\ }\textbf {\bibinfo {volume} {108}},\ \bibinfo
  {pages} {195148} (\bibinfo {year} {2023})}\BibitemShut {NoStop}%
\bibitem [{\citenamefont {Cao}\ \emph {et~al.}(2018{\natexlab{a}})\citenamefont
  {Cao}, \citenamefont {Fatemi}, \citenamefont {Demir}, \citenamefont {Fang},
  \citenamefont {Tomarken}, \citenamefont {Luo}, \citenamefont
  {Sanchez-Yamagishi}, \citenamefont {Watanabe}, \citenamefont {Taniguchi},
  \citenamefont {Kaxiras}, \citenamefont {Ashoori},\ and\ \citenamefont
  {Jarillo-Herrero}}]{cao1}%
  \BibitemOpen
  \bibfield  {author} {\bibinfo {author} {\bibfnamefont {Y.}~\bibnamefont
  {Cao}}, \bibinfo {author} {\bibfnamefont {V.}~\bibnamefont {Fatemi}},
  \bibinfo {author} {\bibfnamefont {A.}~\bibnamefont {Demir}}, \bibinfo
  {author} {\bibfnamefont {S.}~\bibnamefont {Fang}}, \bibinfo {author}
  {\bibfnamefont {S.~L.}\ \bibnamefont {Tomarken}}, \bibinfo {author}
  {\bibfnamefont {J.~Y.}\ \bibnamefont {Luo}}, \bibinfo {author} {\bibfnamefont
  {J.~D.}\ \bibnamefont {Sanchez-Yamagishi}}, \bibinfo {author} {\bibfnamefont
  {K.}~\bibnamefont {Watanabe}}, \bibinfo {author} {\bibfnamefont
  {T.}~\bibnamefont {Taniguchi}}, \bibinfo {author} {\bibfnamefont
  {E.}~\bibnamefont {Kaxiras}}, \bibinfo {author} {\bibfnamefont {R.~C.}\
  \bibnamefont {Ashoori}},\ and\ \bibinfo {author} {\bibfnamefont
  {P.}~\bibnamefont {Jarillo-Herrero}},\ }\bibfield  {title} {\bibinfo {title}
  {Correlated insulator behaviour at half-filling in magic-angle graphene
  superlattices},\ }\href {https://doi.org/10.1038/nature26154} {\bibfield
  {journal} {\bibinfo  {journal} {Nature}\ }\textbf {\bibinfo {volume} {556}},\
  \bibinfo {pages} {80 EP } (\bibinfo {year} {2018}{\natexlab{a}})}\BibitemShut
  {NoStop}%
\bibitem [{\citenamefont {Cao}\ \emph {et~al.}(2018{\natexlab{b}})\citenamefont
  {Cao}, \citenamefont {Fatemi}, \citenamefont {Fang}, \citenamefont
  {Watanabe}, \citenamefont {Taniguchi}, \citenamefont {Kaxiras},\ and\
  \citenamefont {Jarillo-Herrero}}]{cao2}%
  \BibitemOpen
  \bibfield  {author} {\bibinfo {author} {\bibfnamefont {Y.}~\bibnamefont
  {Cao}}, \bibinfo {author} {\bibfnamefont {V.}~\bibnamefont {Fatemi}},
  \bibinfo {author} {\bibfnamefont {S.}~\bibnamefont {Fang}}, \bibinfo {author}
  {\bibfnamefont {K.}~\bibnamefont {Watanabe}}, \bibinfo {author}
  {\bibfnamefont {T.}~\bibnamefont {Taniguchi}}, \bibinfo {author}
  {\bibfnamefont {E.}~\bibnamefont {Kaxiras}},\ and\ \bibinfo {author}
  {\bibfnamefont {P.}~\bibnamefont {Jarillo-Herrero}},\ }\bibfield  {title}
  {\bibinfo {title} {Unconventional superconductivity in magic-angle graphene
  superlattices},\ }\href {https://doi.org/10.1038/nature26160} {\bibfield
  {journal} {\bibinfo  {journal} {Nature}\ }\textbf {\bibinfo {volume} {556}},\
  \bibinfo {pages} {43} (\bibinfo {year} {2018}{\natexlab{b}})}\BibitemShut
  {NoStop}%
\bibitem [{\citenamefont {Yankowitz}\ \emph {et~al.}(2019)\citenamefont
  {Yankowitz}, \citenamefont {Chen}, \citenamefont {Polshyn}, \citenamefont
  {Zhang}, \citenamefont {Watanabe}, \citenamefont {Taniguchi}, \citenamefont
  {Graf}, \citenamefont {Young},\ and\ \citenamefont
  {Dean}}]{yankowitz2019tuning}%
  \BibitemOpen
  \bibfield  {author} {\bibinfo {author} {\bibfnamefont {M.}~\bibnamefont
  {Yankowitz}}, \bibinfo {author} {\bibfnamefont {S.}~\bibnamefont {Chen}},
  \bibinfo {author} {\bibfnamefont {H.}~\bibnamefont {Polshyn}}, \bibinfo
  {author} {\bibfnamefont {Y.}~\bibnamefont {Zhang}}, \bibinfo {author}
  {\bibfnamefont {K.}~\bibnamefont {Watanabe}}, \bibinfo {author}
  {\bibfnamefont {T.}~\bibnamefont {Taniguchi}}, \bibinfo {author}
  {\bibfnamefont {D.}~\bibnamefont {Graf}}, \bibinfo {author} {\bibfnamefont
  {A.~F.}\ \bibnamefont {Young}},\ and\ \bibinfo {author} {\bibfnamefont
  {C.~R.}\ \bibnamefont {Dean}},\ }\bibfield  {title} {\bibinfo {title} {Tuning
  superconductivity in twisted bilayer graphene},\ }\href
  {https://science.sciencemag.org/content/363/6431/1059.abstract} {\bibfield
  {journal} {\bibinfo  {journal} {Science}\ }\textbf {\bibinfo {volume}
  {363}},\ \bibinfo {pages} {1059} (\bibinfo {year} {2019})}\BibitemShut
  {NoStop}%
\bibitem [{\citenamefont {Lu}\ \emph {et~al.}(2019)\citenamefont {Lu},
  \citenamefont {Stepanov}, \citenamefont {Yang}, \citenamefont {Xie},
  \citenamefont {Aamir}, \citenamefont {Das}, \citenamefont {Urgell},
  \citenamefont {Watanabe}, \citenamefont {Taniguchi}, \citenamefont {Zhang},
  \citenamefont {Bachtold}, \citenamefont {MacDonald},\ and\ \citenamefont
  {Efetov}}]{efetov2019}%
  \BibitemOpen
  \bibfield  {author} {\bibinfo {author} {\bibfnamefont {X.}~\bibnamefont
  {Lu}}, \bibinfo {author} {\bibfnamefont {P.}~\bibnamefont {Stepanov}},
  \bibinfo {author} {\bibfnamefont {W.}~\bibnamefont {Yang}}, \bibinfo {author}
  {\bibfnamefont {M.}~\bibnamefont {Xie}}, \bibinfo {author} {\bibfnamefont
  {M.~A.}\ \bibnamefont {Aamir}}, \bibinfo {author} {\bibfnamefont
  {I.}~\bibnamefont {Das}}, \bibinfo {author} {\bibfnamefont {C.}~\bibnamefont
  {Urgell}}, \bibinfo {author} {\bibfnamefont {K.}~\bibnamefont {Watanabe}},
  \bibinfo {author} {\bibfnamefont {T.}~\bibnamefont {Taniguchi}}, \bibinfo
  {author} {\bibfnamefont {G.}~\bibnamefont {Zhang}}, \bibinfo {author}
  {\bibfnamefont {A.}~\bibnamefont {Bachtold}}, \bibinfo {author}
  {\bibfnamefont {A.~H.}\ \bibnamefont {MacDonald}},\ and\ \bibinfo {author}
  {\bibfnamefont {D.~K.}\ \bibnamefont {Efetov}},\ }\bibfield  {title}
  {\bibinfo {title} {Superconductors, orbital magnets and correlated states in
  magic-angle bilayer graphene},\ }\href
  {https://doi.org/10.1038/s41586-019-1695-0} {\bibfield  {journal} {\bibinfo
  {journal} {Nature}\ }\textbf {\bibinfo {volume} {574}},\ \bibinfo {pages}
  {653} (\bibinfo {year} {2019})}\BibitemShut {NoStop}%
\bibitem [{\citenamefont {Oh}\ \emph {et~al.}(2021)\citenamefont {Oh},
  \citenamefont {Nuckolls}, \citenamefont {Wong}, \citenamefont {Lee},
  \citenamefont {Liu}, \citenamefont {Watanabe}, \citenamefont {Taniguchi},\
  and\ \citenamefont {Yazdani}}]{Oh2021}%
  \BibitemOpen
  \bibfield  {author} {\bibinfo {author} {\bibfnamefont {M.}~\bibnamefont
  {Oh}}, \bibinfo {author} {\bibfnamefont {K.~P.}\ \bibnamefont {Nuckolls}},
  \bibinfo {author} {\bibfnamefont {D.}~\bibnamefont {Wong}}, \bibinfo {author}
  {\bibfnamefont {R.~L.}\ \bibnamefont {Lee}}, \bibinfo {author} {\bibfnamefont
  {X.}~\bibnamefont {Liu}}, \bibinfo {author} {\bibfnamefont {K.}~\bibnamefont
  {Watanabe}}, \bibinfo {author} {\bibfnamefont {T.}~\bibnamefont
  {Taniguchi}},\ and\ \bibinfo {author} {\bibfnamefont {A.}~\bibnamefont
  {Yazdani}},\ }\bibfield  {title} {\bibinfo {title} {Evidence for
  unconventional superconductivity in twisted bilayer graphene},\ }\href
  {https://doi.org/10.1038/s41586-021-04121-x} {\bibfield  {journal} {\bibinfo
  {journal} {Nature}\ }\textbf {\bibinfo {volume} {600}},\ \bibinfo {pages}
  {240–245} (\bibinfo {year} {2021})}\BibitemShut {NoStop}%
\bibitem [{\citenamefont {Sharpe}\ \emph {et~al.}(2019)\citenamefont {Sharpe},
  \citenamefont {Fox}, \citenamefont {Barnard}, \citenamefont {Finney},
  \citenamefont {Watanabe}, \citenamefont {Taniguchi}, \citenamefont
  {Kastner},\ and\ \citenamefont {Goldhaber-Gordon}}]{Sharpe605}%
  \BibitemOpen
  \bibfield  {author} {\bibinfo {author} {\bibfnamefont {A.~L.}\ \bibnamefont
  {Sharpe}}, \bibinfo {author} {\bibfnamefont {E.~J.}\ \bibnamefont {Fox}},
  \bibinfo {author} {\bibfnamefont {A.~W.}\ \bibnamefont {Barnard}}, \bibinfo
  {author} {\bibfnamefont {J.}~\bibnamefont {Finney}}, \bibinfo {author}
  {\bibfnamefont {K.}~\bibnamefont {Watanabe}}, \bibinfo {author}
  {\bibfnamefont {T.}~\bibnamefont {Taniguchi}}, \bibinfo {author}
  {\bibfnamefont {M.~A.}\ \bibnamefont {Kastner}},\ and\ \bibinfo {author}
  {\bibfnamefont {D.}~\bibnamefont {Goldhaber-Gordon}},\ }\bibfield  {title}
  {\bibinfo {title} {Emergent ferromagnetism near three-quarters filling in
  twisted bilayer graphene},\ }\href {https://doi.org/10.1126/science.aaw3780}
  {\bibfield  {journal} {\bibinfo  {journal} {Science}\ }\textbf {\bibinfo
  {volume} {365}},\ \bibinfo {pages} {605} (\bibinfo {year}
  {2019})}\BibitemShut {NoStop}%
\bibitem [{\citenamefont {Serlin}\ \emph {et~al.}(2020)\citenamefont {Serlin},
  \citenamefont {Tschirhart}, \citenamefont {Polshyn}, \citenamefont {Zhang},
  \citenamefont {Zhu}, \citenamefont {Watanabe}, \citenamefont {Taniguchi},
  \citenamefont {Balents},\ and\ \citenamefont {Young}}]{Serlin2020}%
  \BibitemOpen
  \bibfield  {author} {\bibinfo {author} {\bibfnamefont {M.}~\bibnamefont
  {Serlin}}, \bibinfo {author} {\bibfnamefont {C.~L.}\ \bibnamefont
  {Tschirhart}}, \bibinfo {author} {\bibfnamefont {H.}~\bibnamefont {Polshyn}},
  \bibinfo {author} {\bibfnamefont {Y.}~\bibnamefont {Zhang}}, \bibinfo
  {author} {\bibfnamefont {J.}~\bibnamefont {Zhu}}, \bibinfo {author}
  {\bibfnamefont {K.}~\bibnamefont {Watanabe}}, \bibinfo {author}
  {\bibfnamefont {T.}~\bibnamefont {Taniguchi}}, \bibinfo {author}
  {\bibfnamefont {L.}~\bibnamefont {Balents}},\ and\ \bibinfo {author}
  {\bibfnamefont {A.~F.}\ \bibnamefont {Young}},\ }\bibfield  {title} {\bibinfo
  {title} {Intrinsic quantized anomalous hall effect in a moiré
  heterostructure},\ }\href {https://doi.org/10.1126/science.aay5533}
  {\bibfield  {journal} {\bibinfo  {journal} {Science}\ }\textbf {\bibinfo
  {volume} {367}},\ \bibinfo {pages} {900–903} (\bibinfo {year}
  {2020})}\BibitemShut {NoStop}%
\bibitem [{\citenamefont {Nuckolls}\ \emph {et~al.}(2020)\citenamefont
  {Nuckolls}, \citenamefont {Oh}, \citenamefont {Wong}, \citenamefont {Lian},
  \citenamefont {Watanabe}, \citenamefont {Taniguchi}, \citenamefont
  {Bernevig},\ and\ \citenamefont {Yazdani}}]{Nuckolls2020}%
  \BibitemOpen
  \bibfield  {author} {\bibinfo {author} {\bibfnamefont {K.~P.}\ \bibnamefont
  {Nuckolls}}, \bibinfo {author} {\bibfnamefont {M.}~\bibnamefont {Oh}},
  \bibinfo {author} {\bibfnamefont {D.}~\bibnamefont {Wong}}, \bibinfo {author}
  {\bibfnamefont {B.}~\bibnamefont {Lian}}, \bibinfo {author} {\bibfnamefont
  {K.}~\bibnamefont {Watanabe}}, \bibinfo {author} {\bibfnamefont
  {T.}~\bibnamefont {Taniguchi}}, \bibinfo {author} {\bibfnamefont {B.~A.}\
  \bibnamefont {Bernevig}},\ and\ \bibinfo {author} {\bibfnamefont
  {A.}~\bibnamefont {Yazdani}},\ }\bibfield  {title} {\bibinfo {title}
  {Strongly correlated chern insulators in magic-angle twisted bilayer
  graphene},\ }\href {https://doi.org/10.1038/s41586-020-3028-8} {\bibfield
  {journal} {\bibinfo  {journal} {Nature}\ }\textbf {\bibinfo {volume} {588}},\
  \bibinfo {pages} {610–615} (\bibinfo {year} {2020})}\BibitemShut {NoStop}%
\bibitem [{\citenamefont {Arora}\ \emph {et~al.}(2020)\citenamefont {Arora},
  \citenamefont {Polski}, \citenamefont {Zhang}, \citenamefont {Thomson},
  \citenamefont {Choi}, \citenamefont {Kim}, \citenamefont {Lin}, \citenamefont
  {Wilson}, \citenamefont {Xu}, \citenamefont {Chu}, \citenamefont {Watanabe},
  \citenamefont {Taniguchi}, \citenamefont {Alicea},\ and\ \citenamefont
  {Nadj-Perge}}]{10.1038/s41586-020-2473-8}%
  \BibitemOpen
  \bibfield  {author} {\bibinfo {author} {\bibfnamefont {H.~S.}\ \bibnamefont
  {Arora}}, \bibinfo {author} {\bibfnamefont {R.}~\bibnamefont {Polski}},
  \bibinfo {author} {\bibfnamefont {Y.}~\bibnamefont {Zhang}}, \bibinfo
  {author} {\bibfnamefont {A.}~\bibnamefont {Thomson}}, \bibinfo {author}
  {\bibfnamefont {Y.}~\bibnamefont {Choi}}, \bibinfo {author} {\bibfnamefont
  {H.}~\bibnamefont {Kim}}, \bibinfo {author} {\bibfnamefont {Z.}~\bibnamefont
  {Lin}}, \bibinfo {author} {\bibfnamefont {I.~Z.}\ \bibnamefont {Wilson}},
  \bibinfo {author} {\bibfnamefont {X.}~\bibnamefont {Xu}}, \bibinfo {author}
  {\bibfnamefont {J.-H.}\ \bibnamefont {Chu}}, \bibinfo {author} {\bibfnamefont
  {K.}~\bibnamefont {Watanabe}}, \bibinfo {author} {\bibfnamefont
  {T.}~\bibnamefont {Taniguchi}}, \bibinfo {author} {\bibfnamefont
  {J.}~\bibnamefont {Alicea}},\ and\ \bibinfo {author} {\bibfnamefont
  {S.}~\bibnamefont {Nadj-Perge}},\ }\bibfield  {title} {\bibinfo {title}
  {Superconductivity in metallic twisted bilayer graphene stabilized by wse2},\
  }\href {https://doi.org/10.1038/s41586-020-2473-8} {\bibfield  {journal}
  {\bibinfo  {journal} {Nature}\ }\textbf {\bibinfo {volume} {583}},\ \bibinfo
  {pages} {379} (\bibinfo {year} {2020})}\BibitemShut {NoStop}%
\bibitem [{\citenamefont {Choi}\ \emph {et~al.}(2019)\citenamefont {Choi},
  \citenamefont {Kemmer}, \citenamefont {Peng}, \citenamefont {Thomson},
  \citenamefont {Arora}, \citenamefont {Polski}, \citenamefont {Zhang},
  \citenamefont {Ren}, \citenamefont {Alicea}, \citenamefont {Refael},
  \citenamefont {von Oppen}, \citenamefont {Watanabe}, \citenamefont
  {Taniguchi},\ and\ \citenamefont {Nadj-Perge}}]{SNP19}%
  \BibitemOpen
  \bibfield  {author} {\bibinfo {author} {\bibfnamefont {Y.}~\bibnamefont
  {Choi}}, \bibinfo {author} {\bibfnamefont {J.}~\bibnamefont {Kemmer}},
  \bibinfo {author} {\bibfnamefont {Y.}~\bibnamefont {Peng}}, \bibinfo {author}
  {\bibfnamefont {A.}~\bibnamefont {Thomson}}, \bibinfo {author} {\bibfnamefont
  {H.}~\bibnamefont {Arora}}, \bibinfo {author} {\bibfnamefont
  {R.}~\bibnamefont {Polski}}, \bibinfo {author} {\bibfnamefont
  {Y.}~\bibnamefont {Zhang}}, \bibinfo {author} {\bibfnamefont
  {H.}~\bibnamefont {Ren}}, \bibinfo {author} {\bibfnamefont {J.}~\bibnamefont
  {Alicea}}, \bibinfo {author} {\bibfnamefont {G.}~\bibnamefont {Refael}},
  \bibinfo {author} {\bibfnamefont {F.}~\bibnamefont {von Oppen}}, \bibinfo
  {author} {\bibfnamefont {K.}~\bibnamefont {Watanabe}}, \bibinfo {author}
  {\bibfnamefont {T.}~\bibnamefont {Taniguchi}},\ and\ \bibinfo {author}
  {\bibfnamefont {S.}~\bibnamefont {Nadj-Perge}},\ }\bibfield  {title}
  {\bibinfo {title} {Electronic correlations in twisted bilayer graphene near
  the magic angle},\ }\href {https://doi.org/10.1038/s41567-019-0606-5}
  {\bibfield  {journal} {\bibinfo  {journal} {Nature Physics}\ }\textbf
  {\bibinfo {volume} {15}},\ \bibinfo {pages} {1174} (\bibinfo {year}
  {2019})}\BibitemShut {NoStop}%
\bibitem [{\citenamefont {Choi}\ \emph {et~al.}(2021)\citenamefont {Choi},
  \citenamefont {Kim}, \citenamefont {Lewandowski}, \citenamefont {Peng},
  \citenamefont {Thomson}, \citenamefont {Polski}, \citenamefont {Zhang},
  \citenamefont {Watanabe}, \citenamefont {Taniguchi}, \citenamefont {Alicea},\
  and\ \citenamefont {Nadj-Perge}}]{Choi2021}%
  \BibitemOpen
  \bibfield  {author} {\bibinfo {author} {\bibfnamefont {Y.}~\bibnamefont
  {Choi}}, \bibinfo {author} {\bibfnamefont {H.}~\bibnamefont {Kim}}, \bibinfo
  {author} {\bibfnamefont {C.}~\bibnamefont {Lewandowski}}, \bibinfo {author}
  {\bibfnamefont {Y.}~\bibnamefont {Peng}}, \bibinfo {author} {\bibfnamefont
  {A.}~\bibnamefont {Thomson}}, \bibinfo {author} {\bibfnamefont
  {R.}~\bibnamefont {Polski}}, \bibinfo {author} {\bibfnamefont
  {Y.}~\bibnamefont {Zhang}}, \bibinfo {author} {\bibfnamefont
  {K.}~\bibnamefont {Watanabe}}, \bibinfo {author} {\bibfnamefont
  {T.}~\bibnamefont {Taniguchi}}, \bibinfo {author} {\bibfnamefont
  {J.}~\bibnamefont {Alicea}},\ and\ \bibinfo {author} {\bibfnamefont
  {S.}~\bibnamefont {Nadj-Perge}},\ }\bibfield  {title} {\bibinfo {title}
  {Interaction-driven band flattening and correlated phases in twisted bilayer
  graphene},\ }\href {https://doi.org/10.1038/s41567-021-01359-0} {\bibfield
  {journal} {\bibinfo  {journal} {Nature Physics}\ }\textbf {\bibinfo {volume}
  {17}},\ \bibinfo {pages} {1375–1381} (\bibinfo {year} {2021})}\BibitemShut
  {NoStop}%
\bibitem [{\citenamefont {Kim}\ \emph {et~al.}(2023)\citenamefont {Kim},
  \citenamefont {Choi}, \citenamefont {Lantagne-Hurtubise}, \citenamefont
  {Lewandowski}, \citenamefont {Thomson}, \citenamefont {Kong}, \citenamefont
  {Zhou}, \citenamefont {Baum}, \citenamefont {Zhang}, \citenamefont {Holleis},
  \citenamefont {Watanabe}, \citenamefont {Taniguchi}, \citenamefont {Young},
  \citenamefont {Alicea},\ and\ \citenamefont {Nadj-Perge}}]{Kim2023}%
  \BibitemOpen
  \bibfield  {author} {\bibinfo {author} {\bibfnamefont {H.}~\bibnamefont
  {Kim}}, \bibinfo {author} {\bibfnamefont {Y.}~\bibnamefont {Choi}}, \bibinfo
  {author} {\bibfnamefont {E.}~\bibnamefont {Lantagne-Hurtubise}}, \bibinfo
  {author} {\bibfnamefont {C.}~\bibnamefont {Lewandowski}}, \bibinfo {author}
  {\bibfnamefont {A.}~\bibnamefont {Thomson}}, \bibinfo {author} {\bibfnamefont
  {L.}~\bibnamefont {Kong}}, \bibinfo {author} {\bibfnamefont {H.}~\bibnamefont
  {Zhou}}, \bibinfo {author} {\bibfnamefont {E.}~\bibnamefont {Baum}}, \bibinfo
  {author} {\bibfnamefont {Y.}~\bibnamefont {Zhang}}, \bibinfo {author}
  {\bibfnamefont {L.}~\bibnamefont {Holleis}}, \bibinfo {author} {\bibfnamefont
  {K.}~\bibnamefont {Watanabe}}, \bibinfo {author} {\bibfnamefont
  {T.}~\bibnamefont {Taniguchi}}, \bibinfo {author} {\bibfnamefont {A.~F.}\
  \bibnamefont {Young}}, \bibinfo {author} {\bibfnamefont {J.}~\bibnamefont
  {Alicea}},\ and\ \bibinfo {author} {\bibfnamefont {S.}~\bibnamefont
  {Nadj-Perge}},\ }\bibfield  {title} {\bibinfo {title} {Imaging inter-valley
  coherent order in magic-angle twisted trilayer graphene},\ }\href
  {https://doi.org/10.1038/s41586-023-06663-8} {\bibfield  {journal} {\bibinfo
  {journal} {Nature}\ }\textbf {\bibinfo {volume} {623}},\ \bibinfo {pages}
  {942–948} (\bibinfo {year} {2023})}\BibitemShut {NoStop}%
\bibitem [{\citenamefont {Nuckolls}\ \emph {et~al.}(2023)\citenamefont
  {Nuckolls}, \citenamefont {Lee}, \citenamefont {Oh}, \citenamefont {Wong},
  \citenamefont {Soejima}, \citenamefont {Hong}, \citenamefont {Călugăru},
  \citenamefont {Herzog-Arbeitman}, \citenamefont {Bernevig}, \citenamefont
  {Watanabe}, \citenamefont {Taniguchi}, \citenamefont {Regnault},
  \citenamefont {Zaletel},\ and\ \citenamefont {Yazdani}}]{Nuckolls2023}%
  \BibitemOpen
  \bibfield  {author} {\bibinfo {author} {\bibfnamefont {K.~P.}\ \bibnamefont
  {Nuckolls}}, \bibinfo {author} {\bibfnamefont {R.~L.}\ \bibnamefont {Lee}},
  \bibinfo {author} {\bibfnamefont {M.}~\bibnamefont {Oh}}, \bibinfo {author}
  {\bibfnamefont {D.}~\bibnamefont {Wong}}, \bibinfo {author} {\bibfnamefont
  {T.}~\bibnamefont {Soejima}}, \bibinfo {author} {\bibfnamefont {J.~P.}\
  \bibnamefont {Hong}}, \bibinfo {author} {\bibfnamefont {D.}~\bibnamefont
  {Călugăru}}, \bibinfo {author} {\bibfnamefont {J.}~\bibnamefont
  {Herzog-Arbeitman}}, \bibinfo {author} {\bibfnamefont {B.~A.}\ \bibnamefont
  {Bernevig}}, \bibinfo {author} {\bibfnamefont {K.}~\bibnamefont {Watanabe}},
  \bibinfo {author} {\bibfnamefont {T.}~\bibnamefont {Taniguchi}}, \bibinfo
  {author} {\bibfnamefont {N.}~\bibnamefont {Regnault}}, \bibinfo {author}
  {\bibfnamefont {M.~P.}\ \bibnamefont {Zaletel}},\ and\ \bibinfo {author}
  {\bibfnamefont {A.}~\bibnamefont {Yazdani}},\ }\bibfield  {title} {\bibinfo
  {title} {Quantum textures of the many-body wavefunctions in magic-angle
  graphene},\ }\href {https://doi.org/10.1038/s41586-023-06226-x} {\bibfield
  {journal} {\bibinfo  {journal} {Nature}\ }\textbf {\bibinfo {volume} {620}},\
  \bibinfo {pages} {525–532} (\bibinfo {year} {2023})}\BibitemShut {NoStop}%
\bibitem [{\citenamefont {Ahn}\ \emph {et~al.}(2020)\citenamefont {Ahn},
  \citenamefont {Guo},\ and\ \citenamefont {Nagaosa}}]{low-freq_divergence}%
  \BibitemOpen
  \bibfield  {author} {\bibinfo {author} {\bibfnamefont {J.}~\bibnamefont
  {Ahn}}, \bibinfo {author} {\bibfnamefont {G.-Y.}\ \bibnamefont {Guo}},\ and\
  \bibinfo {author} {\bibfnamefont {N.}~\bibnamefont {Nagaosa}},\ }\bibfield
  {title} {\bibinfo {title} {Low-frequency divergence and quantum geometry of
  the bulk photovoltaic effect in topological semimetals},\ }\href
  {https://doi.org/10.1103/PhysRevX.10.041041} {\bibfield  {journal} {\bibinfo
  {journal} {Phys. Rev. X}\ }\textbf {\bibinfo {volume} {10}},\ \bibinfo
  {pages} {041041} (\bibinfo {year} {2020})}\BibitemShut {NoStop}%
\bibitem [{\citenamefont {Ahn}\ \emph {et~al.}(2021)\citenamefont {Ahn},
  \citenamefont {Guo}, \citenamefont {Nagaosa},\ and\ \citenamefont
  {Vishwanath}}]{Ahn2021}%
  \BibitemOpen
  \bibfield  {author} {\bibinfo {author} {\bibfnamefont {J.}~\bibnamefont
  {Ahn}}, \bibinfo {author} {\bibfnamefont {G.-Y.}\ \bibnamefont {Guo}},
  \bibinfo {author} {\bibfnamefont {N.}~\bibnamefont {Nagaosa}},\ and\ \bibinfo
  {author} {\bibfnamefont {A.}~\bibnamefont {Vishwanath}},\ }\bibfield  {title}
  {\bibinfo {title} {Riemannian geometry of resonant optical responses},\
  }\href {https://doi.org/10.1038/s41567-021-01465-z} {\bibfield  {journal}
  {\bibinfo  {journal} {Nature Physics}\ }\textbf {\bibinfo {volume} {18}},\
  \bibinfo {pages} {290–295} (\bibinfo {year} {2021})}\BibitemShut {NoStop}%
\bibitem [{\citenamefont {Huhtinen}\ \emph {et~al.}(2022)\citenamefont
  {Huhtinen}, \citenamefont {Herzog-Arbeitman}, \citenamefont {Chew},
  \citenamefont {Bernevig},\ and\ \citenamefont {T\"orm\"a}}]{Huhtinen2022}%
  \BibitemOpen
  \bibfield  {author} {\bibinfo {author} {\bibfnamefont {K.-E.}\ \bibnamefont
  {Huhtinen}}, \bibinfo {author} {\bibfnamefont {J.}~\bibnamefont
  {Herzog-Arbeitman}}, \bibinfo {author} {\bibfnamefont {A.}~\bibnamefont
  {Chew}}, \bibinfo {author} {\bibfnamefont {B.~A.}\ \bibnamefont {Bernevig}},\
  and\ \bibinfo {author} {\bibfnamefont {P.}~\bibnamefont {T\"orm\"a}},\
  }\bibfield  {title} {\bibinfo {title} {Revisiting flat band
  superconductivity: Dependence on minimal quantum metric and band touchings},\
  }\href {https://doi.org/10.1103/PhysRevB.106.014518} {\bibfield  {journal}
  {\bibinfo  {journal} {Phys. Rev. B}\ }\textbf {\bibinfo {volume} {106}},\
  \bibinfo {pages} {014518} (\bibinfo {year} {2022})}\BibitemShut {NoStop}%
\bibitem [{\citenamefont {Lewandowski}\ and\ \citenamefont
  {Levitov}(2019)}]{lewandowski2019intrinsically}%
  \BibitemOpen
  \bibfield  {author} {\bibinfo {author} {\bibfnamefont {C.}~\bibnamefont
  {Lewandowski}}\ and\ \bibinfo {author} {\bibfnamefont {L.}~\bibnamefont
  {Levitov}},\ }\bibfield  {title} {\bibinfo {title} {Intrinsically undamped
  plasmon modes in narrow electron bands},\ }\href
  {https://www.pnas.org/doi/full/10.1073/pnas.1909069116} {\bibfield  {journal}
  {\bibinfo  {journal} {Proceedings of the National Academy of Sciences}\
  }\textbf {\bibinfo {volume} {116}},\ \bibinfo {pages} {20869} (\bibinfo
  {year} {2019})}\BibitemShut {NoStop}%
\bibitem [{\citenamefont {Arora}\ \emph {et~al.}(2022)\citenamefont {Arora},
  \citenamefont {Rudner},\ and\ \citenamefont {Song}}]{arora2022quantum}%
  \BibitemOpen
  \bibfield  {author} {\bibinfo {author} {\bibfnamefont {A.}~\bibnamefont
  {Arora}}, \bibinfo {author} {\bibfnamefont {M.~S.}\ \bibnamefont {Rudner}},\
  and\ \bibinfo {author} {\bibfnamefont {J.~C.}\ \bibnamefont {Song}},\
  }\bibfield  {title} {\bibinfo {title} {Quantum plasmonic nonreciprocity in
  parity-violating magnets},\ }\href
  {https://pubs.acs.org/doi/10.1021/acs.nanolett.2c03126} {\bibfield  {journal}
  {\bibinfo  {journal} {Nano Letters}\ }\textbf {\bibinfo {volume} {22}},\
  \bibinfo {pages} {9351} (\bibinfo {year} {2022})}\BibitemShut {NoStop}%
\bibitem [{\citenamefont {Dutta}\ \emph {et~al.}(2023)\citenamefont {Dutta},
  \citenamefont {Chakraborty},\ and\ \citenamefont {Agarwal}}]{Dutta2023}%
  \BibitemOpen
  \bibfield  {author} {\bibinfo {author} {\bibfnamefont {D.}~\bibnamefont
  {Dutta}}, \bibinfo {author} {\bibfnamefont {A.}~\bibnamefont {Chakraborty}},\
  and\ \bibinfo {author} {\bibfnamefont {A.}~\bibnamefont {Agarwal}},\
  }\bibfield  {title} {\bibinfo {title} {Intrinsic nonreciprocal bulk plasmons
  in noncentrosymmetric magnetic systems},\ }\href
  {https://doi.org/10.1103/PhysRevB.107.165404} {\bibfield  {journal} {\bibinfo
   {journal} {Phys. Rev. B}\ }\textbf {\bibinfo {volume} {107}},\ \bibinfo
  {pages} {165404} (\bibinfo {year} {2023})}\BibitemShut {NoStop}%
\bibitem [{\citenamefont {Kitamura}\ \emph {et~al.}(2020)\citenamefont
  {Kitamura}, \citenamefont {Nagaosa},\ and\ \citenamefont
  {Morimoto}}]{Kitamura2020}%
  \BibitemOpen
  \bibfield  {author} {\bibinfo {author} {\bibfnamefont {S.}~\bibnamefont
  {Kitamura}}, \bibinfo {author} {\bibfnamefont {N.}~\bibnamefont {Nagaosa}},\
  and\ \bibinfo {author} {\bibfnamefont {T.}~\bibnamefont {Morimoto}},\
  }\bibfield  {title} {\bibinfo {title} {Nonreciprocal landau–zener
  tunneling},\ }\bibfield  {journal} {\bibinfo  {journal} {Communications
  Physics}\ }\textbf {\bibinfo {volume} {3}},\ \href
  {https://doi.org/10.1038/s42005-020-0328-0} {10.1038/s42005-020-0328-0}
  (\bibinfo {year} {2020})\BibitemShut {NoStop}%
\bibitem [{\citenamefont {Xie}\ \emph {et~al.}(2020{\natexlab{b}})\citenamefont
  {Xie}, \citenamefont {Song}, \citenamefont {Lian},\ and\ \citenamefont
  {Bernevig}}]{PhysRevLett.124.167002}%
  \BibitemOpen
  \bibfield  {author} {\bibinfo {author} {\bibfnamefont {F.}~\bibnamefont
  {Xie}}, \bibinfo {author} {\bibfnamefont {Z.}~\bibnamefont {Song}}, \bibinfo
  {author} {\bibfnamefont {B.}~\bibnamefont {Lian}},\ and\ \bibinfo {author}
  {\bibfnamefont {B.~A.}\ \bibnamefont {Bernevig}},\ }\bibfield  {title}
  {\bibinfo {title} {Topology-bounded superfluid weight in twisted bilayer
  graphene},\ }\href {https://doi.org/10.1103/PhysRevLett.124.167002}
  {\bibfield  {journal} {\bibinfo  {journal} {Phys. Rev. Lett.}\ }\textbf
  {\bibinfo {volume} {124}},\ \bibinfo {pages} {167002} (\bibinfo {year}
  {2020}{\natexlab{b}})}\BibitemShut {NoStop}%
\bibitem [{\citenamefont {Su}\ \emph {et~al.}(1979)\citenamefont {Su},
  \citenamefont {Schrieffer},\ and\ \citenamefont
  {Heeger}}]{PhysRevLett.42.1698}%
  \BibitemOpen
  \bibfield  {author} {\bibinfo {author} {\bibfnamefont {W.~P.}\ \bibnamefont
  {Su}}, \bibinfo {author} {\bibfnamefont {J.~R.}\ \bibnamefont {Schrieffer}},\
  and\ \bibinfo {author} {\bibfnamefont {A.~J.}\ \bibnamefont {Heeger}},\
  }\bibfield  {title} {\bibinfo {title} {Solitons in polyacetylene},\ }\href
  {https://doi.org/10.1103/PhysRevLett.42.1698} {\bibfield  {journal} {\bibinfo
   {journal} {Phys. Rev. Lett.}\ }\textbf {\bibinfo {volume} {42}},\ \bibinfo
  {pages} {1698} (\bibinfo {year} {1979})}\BibitemShut {NoStop}%
\bibitem [{\citenamefont {Zhang}\ \emph {et~al.}(2021)\citenamefont {Zhang},
  \citenamefont {Polski}, \citenamefont {Lewandowski}, \citenamefont {Thomson},
  \citenamefont {Peng}, \citenamefont {Choi}, \citenamefont {Kim},
  \citenamefont {Watanabe}, \citenamefont {Taniguchi}, \citenamefont {Alicea}
  \emph {et~al.}}]{zhang2021ascendance}%
  \BibitemOpen
  \bibfield  {author} {\bibinfo {author} {\bibfnamefont {Y.}~\bibnamefont
  {Zhang}}, \bibinfo {author} {\bibfnamefont {R.}~\bibnamefont {Polski}},
  \bibinfo {author} {\bibfnamefont {C.}~\bibnamefont {Lewandowski}}, \bibinfo
  {author} {\bibfnamefont {A.}~\bibnamefont {Thomson}}, \bibinfo {author}
  {\bibfnamefont {Y.}~\bibnamefont {Peng}}, \bibinfo {author} {\bibfnamefont
  {Y.}~\bibnamefont {Choi}}, \bibinfo {author} {\bibfnamefont {H.}~\bibnamefont
  {Kim}}, \bibinfo {author} {\bibfnamefont {K.}~\bibnamefont {Watanabe}},
  \bibinfo {author} {\bibfnamefont {T.}~\bibnamefont {Taniguchi}}, \bibinfo
  {author} {\bibfnamefont {J.}~\bibnamefont {Alicea}}, \emph {et~al.},\
  }\bibfield  {title} {\bibinfo {title} {Ascendance of superconductivity in
  magic-angle graphene multilayers},\ }\href {https://arxiv.org/abs/2112.09270}
  {\bibfield  {journal} {\bibinfo  {journal} {arXiv:2112.09270}\ } (\bibinfo
  {year} {2021})}\BibitemShut {NoStop}%
\end{thebibliography}%

\newpage 
 
\onecolumngrid
\appendix
\section*{Supplementary Materials}

\renewcommand{\theequation}{S.\arabic{equation}}
\setcounter{equation}{0}

\renewcommand\thefigure{\thesection S\arabic{figure}}    
\setcounter{figure}{0}    

\subsection{Deriving shift current integrand expression} \label{deriving_integrand}

For completeness, we include the derivation of the shift current integrand used in the main text. First derived in \cite{sipe2000}, the expression of shift current can be written in terms of dipole matrix element $r_{mn}^\alpha$ and its generalized derivative $r_{mn;\beta}^\alpha$ is given by 
\begin{align}
    \sigma^{\alpha\alpha}_\beta(0;\omega, -\omega) = \frac{-i \pi e^3}{\hbar^2}\int_{BZ}\sum_{mn}f_{mn} r_{mn;\beta}^\alpha r_{mn}^\alpha \delta(\omega_{mn} - \omega),
\end{align}
where 
\begin{align}
    r_{mn}^\alpha &= A_{mn}^\alpha \qquad m\neq n \text{ and $0$ otherwise} \\
    r_{mn;\beta}^\alpha &= \frac{\partial r_{mn}^\alpha}{\partial k^\beta} - i (A_{mm}^\beta - A_{nn}^\beta) r_{mn}^\alpha.
\end{align}

The generalized sum rule for generalized derivative as provided in \cite{cook2017design} is 
\begin{align}
    r_{mn;\beta}^\alpha = -\frac{1}{i\epsilon_{mn}}\left[
    \frac{h_{mn}^\alpha \Delta_{mn}^\beta + h_{mn}^\beta \Delta_{mn}^\alpha}{\epsilon_{mn}} - w_{mn}^{\alpha\beta} + \sum_{l\neq m, n}\left(\frac{h_{ml}^\alpha h_{ln}^\beta}{\epsilon_{ln}} - \frac{h_{ml}^\beta h_{ln}^\alpha}{\epsilon_{ml}} \right)
    \right],
\end{align}
where $h_{mn}^\alpha = \mel{m}{\partial_{k_\alpha}H}{n}$, $w_{mn}^{\alpha\beta} = \mel{m}{\partial_{k_\alpha}\partial_{k_\beta}H}{n}$. Using $r_{nm}^\alpha = -i h_{nm}^\alpha/\epsilon_{nm}$,
the shift current integrand as defined in the main text is then
\begin{align}
    R_{mn}^{\alpha\alpha\beta} &= \text{Im}\left[{r_{mn;\beta}^\alpha r_{nm}^\alpha}\right]  \\
    &= \frac{1}{\epsilon_{mn}^2}\text{Im} \left[-\frac{h_{nm}^\alpha h_{mn}^\alpha \Delta_{mn}^\beta + h_{nm}^\alpha h_{mn}^\beta \Delta_{mn}^\alpha}{\epsilon_{mn}} + w_{mn}^{\alpha\beta} h_{mn}^\alpha \right]  \\
    &+ \frac{1}{\epsilon_{mn}^2}\text{Im}\left[\sum_{l\neq mn}\left(\frac{h_{nm}^\alpha h_{ml}^\beta h_{ln}^\alpha}{\epsilon_{ml}} - \frac{h_{nm}^\alpha h_{ml}^\alpha h_{ln}^\beta}{\epsilon_{ln}} \right)\right].
\end{align}

 Noting that $h_{nm}^\alpha h_{mn}^\alpha = |h_{mn}^\alpha|^2$ and $ \text{Im}[h_{mn}^\alpha h_{nm}^\beta] = -\text{Im}[h_{nm}^\alpha h_{mn}^\beta]$, we recover the expression in the main text:
 \begin{align}
     R_{mn}^{\alpha\alpha\beta} &= \frac{1}{\epsilon_{mn}^2}\text{Im} \left[\frac{h_{mn}^\alpha h_{nm}^\beta \Delta_{mn}^\alpha}{\epsilon_{mn}} + w_{mn}^{\alpha\beta} h_{mn}^\alpha \right]  + \frac{1}{\epsilon_{mn}^2}\text{Im}\left[\sum_{l\neq mn}\left(\frac{h_{nm}^\alpha h_{ml}^\beta h_{ln}^\alpha}{\epsilon_{ml}} - \frac{h_{nm}^\alpha h_{ml}^\alpha h_{ln}^\beta}{\epsilon_{ln}} \right)\right].
 \end{align}

\subsection{Fubini-Study metric} \label{FS}

The quantum geometric tensor provide a tool for us to describe the geometry of eigenstate. In the context of bandstructure, the quantum geometric tensor\cite{Provost1980, Torma2023} for a single band $m$ is
\begin{align}
    G^m_{\mu\nu}(k) &= \bra{\p_\mu m(k)} (1 - \ket{m(k)}\bra{m(k)}) \ket{\p_\nu m(k)} \\
    &= \sum_{m\neq n} \bra{\p_\mu m(k)}\ket{n(k)}\bra{n(k)}\ket{\p_\nu m(k)} \\
    &= \sum_{m\neq n} \frac{\mel{m}{\p_\mu H}{n}\mel{n}{\p_\nu H}{m}}{(E_m - E_n)^2}
\end{align}
Under this definition, we find 
\begin{align}
    G^m_{\mu \nu}(k) = \text{Re}[G^m_{\mu \nu}(k)] + i\text{Im}[G^m_{\mu \nu}(k)] = g^m_{\mu\nu}(k) - \frac{i}{2}F^m_{\mu\nu},
\end{align}
with the real part being the symmetric Fubini-Study (FS) metric and the imaginary part being the antisymmetric Berry curvature\cite{Provost1980, Torma2023}, given by
\begin{align}
    g_{\mu \nu}^m &= \frac{1}{2} \sum_{m\neq n} \frac{\mel{m}{\p_\mu H}{n}\mel{n}{\p_\nu H}{m} + \mel{m}{\p_\nu H}{n}\mel{n}{\p_\mu H}{m} }{(E_m - E_n)^2}, \\
    F_{\mu \nu}^m &= i\sum_{m\neq n} \frac{\mel{m}{\p_\mu H}{n}\mel{n}{\p_\nu H}{m} - \mel{m}{\p_\nu H}{n}\mel{n}{\p_\mu H}{m} }{(E_m - E_n)^2}
\end{align}

As pointed out in \cite{PhysRevLett.124.167002}, the FS metric characterizes the spread of Wannier wavefunction. Following the definition of Ref.\cite{PhysRevLett.124.167002}, we define the localization of Wannier wavefunciton as 
\begin{align}
    F = \sum_n \left[ \mel{0n}{\mathbf{\hat{r}}^2}{0n} -  |\mel{0n}{\mathbf{\hat{r}}}{0n}|^2\right],
\end{align}
where $\mathbf{\hat{r}}$ is the position operator and $\ket{\mathbf{R}n} $ is the Wannier state given by $\ket{\mathbf{R}n} = \frac{1}{\sqrt{N}} \sum_{\mathbf{k}} e^{-i\mathbf{k} \cdot \mathbf{R}} \ket{\psi_{n\mathbf{k}}} $.
Then, FS metric provides a lower bound on $F$ given by 
\begin{align} \label{trace_g}
    F \geq \frac{\Omega_c}{(2\pi)^2} \int d^2k \tr g(k),
\end{align}
with $\Omega_c$ being the area of the unit cell and $g_{\mu \nu}(k) = \sum_{m}g^m_{\mu \nu}(k)$.

The FS metric defined above refers to a single band $m$ and sums over contribution from all intermediate band $n$. Without summing over $n$, we define
\begin{align}
    Q^{mn}_{\mu\nu}(k) &= \bra{\p_\mu m(k)}\ket{n(k)}\bra{n(k)}\ket{\p_\nu m(k)}  \\
    &= \frac{\mel{m}{\p_\mu H}{n}\mel{n}{\p_\nu H}{m}}{(E_m - E_n)^2},
\end{align}
we can consequently define interband FS metric and interband Berry curvature as 
\begin{align}
    g^{mn}_{\mu\nu}(k) = \text{Re}[Q_{\mu\nu}^{mn}(k)], \qquad F^{mn}_{\mu\nu} = -2\text{Im}[G^{mn}_{\mu\nu}(k)].
\end{align}
In terms of projector notations, we can write 
\begin{align}
    G_{\mu \nu}^{mn} = \Tr[P_n \p_\mu P_m \p_{\nu} P_m] = \Tr [P_m \p_{\mu}P_{nm} \p_{\nu}P_{mn}],
\end{align}
where $P_m = \ket{m}\bra{n} $, and $P_{mn}=\ket{m}\bra{n}$. 

In 1D, there is no interband Berry curvature since $G_{xx}^{mn}$ is real. Furthermore, the transition probability amplitude can be identified with interband FS metric in the direction of transition:
\begin{align}
    |A_{mn}^\alpha|^2 = -\bra{m(k)}\ket{\p_{\alpha} n(k)} \bra{n(k)} \ket{\p_{\alpha} m(k)} = g_{mn}^{\alpha \alpha}.
\end{align}
This allows us to write the shift current expression of Eq. \ref{sigma} as
\begin{align}
    \sigma^\beta_{\alpha \alpha}(\omega) = \frac{\pi e^3}{\hbar^2}\sum_{m,n}\int d^2 \mathbf{k} f_{mn}S^{\beta \alpha}_{mn} g_{mn}^{\alpha \alpha} \delta(\omega - \epsilon_{mn}),
\end{align}
where we establish a direct connection between shift current response and FS metric. Since $ \sum_{mn \alpha} \int d\mathbf{k} g_{mn}^{\alpha \alpha}$ gives a lower bound for the spread of Wannier wavefunction, this confirms many of the experimental evidence\cite{Dai_Rappe_2023,Fregoso2017} that materials with delocalized wavefunction tend to exhibit large shift current response. In addition, viewing $g_{mn}^{\alpha \alpha}$ as the $\alpha$ component of the metric, we can view shift current conductivity tensor as the sum of shift vector at each k-state over the BZ weighted by the interband metric along the direction of transition.

\subsection{Resonance of the Quantum Geometric Tensor in Parameter Space} \label{resonance}
Near the parameter region where the band gap vanishes at the band edge, we expect $g^{mn}_{xx}$ to diverge. We describe the rate of divergence by a Lorentzian function 
\begin{align}
    L(x) = \frac{2A}{\pi} \frac{\Gamma}{4(x-x_0)^2 + \Gamma^2} + c,
\end{align}
with constant $A, x_0, \Gamma, c$. For the 2RM model considered in the main text, we fix parameter $(t, \delta_1, \Delta_1, \Delta_2, \epsilon)$ and vary $\delta_2$ to optimize for shift current enhancement via virtual transitions, and we note that the divergence of FS metric at the band edge as $\delta_2 \rightarrow \delta_{2,0}$ is well-characterized by a Lorentzian function, where $\delta_{2,0}$ is where the gap is minimized.

\begin{figure}[h]
    \centering
    \includegraphics[width=\linewidth]{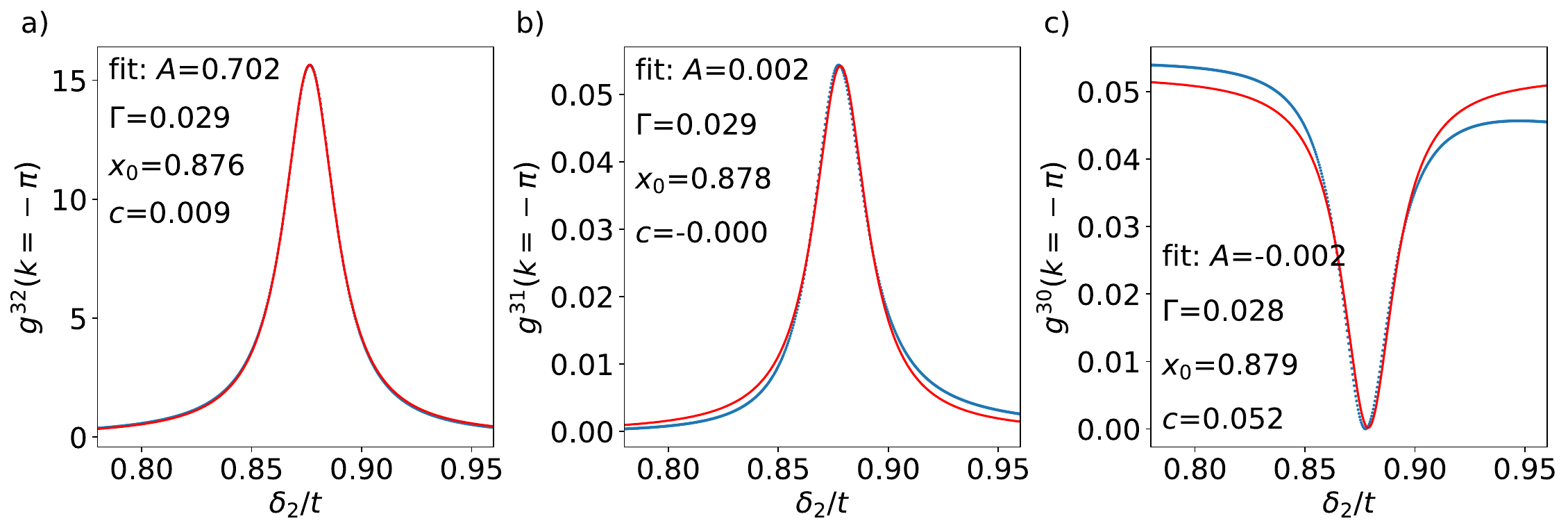}
    \caption{The FS metric $g^{mn}_{xx}(k)$ at the band edge $k a=-\pi$ for $m=3$, $n=2, 1, 0$, and the fit using a Lorentzian function, with fitting curve in red and exact calculation in blue dots. Here we used the same parameters as in the main text, namely $(t, \delta_1, \Delta_1, \Delta_2, \epsilon) = (1, 0.8, 0.7, 0.6, 0.1)$. }
    \label{figS1}
\end{figure}

We plot the component of $g^{mn}$ at the band edge for the upper conduction band $m=3$ and $n=0,1,2$ in Fig. \ref{figS1}a,b,c. For $g^{32}(k)$, the Lorentzian function well-characterizes the divergence at the band edge as  $\delta_{2,0}$ is approached. For $g^{31}(k)$ and $g^{30}(k)$, the tails show deviations from Lorentzian behavior, but near the resonance pole, they can be described by the Lorentzian function. It is this divergence structure that gives rise to enhanced $g_3(k) = \sum_{n\neq 3}g^{3n}(k) $ near $\delta_{2,0}$.

This evokes the resonance phenomena but in parameter space. As we vary the hopping parameters, there exists a critical hopping strength such that the FS metric diverges at the band edge. This enhances the sum of FS over all k-states, leading to  Wannier wavefunction being ``resonantly'' more delocalized. Consequently, we observe an enhancement in shift current response via virtual transitions.

Now, we demonstrate this connection more precisely by explicitly putting $g^{mn}$ into the form of a Lorentzian. The exact energy eigenvalues of 2RM model is given by 
\begin{align}
a &= \Delta_{1}^{2} + \Delta_{2}^{2} + \delta_{1}^{2} \sin^{2}{\left(\frac{k}{2} \right)} + \delta_{2}^{2} \sin^{2}{\left(\frac{k}{2} \right)} + 2 t^{2} \cos^{2}{\left(\frac{k}{2} \right)}, \\
b &=  4 t^{2} \cos^{2}{\left(\frac{k}{2} \right)} + \left(\Delta_{1} - \Delta_{2}\right)^{2}   
    + \left(\delta_{1} - \delta_{2}\right)^{2} \sin^{2}{\left(\frac{k}{2} \right)},   \\
c &= \left(\Delta_{1} - \Delta_{2}\right) \left(\Delta_{1} + \Delta_{2}\right) + \left(\delta_{1} - \delta_{2}\right) \left(\delta_{1} + \delta_{2}\right) \sin^{2}{\left(\frac{k}{2} \right)}, \\
E & = \pm \frac{\sqrt{2}}{2} \sqrt{a + 2 \epsilon^{2} \pm \sqrt{4 \epsilon^{2}b + c^2}}.
\end{align}
Recall that the FS metric tensor defined in Appendix \ref{FS} in 1D is given by
\begin{align}
    g^{mn}(k) = \frac{\mel{m}{\p_k H}{n}\mel{n}{\p_k H}{m}}{(E_m - E_n)^2}.
\end{align}
Using $m=3$, $n=2$ as an example and focus on the band edge at $k=\pm \pi$, we put the above FS in the form of Lorentzian explicitly. Noting that there are no poles in the numerator, so we focus on the denominator term, which gives 
\begin{align}
    (E_3 - E_2)^2 = a+2\epsilon^2 - \sqrt{(a+2\epsilon^2)^2 - 4 \epsilon^2 b + c^2 } \approx \frac{1}{2}\frac{4\epsilon^2 b + c^2}{(a+2\epsilon^2)},
\end{align}
where we note that $4\epsilon^2 b + c^2 \ll (a+2\epsilon^2)^2$ near the pole. The pole occurs in $4\epsilon^2 b + c^2$. We expand $\delta_2$ around the maximum of $g^{32}(k)$, denoted by $\delta_{2,0}$, to second order, and we find
\begin{align}
    4\epsilon^2 b + c^2 = \Tilde{a}\left(\delta_2 - \delta_{2,0} + \frac{\Tilde{b}}{2\Tilde{a}}\right)^2 - \frac{\Tilde{b}^2}{4\Tilde{a}} + \Tilde{c} + \mathcal{O}((\delta_2 - \delta_{2,0})^2),
\end{align}
with 
\begin{align}
    \Tilde{a} &= -2(\delta_1^2 + \Delta_1^2 - \Delta_2^2 - 3\delta_{2,0}^2 - 2\epsilon^2) \\
    \Tilde{b} &= -4(\delta_{2,0} (\delta_1^2 + \Delta_1^2 - \Delta_2^2 - \delta_{2,0}^2) + 2(\delta_1 - \delta_{2,0})\epsilon^2 ) \\
    \Tilde{c} &= (\delta_1^2 + \Delta_1^2 - \Delta_2^2 - \delta_{2,0}^2)^2 + 4((\Delta_1 - \Delta_2)^2 + (\delta_1 - \delta_{2,0})^2) \epsilon^2.
\end{align}
By the assumption that $\delta_{2,0}$ is the pole, we require $\frac{\Tilde{b}}{2\Tilde{a}} = 0$. Alternatively, the value of $\delta_{2,0}$ is obtained by solving for $\frac{\Tilde{b}}{2\Tilde{a}} = 0$. The FS metric $g^{32}$ at $k=\pm \pi$ as a function of hopping strength $\delta_2$ can be written in the form of a Lorenzian 
\begin{align}
    g^{32}(\delta_2) = \frac{2A}{\pi} \frac{\Gamma}{4(\delta_2 - \delta_{2,0})^2 + \Gamma^2},
\end{align}
with the width and amplitude given by 
\begin{align}
    \Gamma = 2 \sqrt{\frac{\Tilde{c}}{\Tilde{a}}}, \qquad A = \frac{4\pi (a+2\epsilon^2) |\mel{3}{\p_k H}{2}|^2 }{\Tilde{a}\Gamma}.
\end{align}
A similar manipulation follows for the case of $n=1$ and $n=0$. The $\Gamma$ terms acts like a damping term that restricts the divergences of FS at the band edge. This $\Gamma$ damping term usually occurs in systems with many parameters, and may be absent in models with fewer parameters. Consider the Su–Schrieffer–Heeger (SSH) model\cite{PhysRevLett.42.1698}, with Hamiltonian given by
\begin{align}
    H(k) = h_x \sigma_x + h_y \sigma_y, \qquad h_x = v+w\cos(k), \qquad h_y = w \sin(k),
\end{align}
where $v$ is the intralayer hopping amplitude and $w$ is the interlayer hopping amplitude. Fixing interlayer hopping $w$ and vary intralayer hopping $v$, the FS metric at $k=\pm \pi$ is given by
\begin{align}
    g(v) &= \frac{w^2}{4(v-w)^2},
\end{align}
where it diverges without damping at $v=w$, the point when energy gap closes. Adding a mass term $\Delta$ to break the chiral symmetry, we obtain a single-layer Rice-Mele Model, and the damping in FS metric becomes $4\Delta$.

\subsection{Analytical Results of 2RM model} \label{2RM_ana}

In the main text, we established that it is the enhancement in virtual transitions near the critical parameter region that leads to enhanced shift current response. Inspecting Eq. \ref{integrand} shows us the two factors that control of the magnitude of the system are energy gap $\epsilon_{mn}$ and matrix element $h_{mn} = \mel{m}{\p_k H}{n}$. In particular, we expect $\epsilon_{32}$ to dominate in the denominator when the gap is small, and we will show that it is divergence in $\text{Im}\{h_{20}\}$ that dominates in the numerator. 

The Hamiltonian of 2RM model takes the form $H = H_0 + V$, with 
\begin{align}
    H_0 = 
\begin{pmatrix}
d^{(1)}_z & d^{(1)}_x - i d^{(1)}_y & 0 & 0\\ 
d^{(1)}_x + i d^{(1)}_y & -d^{(1)}_z & 0 & 0 \\
0 & 0 & d^{(2)}_z & d^{(2)}_x - i d^{(2)}_y \\ 
0 & 0 & d^{(2)}_x + i d^{(2)}_y & -d^{(2)}_z
\end{pmatrix}, \quad 
V = 
\begin{pmatrix}
    0 & 0 & 0 & \epsilon \\
    0 & 0 & \epsilon & 0 \\
    0 & \epsilon & 0 & 0 \\
    \epsilon & 0 & 0 & 0
\end{pmatrix},
\end{align}
where $H$ is in the layer basis $\{\ket{1A},\ket{1B},\ket{2A},\ket{2B}\}$, and superscript $(1/2)$ denotes the first/second RM model. Exactly solving for the eigenstate and computing matrix element will result in complicated expressions that shed little intuition. Instead, we will use perturbation theory and expand near the band edge. A direct application of non-degenerate perturbation theory leads to inaccurate results near the band edge; hence, we first perform a basis transformation to the unperturbed eigenstate $\{ \ket{\psi_{1,A}}, \ket{\psi_{2,A}}, \ket{\psi_{1,B}}, \ket{\psi_{2,B}} \} $. The Hamiltonian $\Tilde{H} = \Tilde{H_0} + \Tilde{V}  $ becomes
\begin{align}
    \Tilde{H_0} = 
    \begin{pmatrix}
        -\epsilon_1(k) & f_A(k) & 0 & 0\\ 
        f_A^*(k) & -\epsilon_2(k) &  0 & 0 \\
        0 & 0 & \epsilon_1(k) & f_{B}(k) \\
        0 & 0 & f^*_B(k) & \epsilon_2(k)
    \end{pmatrix},  \quad
    \Tilde{V} = 
\begin{pmatrix}
    0 & 0 & 0 & g_{1}(k) \\
    0 & 0 & g_{2}(k) & 0 \\
    0 & g_{2}^*(k) & 0 & 0 \\
    g_1^*(k) & 0 & 0 & 0
\end{pmatrix},
\end{align}
where $\epsilon_{1/2}$ is the eigenvalue of the first/second unperturbed RM model, and
\begin{align}
    f_A(k) &= \mel{\psi_{1,A}}{V}{\psi_{2A}} = \frac{\epsilon}{2\sqrt{\epsilon_1 \epsilon_2}}\left(-e^{i\phi_k^{(2)}}\sqrt{\epsilon_1 - d_z^{(1)}}\sqrt{\epsilon_2 + d_z^{(2)}} - e^{-i\phi_k^{(1)}}\sqrt{\epsilon_1 + d_z^{(1)}}\sqrt{\epsilon_2 - d_z^{(2)}}  \right)  , \\
    f_B(k) &= \mel{\psi_{1,B}}{V}{\psi_{2B}} = \frac{\epsilon}{2\sqrt{\epsilon_1 \epsilon_2}}
    \left(e^{-i\phi_k^{(1)}}\sqrt{\epsilon_1 - d_z^{(1)}}\sqrt{\epsilon_2 + d_z^{(2)}} +e^{i\phi_k^{(2)}}\sqrt{\epsilon_1 + d_z^{(1)}}\sqrt{\epsilon_2 - d_z^{(2)}}  \right),\\
    g_{1}(k) &= \mel{\psi_{1,A}}{V}{\psi_{2B}} = \frac{\epsilon}{2\sqrt{\epsilon_1 \epsilon_2}}
    \left(-e^{i\phi_k^{(2)}}\sqrt{\epsilon_1 - d_z^{(1)}}\sqrt{\epsilon_2 - d_z^{(2)}} + e^{-i\phi_k^{(1)}}\sqrt{\epsilon_1 + d_z^{(1)}}\sqrt{\epsilon_2 + d_z^{(2)}}  \right), \\
    g_{2}(k) &= \mel{\psi_{2,A}}{V}{\psi_{1B}} = \frac{\epsilon}{2\sqrt{\epsilon_1 \epsilon_2}}
    \left(-e^{i\phi_k^{(1)}}\sqrt{\epsilon_1 - d_z^{(1)}}\sqrt{\epsilon_2 - d_z^{(2)}} + e^{-i\phi_k^{(2)}}\sqrt{\epsilon_1 + d_z^{(1)}}\sqrt{\epsilon_2 + d_z^{(2)}}  \right),
\end{align}
where $\phi^{(1/2)}_k = \arctan(d^{(1/2)}_y/d^{(1/2)}_x)$. Now, we can apply the non-degenerate perturbation theory safely. In particular, the first order correction in state is given by
\begin{align}
    \ket{n} = \ket{n_0} + \sum_{k\neq n}\frac{\mel{k_0}{V}{n_0}}{E_{n} - E_k} \ket{k_0},
\end{align}
with nonzero matrix element only in 
\begin{align}
    \mel{1_0}{\Tilde{V}}{3_0} = \frac{1}{2\sqrt{E_d^{(A)}E_d^{(B)}}}  \Big( &-g_1(k) e^{i\phi_k^{(B)}} \sqrt{E_d^{(A)} + d_3^{(A)}} \sqrt{E_d^{(B)} + d_3^{(B)}}   -g_2(k) e^{-i\phi_k^{(A)}} \sqrt{E_d^{(A)} - d_3^{(A)}} \sqrt{E_d^{(B)} - d_3^{(B)}}  \Big),\\
    \mel{2_0}{\Tilde{V}}{3_0} = \frac{1}{2\sqrt{E_d^{(A)}E_d^{(B)}}}  \Big( &g_1(k) e^{i\phi_k^{(B)}} \sqrt{E_d^{(A)} - d_3^{(A)}} \sqrt{E_d^{(B)} + d_3^{(B)}}  -g_2(k) e^{-i\phi_k^{(A)}} \sqrt{E_d^{(A)} + d_3^{(A)}} \sqrt{E_d^{(B)} - d_3^{(B)}}  \Big), \\
    \mel{1_0}{\Tilde{V}}{4_0} = \frac{1}{2\sqrt{E_d^{(A)}E_d^{(B)}}}  \Big( &
    -g_1(k) e^{i\phi_k^{(B)}} \sqrt{E_d^{(A)} + d_3^{(A)}} \sqrt{E_d^{(B)} - d_3^{(B)}}  + g_2(k) e^{-i\phi_k^{(A)}} \sqrt{E_d^{(A)} - d_3^{(A)}} \sqrt{E_d^{(B)} + d_3^{(B)}}  \Big), \\
    \mel{2_0}{\Tilde{V}}{4_0} = \frac{1}{2\sqrt{E_d^{(A)}E_d^{(B)}}}  \Big( &g_1(k) e^{i\phi_k^{(B)}} \sqrt{E_d^{(A)} - d_3^{(A)}} \sqrt{E_d^{(B)} - d_3^{(B)}}  + g_2(k) e^{-i\phi_k^{(A)}} \sqrt{E_d^{(A)} + d_3^{(A)}} \sqrt{E_d^{(B)} + d_3^{(B)}}  \Big),
\end{align}
where 
\begin{align}
    d_0 = \frac{\epsilon_1(k) + \epsilon_2(k)}{2} &, \quad d_1^{(A/B)} = \text{Re}\{f_{A/B}(k)\}, \quad d_2^{(A/B)}  = \text{Im}\{f_{A/B}(k)\}, \quad d_3 = \frac{\epsilon_1(k) - \epsilon_2(k)}{2}, \\
     E_d^{(A/B)} &= \sqrt{\left(d_1^{(A/B)}\right)^2 + \left(d_2^{(A/B)}\right)^2 + d_3^2}, \quad \phi_k^{(A/B)} = \arctan(d_2^{(A/B)} / d_1^{(A/B)})
\end{align}
In particular, there is no matrix element connecting the same block of $\Tilde{H}_0$ where the difference in energy is small, thus avoiding the divergence in a small energy gap near gap closing. Comparing the perturbation theory result with exact diagonalization indeed shows the validity of this approach.  The $\Tilde{H}_0$ now becomes an effective 2-band model, which shift current response is shown in Fig. \ref{fig2}d. Comparing with tight-binding 2-band model, this effective 2-band model can achieve a large joint density of states and dispersive band structure simultaneously. 

Now, we have the ingredients necessary to compute matrix element $h_{mn}$. For our consideration, take $m=3, n=0$. At the band edge, comparing the magnitude of direct and virtual transition contribution to shift current amounts to comparing $\text{Re}\{w_{30}\}$ and $\text{Re}\{h_{32}h_{20}\} = \text{Im}\{h_{32}\} \text{Im}\{h_{20}\}$ (in the integrand we are taking the Im of the expression, and $\text{Re}\{h_{03}\} = 0$ at band edge). We find 
\begin{align}
    \text{Re}\{w_{30}\} &= \mel{3_0}{\p_k^2 \Tilde{H}}{0_0} + \frac{\mel{3_0}{\Tilde{V}}{0_0}}{E_{30}} \left[ \mel{0_0}{\p_k^2 \Tilde{H}}{0_0} - \mel{3_0}{\p_k^2 \Tilde{H}}{3_0} \right], \\
    \text{Im}\{h_{32}\} &= \frac{\mel{1_0}{\p_k \Tilde{H}}{2_0} \mel{3_0}{\Tilde{V}}{1_0}}{E_{31}} + \frac{\mel{0_0}{\p_k \Tilde{H}}{2_0} \mel{3_0}{\Tilde{V}}{0_0}}{E_{30}} 
    + \frac{\mel{3_0}{\p_k \Tilde{H}}{1_0} \mel{1_0}{\Tilde{V}}{2_0}}{E_{21}} + \frac{\mel{3_0}{\p_k \Tilde{H}}{0_0} \mel{0_0}{\Tilde{V}}{2_0}}{E_{20}}, \\
    \text{Im}\{h_{20}\} &= \mel{2_0}{\p_k \Tilde{H}}{0_0} + \mathcal{O}(1/E^2),
\end{align}
where we see $\text{Im}\{h_{20}\}$ dominate over $\text{Im}\{h_{32}\}$ since it is not suppressed by energy. In addition, find $\text{Im}\{h_{32}\}$ is maximized near minimum gap point 
\begin{align}
    \delta_{2,0} = \pm \sqrt{\delta_1^2 + \Delta_1^2 - \Delta_2^2}. 
\end{align}
The negative pole is suppressed by other factors, such as direct transition, and it is only near the positive pole where we see enhancement in the overall shift current response.

Near the point of minimum energy gap at the band edge, the small $\epsilon_{32}$ in the denominator and diverging $\text{Im}\{h_{20}\}$ in the numerator of Eq. \ref{integrand} combine to generate an enhanced response in virtual transition. Nevertheless, we remark that the peak shift current response is not located precisely at gap-closing; rather, it is in close proximity to this point, due to the vanishing of transition probability amplitude at the gap-closing point.

\subsection{Total Shift Current Response in 2RM} \label{2RM_total_shift}

The presence of multiple bands allows for additional transition pairs to happen. The total number of transition pairs is the combination of 2 bands out of all the possible bands, and their contributions need to be summed over when computing the total shift current response. The total shift current response for $\nu = 0.5$ is given in Fig. \ref{figS2}:

\begin{figure}[h]
    \centering
    \includegraphics[width=\linewidth]{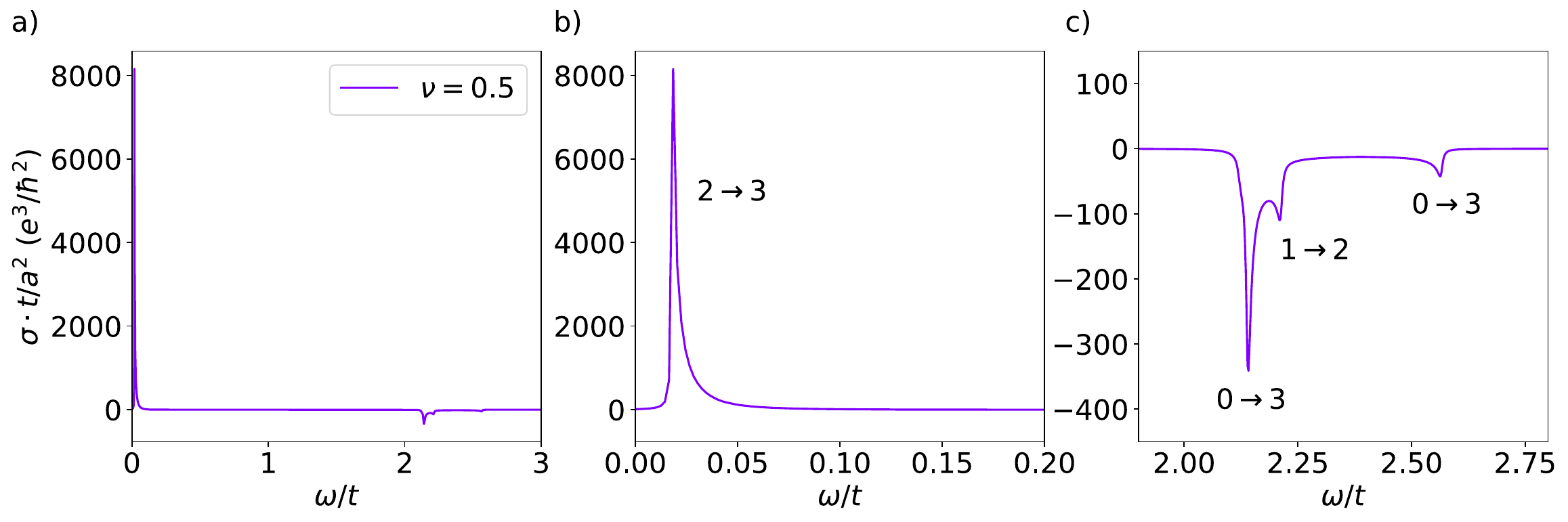}
    \caption{Total shift current response for $\nu=2$ summing up all transition pairs.}
    \label{figS2}
\end{figure}

The band index is numbered based on Fig. \ref{fig2}a in the main text. Most of the contribution from $\nu=0.5$ arises from the transition between band $2$ and $3$, enhanced due to the small band gap. In addition, there are transitions from the first RM model (band $0$ to band $3$), giving rise to two peaks, corresponding to transition mostly from $k a = \pm \pi$ and $ka = 0$, respectively. The transition from the second RM model (band $1$ to band $2$) contributes only one peak with contribution mostly from $ka=0$, which is due to suppression in joint density of state. Summing up all transition pairs also leads to an enhancement in shift current response, compared to the case of simply stacking two pairs of RM chains together (without band mixing).

\subsection{Band Mixing in TMG} \label{band_mixing}
The multilayer Hamiltonian constructed using ATMG model decouples into pairs of TBG-like bands, but we expect band mixing to occur in actual materials \cite{zhang2021ascendance,kolar2023}. There are two ways to induce band mixing, either through the application of an external displacement field or through a self-generated displacement field. 

\textit{External displacement field}
The effect of the external displacement field is applied through tuning external gate on the device. It modifies the hamiltonian through the addition of the following term in the layer basis: 
\begin{align}
    \Delta H = \frac{U}{N-1}
\begin{pmatrix}
\frac{N-1}{2} & & & & \\
& \frac{N-1}{2}-1 & & & \\
& & \ddots & & \\
& & & -\frac{N-1}{2}+1 & \\
& & & & -\frac{N-1}{2}
\end{pmatrix}
\end{align}
where $U$ is the strength of the displacement field, in unit of $meV$. This leads to band mixing among TBG-like sectors, allowing transition to take place.

\subsection{T5G Shift Current and Dependence on Twist Angles}\label{twist_angle}

\begin{figure}[ht]
    \centering
    \includegraphics[width=0.4\linewidth]{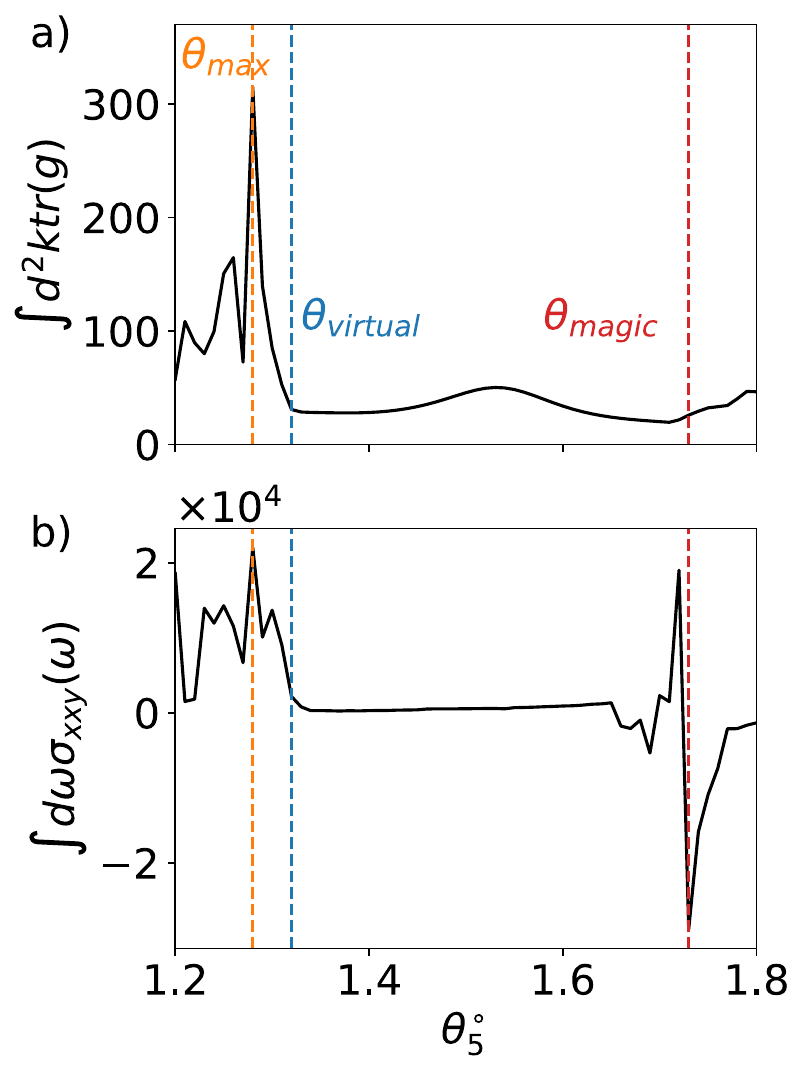}
    \caption{ T5G FS metric of top flat bands and overall shift current response and their dependence on physical twist angle. The orange dashed line shows the angle that leads to max FS, and correspondingly largest overall shift current measured by metric $M$. The blue dashed line shows the angle that is used in the main text to illustrate enhancement via virtual transitions. The red line denotes the magic angle, where we also observe an enhancement in the overall shift current response, which we attribute to the increased joint density of states.}
    \label{figS3}
\end{figure}

The physical twist angle of the TMG system controls the energy gap of nearby flat bands. At the magic angle where the perfect flat band pair emerges, we expect a large shift current response due to the vanishing of the energy gap. In addition, away from the magic angle in multilayer systems, we anticipate a substantial shift current response stemming from enhanced virtual transition via nearby flat bands. Here we show the dependence of total shift current response $\int dw \sigma_{xxy}(\omega)$ on the physical angle $\theta_5$ of T5G. In addition, we compute the integral of FS metric over k-state as a function of twist angle, as shown in Fig. \ref{figS3}. We indeed observe an enhanced shift current response near angles where FS metric diverges, corresponding to a more delocalized Wannier wavefunction. 

In the main text, we illustrate virtual transition enhancement in T5G with $\theta_5 = 1.32^\circ$ since the flat bands belong to distinct TBG-like sectors, allowing us to better compare the effect of band mixing on transitions within the same band.  As shown in Fig. \ref{figS3}b, this angle is not the physical angle that leads to maximized response. The ideal angle occurs when $\theta_5 = 1.28^\circ$, where the two flat bands are tightly intertwined. 

\begin{figure}[h]
    \centering
    \includegraphics[width=\linewidth]{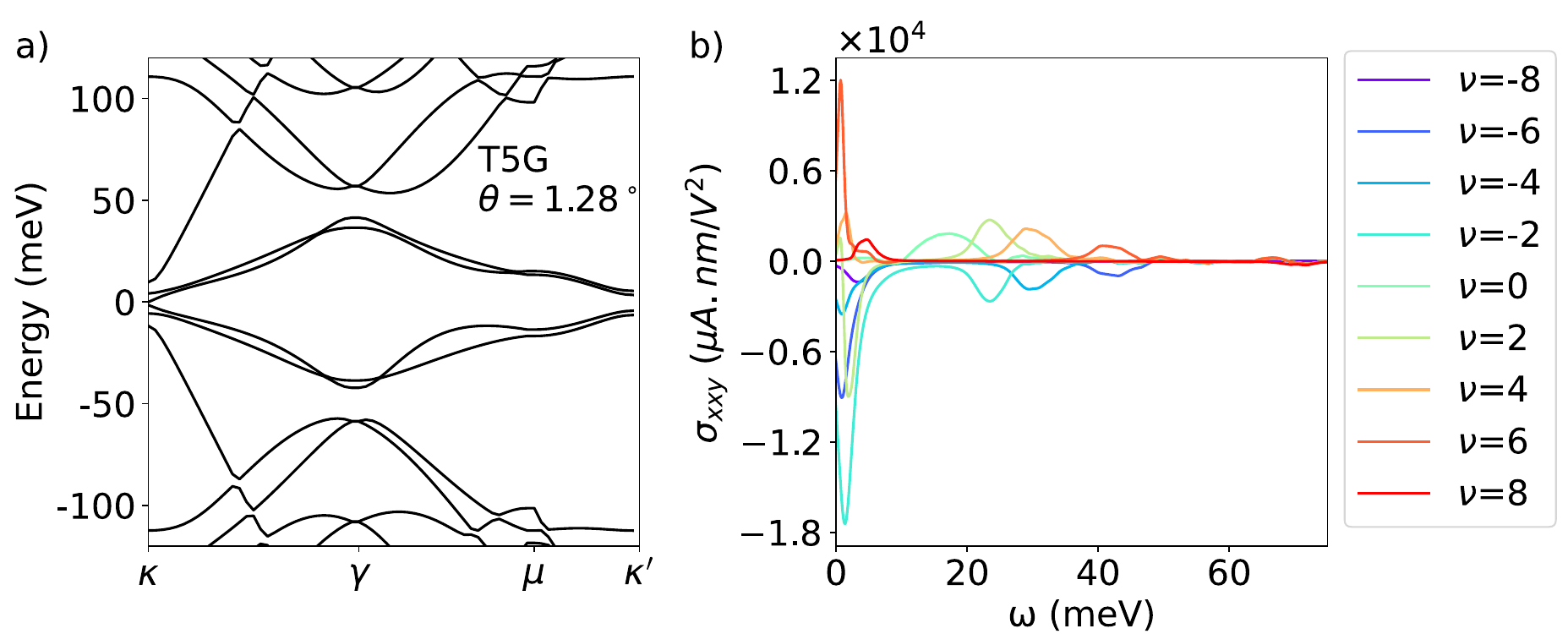}
    \caption{T5G band structure occurring at $\theta = 1.28^\circ$ with $U=20meV$ and the corresponding shift current response at various fillings.}
    \label{figS4}
\end{figure}

At $\theta = 1.28^\circ$, the two pairs of TBG have a similar effective twist angle, but with the top and bottom flat bands flipped with respect to each other. The slight difference in dispersion relations of the top and bottom flat bands results in the intertwining of the bands, as shown in Fig. \ref{figS4}. Instead of studying transitions between single band pairs, observing an enhancement in shift current necessitates summing over all transition pairs. Fig. \ref{figS4} b) shows that virtual transitions among the flat bands indeed lead to an enhancement response, and we also observe that $\theta = 1.28^\circ$ generates a larger shift current response than $\theta = 1.32^\circ$. 

\begin{figure}[!h]
    \centering
    \includegraphics[width=\linewidth]{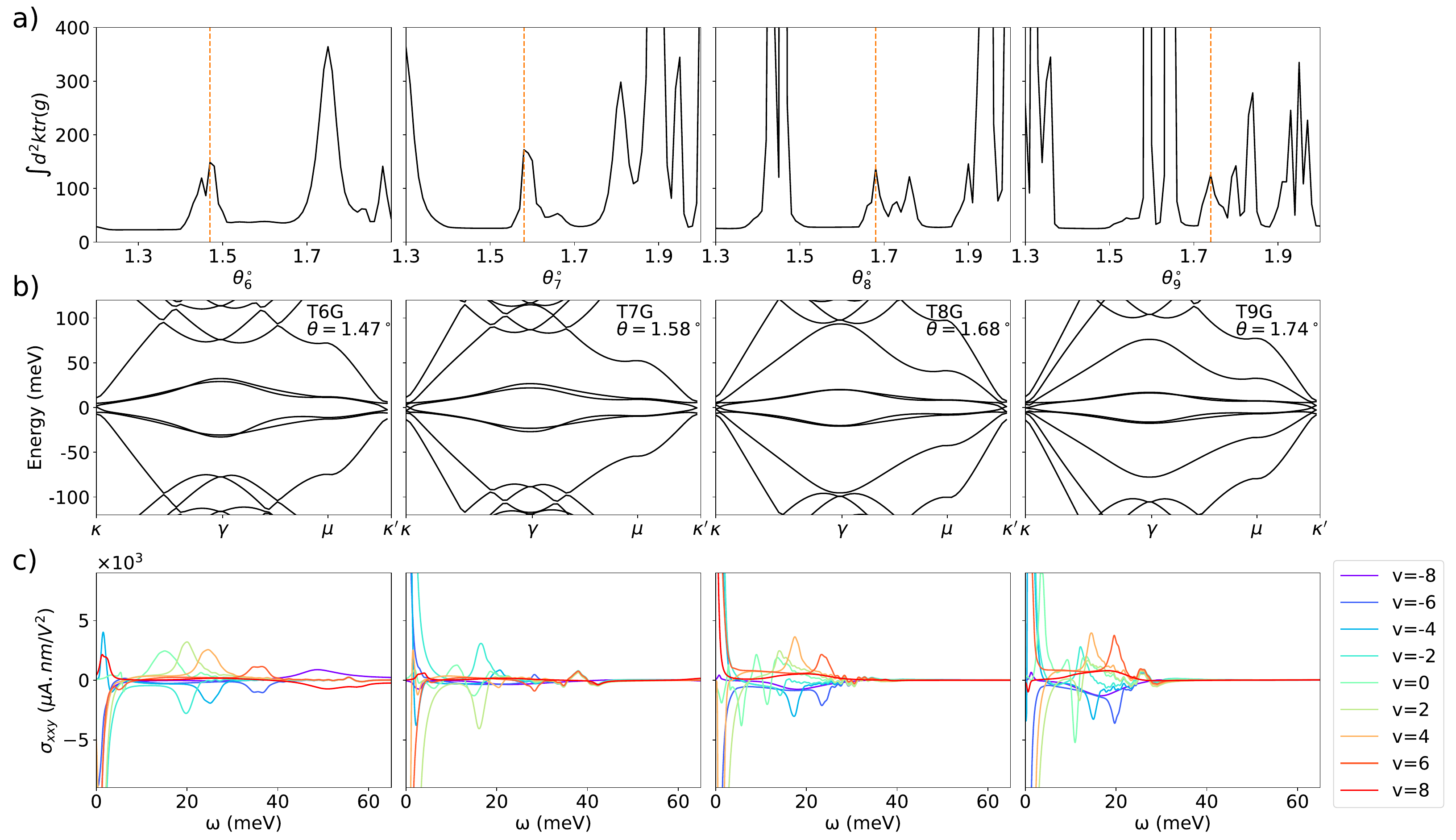}
    \caption{Optimal angle for shift current in TMG a) Integral of trace of FS metric for the top flat band of T6G, T7G, T8G, T9G, with the orange dashed line indicating the optimal angle. b) Their corresponding band structure at optimal angles and c) their conductivity $\sigma_{xxy}(\omega)$ for a wide range of fillings.}
    \label{figS5}
\end{figure}

To search for the optimal twist angles for a higher number of layers, we would want the TBG band pairs to have similar effective angles. A more systematic search can be done by computing the FS metric for the flat bands and pinpointing the angle that gives rise to the largest $\int d^2 k \tr g$. We adopt this procedure and obtained the optimal angles for $N=6, 7, 8, 9$ layers of twisted graphene, and the result is shown in Fig. \ref{figS5}.

The conductivity includes transitions among flat bands and include transition to dispersive bands via virtual transitions. We see that peaks in $\int d^2k \tr(g) $ indeed corresponds to the angles where two flat bands are the closest. 

\clearpage

\subsection{T5G Symmetry} \label{symmetry}
The TMG model has $C_{3z}$ symmetry, which restricted $\sigma^\beta_{\alpha \alpha}$ to two independent components. A detailed argument is discussed in \cite{shift_current} for TBG, so here we simply restate the result for TMG. We have
\begin{align}
    \sigma^y_{xx} = -\sigma^y_{yy} = \sigma^y_{xy} = \sigma^x_{yx} \neq 0, \\
    \sigma^x_{yy} = -\sigma^x_{xx} = \sigma^y_{yx} = \sigma^y_{xy} \neq 0.
\end{align}
In the main text, we focused on $\sigma^y_{xx}$ component.

\subsection{Finite Broadening at Low-Frequency}\label{lorentzian}
We point out that the non-zero shift current conductivity at low-frequency is due to finite Lorentzian broadening. In computing shift current conductivity, we need to evaluate $\delta(\omega - \epsilon_{mn})$ in Eq. \ref{sigma}. We choose to approximate the delta function using 
\begin{align}
    \delta(x) \approx \frac{\gamma}{\pi(x^2 + \gamma^2)},
\end{align}
where $\gamma$ characterizes the width of the function. We choose $\gamma$ to be the average energy spacing of the participating bands between nearby k-points in the Brillouin zone. This large choice of $\gamma$ ensures lack of spurious (due to a finite mesh size) oscillations in the conductivity from Fig. \ref{fig1}e at finite frequencies, however it also renders our conductivity finate as $\omega\to 0$ where the contributions stem from the two closely spaced flat bands. With higher mesh resolution, conductivity as $\omega\to 0$ vanishes as required.

\end{document}